\documentclass[12pt]{iopart}

\usepackage[utf8]{inputenc}

\usepackage{graphicx}% Include figure files
\usepackage{hyperref}
\usepackage{amssymb}
\usepackage{accents}

\usepackage{etoolbox}
\makeatletter
\def\@mkboth#1#2{}
\newlength\appendixwidth
\preto\appendix{\addtocontents{toc}{\protect\patchl@section}}
\newcommand{\patchl@section}{%
  \settowidth{\appendixwidth}{\textbf{Appendix }}%
  \addtolength{\appendixwidth}{1.5em}%
  \patchcmd{\l@section}{1.5em}{\appendixwidth}{}{\ddt}%
}
\makeatother
\appto\appendix{\addtocontents{toc}{\protect\setcounter{tocdepth}{1}}}
\appto\listoffigures{\addtocontents{lof}{\protect\setcounter{tocdepth}{1}}}
\appto\listoftables{\addtocontents{lot}{\protect\setcounter{tocdepth}{1}}}
\begin{document}

\title[The initial value problem as it relates to numerical relativity]
{The initial value problem as it relates to numerical relativity}

\author{Wolfgang Tichy}

\address{Department of Physics, Florida Atlantic University,
         Boca Raton, FL 33431, USA}

%\date{$$Id$$}

%-------------------------------------------------------------------------
%Useful Definitions
%------------------------------------------------------------------------
%
\newcommand\be{\begin{equation}}
\newcommand\ba{\begin{eqnarray}}

\newcommand\ee{\end{equation}}
\newcommand\ea{\end{eqnarray}}

\newcommand\p{{\partial}}
\newcommand{\ut}[1]{\undertilde{#1}}

\newcommand\remove{{{\bf{THIS FIG. OR EQS. COULD BE REMOVED}}}}
%

%-------------------------------------------------------------------------
\begin{abstract}
%-----------------------------------------------------------------------

Spacetime is foliated by spatial
hypersurfaces in the 3+1 split of General Relativity.
The initial value problem then consists of specifying initial data
for all fields on one such a spatial hypersurface, such that the subsequent
evolution forward in time is fully determined. On each hypersurface the
3-metric and extrinsic curvature describe the geometry. Together with matter
fields such as fluid velocity, energy density and rest mass density, the
3-metric and extrinsic curvature then constitute the initial data. There is
a lot of freedom in choosing such initial data. This freedom corresponds to
the physical state of the system at the initial time. At the same time the
initial data have to satisfy the Hamiltonian and momentum constraint
equations of General Relativity and can thus not be chosen completely
freely. We discuss the conformal transverse traceless and conformal thin
sandwich decompositions that are commonly used in the construction of
constraint satisfying initial data. These decompositions allow us to specify
certain free data that describe the physical nature of the system. The
remaining metric fields are then determined by solving elliptic equations
derived from the constraint equations. We describe initial data for
single black holes and single neutron stars, and how we can use conformal
decompositions to construct initial data for binaries made up of black holes
or neutron stars. Orbiting binaries will emit gravitational radiation and
thus lose energy. Since the emitted radiation tends to circularize the
orbits over time, one can thus expect that the objects in a typical binary
move on almost circular orbits with slowly shrinking radii. This leads us to
the concept of quasi-equilibrium which essentially assumes that time
derivatives are negligible in corotating coordinates, for binaries on almost
circular orbits. We review how quasi-equilibrium assumptions can be used to
make physically well motivated approximations that simplify the elliptic
equations we have to solve. 
\\


%-----------------------------------------------------------------------
\end{abstract}
%-----------------------------------------------------------------------


% Sometimes we want to include preprint numbers, let's put them here
%\preprint{???}

\maketitle

\tableofcontents

%%%%%%%%%%%%%%%%%%%%%%%%%%%%%%%%%%%%%%%%%%%%%%%%%%%%%%%%%%%
\section{Introduction}
%%%%%%%%%%%%%%%%%%%%%%%%%%%%%%%%%%%%%%%%%%%%%%%%%%%%%%%%%%%
\label{Intro}

After the first direct detection of a gravitational wave signal emitted by a
binary black holes system~\cite{Abbott:2016blz,TheLIGOScientific:2016qqj},
the problem of faithfully simulating the evolution of binary systems of
compact objects has become increasingly important. The detectors
(LIGO~\cite{LIGO:2007kva,advLIGO2015},
VIRGO~\cite{VIRGO_FAcernese_etal2008,advVIRGO2015},
GEO600~\cite{GEO_main_ref_CQG2008}) use laser interferometry to
measure the strains associated with passing gravitational
waves~\cite{Schutz99}, and offer much higher sensitivity than previous
experiments aimed at direct detection of these waves. Additional detectors
such as KAGRA~\cite{Somiya:2011np} or the space borne
eLISA/NGO~\cite{AmaroSeoane:2012je} and
DECIGO~\cite{Kawamura:2011zz} are in planning and construction stages. In
all these detectors, the measured strains are tiny and contaminated by
noise. For detection and especially parameter estimation,
it is thus necessary to compare the observed signals to theoretical
gravitational wave templates that describe different systems.

One of the most promising sources for gravitational wave detectors is the
inspiral and merger of compact objects such as black holes or neutron stars.
Due to the emission of gravitational waves, the binary loses energy and
the orbit tightens until finally the two objects merge. As long as the two
compact objects are far apart, post-Newtonian
calculations~\cite{Futamase:2007zz} can give highly accurate approximations
for the orbital motion and the gravitational waves emitted by the binary.
When the two objects get closer, the post-Newtonian expansion becomes more
and more inaccurate and eventually breaks down. Thus the full
non-linear equations of General Relativity have to be solved and computer
simulations are used to obtain numerical answers~\cite{Pretorius:2007nq}.
From each of these computer simulations we can then extract a gravitational
wave template for the simulated system. In order to start such numerical
simulations we need to specify initial data that describe the initial state
of the binary system. These initial data should be chosen as accurately and
as realistically as possible. Otherwise the subsequent numerical evolution may
not simulate the kind of system the detectors observe.

In this paper we will concentrate on how one can construct reliable initial
data.
In Sec.~\ref{NumRel} we describe the basic equations that are used in
numerical relativity. This is followed in Sec.~\ref{Conformal_decomp} by a
discussion of several conformal decompositions that are used to derive the
equations commonly used in the construction of initial data.
Sections~\ref{BH_data} and \ref{BBH_data} describe various methods to
construct initial data for single and binary black holes, and
Sec.~\ref{Realistic_BBH_data} presents possible improvements. We proceed in
Sec.~\ref{Matter} by introducing equations that are needed to describe
neutron star matter. In Secs.~\ref{NS_data} and \ref{BNS_data} we discuss
single and binary neutron star data, while Sec.~\ref{BH-NS_data} deals with
mixed binaries. Section~\ref{Codes} provides a short overview of the most
common computer codes that are used to compute initial data.
We provide some conclusions in Sec.~\ref{Conclusions}.
Throughout we will use geometric units where $G=c=1$.

%%%%%%%%%%%%%%%%%%%%%%%%%%%%%%%%%%%%%%%%%%%%%%%%%%%%%%%%%%%
\section{Numerical relativity}
%%%%%%%%%%%%%%%%%%%%%%%%%%%%%%%%%%%%%%%%%%%%%%%%%%%%%%%%%%%
\label{NumRel}

In this section we introduce the basic equations of numerical relativity. We
start by discussing some equations of General Relativity and explain how they
are reformulated for use in numerical simulations.

%%%%%%%%%%%%%%%%%%%%%%%%%%%%%%%%%%%%%%%%%%%%%%%%%%%%%%%%%%%
\subsection{General Relativity}
%%%%%%%%%%%%%%%%%%%%%%%%%%%%%%%%%%%%%%%%%%%%%%%%%%%%%%%%%%%
 
We recall here some of the basic equations of General Relativity
with the purpose of illustrating our notation.

In Relativity theory 3-dimensional space and time are unified into
4-dimensional spacetime. In the simplest case of a flat spacetime
distances in this 4-dimensional spacetime are measured as
\be
ds^2 = (dx^1)^2 + (dx^2)^2 + (dx^3)^2 - (dx^0)^2 ,
\ee
where $x^1$, $x^2$, $x^3$ are the usual Cartesian coordinates of Euclidean
space, and $x^0$ is time.
Usually this distance is written as
\be
\label{lineelement4}
ds^2 = g_{\mu\nu} dx^{\mu}dx^{\nu} ,
\ee
where the indices $\mu$ and $\nu$ run from 0 to 3, and we use the Einstein
summation convention and sum over repeated indices. Here $g_{\mu\nu}$ are
the components of a symmetric tensor called the spacetime metric or
4-metric. For flat spacetime in Cartesian coordinates we have 
$g_{\mu\nu} = \mbox{diag}(-1,1,1,1)$,
but if we change coordinates $g_{\mu\nu}$ will have different values.
The inverse of the 4-metric will be denoted as $g^{\mu\nu}$.
Throughout we will use Greek letters to denote 4-dimensional indices that
run from 0 to 3. Later when we refer to the spatial components of tensors
we will use letters from the middle of the Latin alphabet. An example is the
spatial 3-metric $\gamma_{ij} = \mbox{diag}(1,1,1)$ of flat space, where
both $i$ and $j$ run from 1 to 3.

We denote the 4-dimensional covariant derivative operator 
by $\nabla_{\alpha}$. 
It is defined by
\be
\nabla_{\alpha} g_{\mu\nu} = 0
\ee
in order to be compatible with the 4-metric. The covariant derivative of any
tensor can be given in terms of partial derivatives and the Christoffel
symbols
\be
\label{Christoffel4}
\Gamma^{\alpha}_{\mu\nu} = \frac{1}{2} g^{\alpha\rho}
(\partial_{\nu} g_{\rho\mu} + \partial_{\mu} g_{\rho\nu} - 
 \partial_{\rho} g_{\mu\nu}) .
\ee
For example, the covariant derivative of a 4-vector $V^{\mu}$
can be computed from
\be
\nabla_{\alpha}  V^{\mu} =
\partial_{\alpha} V^{\mu} + \Gamma^{\mu}_{\alpha\nu} V^{\nu} .
\ee

If matter or energy is present, spacetime will become curved and the
4-metric has to be determined from the Einstein equations
\be
\label{EinsteinEqs}
^4R_{\mu\nu} - \frac{1}{2} g_{\mu\nu} { }^4R = 8\pi T_{\mu\nu} .
\ee
Here $T_{\mu\nu}$ is the stress-energy tensor of matter. The Ricci tensor
$^4R_{\mu\nu}={}^4R_{\mu\rho\nu}^{\ \ \ \ \rho}$ and the Ricci Scalar 
$^4R={ }^4R_{\mu\nu}g^{\mu\nu}$ are both related to the Riemann
tensor $^4R^{\mu}_{\ \nu\alpha\beta}$, defined by~\cite{Wald84}
\be
\label{Riemann-def}
{}^4R_{\alpha\beta\mu}^{\ \ \ \ \nu}\omega_{\nu} :=
(\nabla_{\alpha}\nabla_{\beta}-\nabla_{\beta}\nabla_{\alpha})\omega_{\mu} ,
\ee
where $\omega_{\mu}$ is any one-form.
Indices on all tensors can be raised and lowered with the
spacetime metric $g_{\mu\nu}$. 
We will follow the sign conventions of
Misner, Thorne and Wheeler~\cite{Misner73}, so that the
Riemann tensor when computed from the Christoffel symbols in
Eq.~(\ref{Christoffel4}) is given by
\be
\label{Riemann4}
^4R^{\mu}_{\ \nu\alpha\beta} = 
 \partial_{\alpha} \Gamma^{\mu}_{\nu\beta} -
 \partial_{\beta}  \Gamma^{\mu}_{\nu\alpha} +
 \Gamma^{\mu}_{\rho\alpha} \Gamma^{\rho}_{\nu\beta} -
 \Gamma^{\mu}_{\rho\beta}  \Gamma^{\rho}_{\nu\alpha} .
\ee

As we have seen any tensor of any rank can be written using the index
notation introduced above. However, sometimes it can be advantageous to use
index free notation. We use $\vec{u}$ to denote a 4-vector $u^{\mu}$ and
$\ut{\omega}$ to denote a one-form $\omega_{\mu}$. The contraction of the
two is then written as $\vec{u}\cdot \ut{\omega} = u^{\mu} \omega_{\mu}$.

%%%%%%%%%%%%%%%%%%%%%%%%%%%%%%%%%%%%%%%%%%%%%%%%%%%%%%%%%%%
\subsection{The 3+1 split of spacetime}
%%%%%%%%%%%%%%%%%%%%%%%%%%%%%%%%%%%%%%%%%%%%%%%%%%%%%%%%%%%

Since both the 4-metric and the Ricci tensor are symmetric, it is clear that 
the Einstein equations in~(\ref{EinsteinEqs}) are ten independent equations
for the ten independent components of the 4-metric. 
From Eq.~(\ref{Riemann4}) we see that these ten 
equations are linear in  
the second derivatives and quadratic in the first derivatives of the 4-metric.
However, only six of these ten equations contain second time derivatives. 
These six second order in time equations represent evolution equations.
The other four are not evolution equations, but rather represent 
constraints that the 4-metric has to satisfy.

If we can solve the Einstein equations in~(\ref{EinsteinEqs}) we
obtain the 4-metric everywhere in space and time. While a solution for all
of spacetime is certainly very useful, in practice it is possible to find such
direct solutions of Eq.~(\ref{EinsteinEqs}) only in rare cases where one
assumes symmetries such as time independence and axisymmetry. In addition,
in many astrophysical problems of interest it may not be necessary to find a
solution for all of time. One usually would like to start from some given
initial conditions at some initial time and then to calculate the state of
the system at some later time. For example, for a particle in classical
mechanics one starts by giving the initial position and velocity, and then
calculates position and velocity at a later time. We would like to do the
same for General Relativity and specify initial data (i.e. some fields and 
their time derivatives) at some initial time and then calculate these
fields at a later time.
In order to do this we use the 3+1 split of spacetime~\cite{Arnowitt62}:
We foliate spacetime by spacelike 3-dimensional hypersurfaces or slices,
i.e. surfaces with a normal vector
$n^{\mu}$ that is timelike. As time coordinate $x^0$ we use a function $t$
which is constant on each hypersurface, but increases as we go from one
hypersurface to the next. We then define the lapse by
\be
\label{lapsedef}
\alpha := \sqrt{\frac{-1}{g^{\mu\nu} \partial_{\mu} t \partial_{\nu} t}}
 = \sqrt{\frac{-1}{g^{00}}} .
\ee
The normal vector thus must satisfy
\be
\label{ndef}
n_{\mu} = -\alpha \partial_{\mu} t = (-\alpha,0,0,0)
\ee
in order for it to be normalized such that
\be
g^{\mu\nu}n_{\mu}n_{\nu}=-1 .
\ee
The minus
sign in Eq.~(\ref{ndef}) is chosen such that $n^{\mu}$ points in the future
direction of increasing $t$.
Notice that while $n^{\mu}$ is orthogonal to each hypersurface, $n^{\mu}$ is
not in general tangent to lines of constant spatial coordinates 
$x^i = const$, since we are still free to choose arbitrary
spatial coordinates on each hypersurface. Indeed
the tangent vector $t^{\mu}$ to lines of constant spatial coordinates
is given by 
\be
\label{tvecdef}
t^{\mu} = (\partial_t)^{\mu} = (1,0,0,0) ,
\ee 
where we have normalized $t^{\mu}$ such that
\be
t^{\mu} \partial_{\mu} t = 1 .
\ee
The latter also implies
\be
t^{\mu} n_{\mu} = -\alpha
\ee
from which it follows that
\be
\label{tvec}
t^{\mu} = \alpha n^{\mu} + \beta^{\mu} ,
\ee
where $\beta^{\mu}$ is an arbitrary spatial vector,
i.e. any vector that satisfies $\beta^{\mu} n_{\mu} = 0$. Once
the foliation is given the lapse $\alpha$ and the normal vector
$n^{\mu}$ are fixed. Yet the direction of the vector
$t^{\mu}$ still depends on the choice of spatial coordinates on each
hypersurface. This freedom to choose spatial coordinates
is encapsulated in the spatial vector $\beta^{\mu} = (0,\beta^i)$, 
which is called the shift vector.

%%%%%%%%%%%%%%%%%%%%%%%%%%%%%%%%%%%%%%%%%%%%%%%%%%%%%%%%%%%%%%%%%%%%%
\subsection{3-metric, spatial covariant derivative and extrinsic curvature}
%%%%%%%%%%%%%%%%%%%%%%%%%%%%%%%%%%%%%%%%%%%%%%%%%%%%%%%%%%%%%%%%%%%%%

We now define the 3-metric as 
\be
\label{metric3def}
\gamma_{\mu\nu} = g_{\mu\nu} + n_{\mu} n_{\nu}
\ee
which is the projection of the 4-metric onto the hypersurface with normal
vector $n^{\mu}$. This is a proper 4-tensor whose indices can be raised and
lowered with the 4-metric $g_{\mu\nu}$.
Since $n_i=0$ we find $\gamma_{ij} = g_{ij}$ and
$\gamma^i_j = g^i_j = \delta^i_j$. From Eq.~(\ref{metric3def}) we also
find $\gamma^{0\nu}=\gamma^{\mu\nu}n_{\mu}/(-\alpha)=0$
and thus 
$\gamma^{ik} \gamma_{kj} = \gamma^{i\nu} \gamma_{\nu j} = \gamma^{i\nu}
g_{\nu j} = \gamma^i_j=\delta^i_j$.

From Eqs.~(\ref{tvecdef}) and (\ref{tvec}) we find that 
\be
g_{00} = g_{\mu\nu} t^{\mu} t^{\nu} 
= -\alpha^2 + g_{\mu\nu}\beta^{\mu}\beta^{\nu}
= -\alpha^2 + \gamma_{ij}\beta^{i}\beta^{j}
\ee
and
\be
g_{0j} = g_{\mu j} t^{\mu} = -\alpha n_{j} + \beta_{j} = \beta_{j} 
\ee
where $\beta_{j} = g_{\mu j} \beta^{\mu} = \gamma_{ij}\beta^{i}$.
Using the latter two equations, the line element 
in Eq.~(\ref{lineelement4}) can now be written as
\be
ds^2 = -\alpha^2 dt^2 + \gamma_{ij} (dx^i + \beta^i dt)(dx^j + \beta^j dt) .
\ee

The 3-metric defined in Eq.~(\ref{metric3def}) can be used to project any
tensor $T^{\alpha_1\alpha_2...}_{\ \ \ \ \ \beta_1\beta_2...}$
onto the spatial hypersurface. The resulting tensor
$
S^{\alpha_1\alpha_2...}_{\ \ \ \ \ \beta_1\beta_2...} =
\gamma^{\alpha_1}_{\gamma_1}\gamma^{\alpha_2}_{\gamma_2}...
\gamma_{\beta_1}^{\delta_1}\gamma_{\beta_2}^{\delta_2}...
T^{\gamma_1\gamma_2...}_{\ \ \ \ \ \delta_1\delta_2...}
$
is called a spatial tensor since it is orthogonal to $n_{\alpha}$ in each
index.
Note that the indices on any spatial tensor can be raised or lowered by
using either the 3-metric $\gamma_{ij}$ or the 4-metric $g_{\mu\nu}$.
The result will be the same since $\gamma_{\mu\nu}n^{\nu}=0$.

We define the spatial covariant derivative of any spatial tensor
$S^{\alpha_1\alpha_2...}_{\ \ \ \ \ \beta_1\beta_2...}$
as
\be
D_{\mu}S^{\alpha_1\alpha_2...}_{\ \ \ \ \ \beta_1\beta_2...} :=
\gamma_{\mu}^{\rho}
\gamma^{\alpha_1}_{\gamma_1}\gamma^{\alpha_2}_{\gamma_2}...
\gamma_{\beta_1}^{\delta_1}\gamma_{\beta_2}^{\delta_2}...
\nabla_{\rho}S^{\gamma_1\gamma_2...}_{\ \ \ \ \ \delta_1\delta_2...},
\ee
where we have projected all free indices onto the spatial hypersurface.
From Eq.~(\ref{metric3def}) it immediately follows that
\be
D_{\mu} \gamma_{\alpha\beta} = 0 ,
\ee
so that the spatial covariant derivative operator $D_{\mu}$ is compatible
with the 3-metric $\gamma_{\alpha\beta}$. This also implies that
the spatial covariant derivative
$D_i S^{k_1 k_2...}_{\ \ \ \ \ l_1 l_2...}$
of any spatial tensor $S^{k_1 k_2...}_{\ \ \ \ \ l_1 l_2...}$ can be
computed with the help of 3-dimensional Christoffel symbols, which can be
computed from the 3-metric $\gamma_{ij}$.

In order to describe the curvature of a $t=const$ hypersurface one
introduces the extrinsic curvature
\be
\label{K-def}
K_{\mu\nu} :=
- \gamma^{\alpha}_{\mu} \gamma^{\beta}_{\nu} \nabla_{\alpha} n_{\beta} .
\ee
It can be shown that $K_{\mu\nu}$ is symmetric and also given by
\be
\label{K_Lie_n}
K_{\mu\nu} = -\frac{1}{2}\pounds_{n}\gamma_{\mu\nu} ,
\ee
where $\pounds_{n}$ is the  Lie derivative (see \ref{Lie-appendix})
along the vector $n^{\mu}$. 
We can thus express $K_{ij}$ also as
(see \ref{LieDerivs_in_3+1})
\be
\label{K-of-g}
K_{ij}
= -\frac{1}{2\alpha}(\partial_t\gamma_{ij} - D_i\beta_j - D_j\beta_i) ,
\ee
which tells us how time derivatives of the 3-metric are related to the
extrinsic curvature.
From its definition in Eq.~(\ref{K-def}) it is clear that the extrinsic
curvature tells us how the normal vector $n^{\mu}$ changes across a
$t=const$ hypersurface. 
It thus measures how this hypersurface is curved with respect to the
4-dimensional space it is embedded in. 
However, the 3-dimensional space within each hypersurface
can be curved as well. This so called intrinsic curvature is described by
the 3-dimensional Riemann tensor $R^{i}_{\ jkl}$ that can be computed from
the 3-metric $\gamma_{ij}$ and its resulting 3-dimensional Christoffel
symbols using Eqs.~(\ref{Christoffel4}) and (\ref{Riemann4}) for
$\gamma_{ij}$ instead of $g_{\mu\nu}$. We will see below how the curvature
of spacetime described by the 4-dimensional Riemann tensor 
$^4R^{\mu}_{\ \nu\alpha\beta}$ is related to the extrinsic 
curvature $K_{ij}$ and the intrinsic curvature $R^{i}_{\ jkl}$.
Also notice that in three dimensions, the Riemann tensor has the same number
of independent components and carries the same information as the
3-dimensional Ricci tensor $R_{jl}=R^{i}_{\ jil}$.

%%%%%%%%%%%%%%%%%%%%%%%%%%%%%%%%%%%%%%%%%%%%%%%%%%%%%%%%%%%
\subsection{Einstein's equations and the 3+1 split}
%%%%%%%%%%%%%%%%%%%%%%%%%%%%%%%%%%%%%%%%%%%%%%%%%%%%%%%%%%%

Going back to our example of a 
particle in classical mechanics, we see that the 3-metric $\gamma_{ij}$
is analogous to the particle position, while the extrinsic curvature
$K_{ij}$ is analogous to its velocity. Of course, in order to compute the
time evolution of a particle we also need the equation of motion 
that tells us the time derivative of the velocity. In our case
we will use Einsteins equations to find additional equations that
tell us how $K_{ij}$ evolves in time. This is done by relating
the 4-dimensional Riemann tensor $^4R_{\alpha\beta\mu\nu}$
to the 3-dimensional Riemann tensor $R_{\alpha\beta\mu\nu}$
with the help of the Gauss-Codazzi and Codazzi-Mainardi relations
given by (see e.g. \cite{Alcubierre_book} or \cite{gourgoulhon20123+1})
\be
\label{Gauss-Codazzi}
R_{\alpha\beta\mu\nu} +
K_{\alpha\mu}K_{\beta\nu} -
K_{\alpha\nu}K_{\beta\mu}
= \gamma_{\alpha}^{\alpha'} \gamma_{\beta}^{\beta'}
\gamma_{\mu}^{\mu'} \gamma_{\nu}^{\nu'}
{}^4R_{\alpha'\beta'\mu'\nu'}
\ee
and
\be
\label{Codazzi-Mainardi}
D_{\beta}K_{\alpha\mu} - D_{\alpha}K_{\beta\mu} =
\gamma_{\alpha}^{\alpha'} \gamma_{\beta}^{\beta'} \gamma_{\mu}^{\mu'}
{}^4R_{\alpha'\beta'\mu'\nu} n^{\nu} .
\ee
One can also show that
\be
\label{nabla_n}
\nabla_{\mu} n_{\nu} 
= -K_{\mu\nu} - 
  n_{\mu} (\nabla_{\nu} + n_{\nu}n^{\rho}\nabla_{\rho})\ln\alpha
= -K_{\mu\nu} - n_{\mu} D_{\nu}\ln\alpha .
\ee
By looking at the different projections of the Einstein equations
(\ref{EinsteinEqs}) onto $n^{\mu}$ and $\gamma^{\mu\nu}$ and
using Eqs.~(\ref{Gauss-Codazzi}), (\ref{Codazzi-Mainardi}), (\ref{nabla_n}) 
and (\ref{Riemann-def}) we find that
Einstein's equations split into the evolution equations
\begin{eqnarray}
\label{gamma_ij-evo}
\partial_t \gamma_{ij} &=& -2\alpha K_{ij} + \pounds_{\beta} \gamma_{ij} \\
\label{K_ij-evo}
\partial_t K_{ij} &=& \alpha (R_{ij} - 2 K_{il} K^l_j + K K_{ij})
 - D_i  D_j \alpha + \pounds_{\beta} K_{ij} \nonumber \\
 &&- 8\pi\alpha S_{ij} + 4\pi\alpha\gamma_{ij}(S-\rho)
\end{eqnarray}
and the so called Hamiltonian and momentum constraint equations
\begin{eqnarray}
\label{ham0}
R -  K_{ij}  K^{ij} + K^2   &=& 16\pi\rho  \\
\label{mom0}
D_j(K^{ij} - \gamma^{ij} K) &=& 8\pi j^i .
\end{eqnarray}
Here $R_{ij}$ and $R$ are the Ricci tensor and scalar computed from
$\gamma_{ij}$, $D_i$ is the derivative operator compatible 
with $\gamma_{ij}$ and all indices here are raised and lowered
with the 3-metric $\gamma_{ij}$.
The source terms $\rho$, $j^i$, $S_{ij}$ and 
$S=\gamma^{ij}S_{ij}$ are projections of the stress-energy 
tensor $T_{\mu\nu}$ given by
\begin{eqnarray}
\label{mattervars}
\rho   &=& T_{\mu\nu} n^{\mu} n^{\nu} \nonumber \\
j^i    &=& -T_{\mu\nu} n^{\mu} \gamma^{\nu i} \nonumber \\
S^{ij} &=& T_{\mu\nu} \gamma^{\mu i} \gamma^{\nu j}
\end{eqnarray}
and correspond to the energy density, flux and stress-tensor seen by an
observer moving with 4-velocity $n^{\mu}$.

%%%%%%%%%%%%%%%%%%%%%%%%%%%%%%%%%%%%%%%%%%%%%%%%%%%%%%%%%%%
\subsection{Constraints, gauge freedom and physical degrees of freedom}
%%%%%%%%%%%%%%%%%%%%%%%%%%%%%%%%%%%%%%%%%%%%%%%%%%%%%%%%%%%

The evolution equations in Eqs.~(\ref{gamma_ij-evo}) and
(\ref{K_ij-evo}) tell us how we can obtain
$\gamma_{ij}$ and $K_{ij}$ at any time, if we are given $\gamma_{ij}$ and
$K_{ij}$ at some initial time. As one can see, the constraint equations
(\ref{ham0}) and (\ref{mom0}) do not contain any time derivatives and thus
are equations that need to be satisfied on each spatial hypersurface. Thus
when we construct initial data, we have to ensure that both $\gamma_{ij}$
and $K_{ij}$ satisfy the constraint equations at the initial time. It can then
be shown that the evolution equations will preserve the constraints. For this
reason constructing initial data is a non-trivial task in the sense that we
cannot freely choose the 12 fields $\gamma_{ij}$ and $K_{ij}$. Rather
these fields are subject to the 4 constraints in Eqs.~(\ref{ham0}) and
(\ref{mom0}). This leaves us with only 8 freely specifiable fields.
However, in General Relativity we are free to choose any coordinates we
like. Since there are 4 coordinates that we can freely choose,
the 8 freely specifiable fields in $\gamma_{ij}$ and $K_{ij}$
can really contain only 4 physical fields that are independent of
our coordinate choice. Since we specify both $\gamma_{ij}$ and its time
derivative $K_{ij}$ (see Eq.~(\ref{K-of-g})) 2 of these 4 fields are merely
time derivatives of the other 2 fields. Thus we are dealing with
only 2 physical degrees of freedom, as expected for gravity.

The freedom to choose coordinates is usually referred to as gauge freedom
and connected to the choice of lapse $\alpha$ shift $\beta^i$. If we only
change the 3 spatial coordinates, all components of $\gamma_{ij}$ and
$K_{ij}$ as well as the shift will change in general, but the spatial
hypersurfaces themselves are unchanged, so that the normal vector $n^{\mu}$
and the lapse $\alpha$ will not change. If we change the time coordinate
$t$, the hypersurfaces themselves will also change so that now $n^{\mu}$
and the lapse $\alpha$ will change as well~\cite{Hilditch:2015qea}.

%%%%%%%%%%%%%%%%%%%%%%%%%%%%%%%%%%%%%%%%%%%%%%%%%%%%%%%%%%%%%%%%%%%
\section{Conformal decompositions and initial data construction}
%%%%%%%%%%%%%%%%%%%%%%%%%%%%%%%%%%%%%%%%%%%%%%%%%%%%%%%%%%%%%%%%%%%
\label{Conformal_decomp}

As already mentioned, the initial data $\gamma_{ij}$ and $K_{ij}$
cannot be freely chosen, since they must satisfy the constraint equations
(\ref{ham0}) and (\ref{mom0}). However, since the constraint equations alone
cannot determine the initial data either, Eqs.~(\ref{ham0}) and (\ref{mom0})
do not have unique solutions and are thus not directly useful in constructing
initial data. For this reason so called conformal decompositions have been
developed. Using these decompositions it is possible to start by specifying 
arbitrary $\gamma_{ij}$ and $K_{ij}$ that may not satisfy the constraints.
In a second step these $\gamma_{ij}$ and $K_{ij}$ are then modified in such
a way that they satisfy the constraints. This second step involves solving
certain elliptic equations for 4 auxiliary quantities that are used
to compute the final constraint satisfying $\gamma_{ij}$ and $K_{ij}$.
These elliptic equations have the great advantage that they
generally have unique solutions once appropriate boundary conditions are
specified (see \ref{Elliptic-appendix}). Except for
simple cases with many symmetries these elliptic equations are usually
solved numerically~\cite{Cook00a}.

Following Lichnerowicz~\cite{Lichnerowicz44} and York~\cite{York71,York72}
we start by decomposing the 3-metric $\gamma_{ij}$ into a conformal
factor $\psi$ and a conformal metric $\bar{\gamma}_{ij}$ such that
\begin{equation}
\label{conf-metric}
\gamma_{ij} = \psi^4 \bar{\gamma}_{ij} .
\end{equation}
Using this decomposition the Ricci scalar can be written as
\be
\label{R-barR}
R = \psi^{-4} \bar{R} - 8\psi^{-5} \bar{D}_k\bar{D}^k \psi .
\ee
Here $\bar{D}_k$ is the derivative operator compatible
with the conformal metric $\bar{\gamma}_{ij}$ and $\bar{R}$ the
Ricci scalar computed from $\bar{\gamma}_{ij}$. 
It is also possible to verify
that for any symmetric and tracefree tensor $M^{ij}$ we
have~\cite{York-Piran-1982-in-Schild-lectures}
\be
\label{DM-barDM}
D_j M^{ij} = \psi^{-10} \bar{D}_j \bar{M}^{ij}
\ee
if we define
\be
\bar{M}^{ij} = \psi^{10} M^{ij} .
\ee
Let us also introduce a symmetric tracefree differential operator $L$.
It is defined by its action on any vector $W^i$ by
\be
\label{LW-def}
(LW)^{ij} :=
D^i W^j + D^j W^i - \frac{2}{3} \gamma^{ij} D_k W^k .
\ee
It is sometimes referred to as conformal Killing form since it is related to
the Lie derivative of the tensor density $\gamma^{-1/3} \gamma_{ij}$
of weight $-2/3$, where $\gamma = \det(\gamma_{ij})$.
Using Eq.~(\ref{Lie_tensordensity}) we find
\be
\pounds_W (\gamma^{-1/3} \gamma_{ij}) = \gamma^{-1/3} (LW)_{ij} .
\ee
So if $W^i$ is a Killing vector of the metric $\gamma^{-1/3} \gamma_{ij}$
with $\pounds_W (\gamma^{-1/3} \gamma_{ij})=0$,
we will have $(LW)^{ij}=0$. Notice that in general 
$\gamma^{-1/3} \gamma_{ij}$ is not the same as the conformal 
metric $\bar{\gamma}_{ij}$.

One can show that 
\be
\label{LW-confLW}
(LW)^{ij} =  \psi^{-4} (\bar{L}W)^{ij},
\ee
where 
$
(\bar{L}W)^{ij} :=
\bar{D}^i W^j + \bar{D}^j W^i - \frac{2}{3} \bar{\gamma}^{ij} \bar{D}_k W^k
$
is computed using the conformal metric.

Next, the extrinsic curvature is split into its trace 
$K=\gamma_{ij}K^{ij}$ and its tracefree part $A^{ij}$ by writing it as
\begin{equation}
\label{tracefreeA}
K^{ij} = A^{ij} + \frac{1}{3} \gamma^{ij} K .
\end{equation}
The tracefree part is rescaled as
\be
\label{A-rescale}
A^{ij} = \psi^{-10} \bar{A}^{ij} .
\ee
Note that the factor of $\psi^{-10}$ in Eq.~(\ref{A-rescale})
has been picked so that Eq.~(\ref{DM-barDM}) applies to $A^{ij}$.

Inserting Eqs.~(\ref{R-barR}), (\ref{tracefreeA})
and (\ref{A-rescale}) into the Hamiltonian constraint (\ref{ham0})
yields
\be
\label{hamCTT}
\label{hamCTS} % we have same eqn in CTS as well
8\bar{D}_k\bar{D}^k \psi - \bar{R}\psi + 
\psi^{-7} \bar{A}^{ij} \bar{A}_{ij} - \frac{2}{3} \psi^5 K
= -16\pi\psi^5\rho ,
\ee
where indices on the barred quantities are raised and lowered with
$\bar{\gamma}_{ij}$. We will use this equation for the Hamiltonian
constraint to compute the conformal factor in the different conformal
decompositions discussed below. Note, however, that these decompositions
differ in how $\bar{A}^{ij}$ is split further.

%%%%%%%%%%%%%%%%%%%%%%%%%%%%%%%%%%%%%%%%%%%%%%%%%%%%%%%%%%%%%%%%%%%
\subsection{The conformal transverse traceless decomposition}
%%%%%%%%%%%%%%%%%%%%%%%%%%%%%%%%%%%%%%%%%%%%%%%%%%%%%%%%%%%%%%%%%%%
\label{CTT_decomp}

We consider first the conformal transverse traceless (CTT) decomposition,
where we split~\cite{Pfeiffer:2002iy}
\be
\label{Abar-split}
\bar{A}^{ij}
= \bar{M}^{ij} + \frac{1}{\bar{\sigma}}(\bar{L}W)^{ij}
\ee
in two pieces.
Here $\bar{M}^{ij}$ is tracefree but otherwise arbitrary 
and $\bar{\sigma}$ is some positive weighting factor that we can choose.
The vector $W^i$ will be computed below.

The Hamiltonian constraint is written as in Eq.~(\ref{hamCTT}) with
$\bar{A}^{ij}$ defined as in Eq.~(\ref{Abar-split}).
When we insert Eqs.~(\ref{tracefreeA}), (\ref{A-rescale}) 
and (\ref{Abar-split}) 
into the momentum constraint (\ref{mom0}) we find
\be
\label{momCTT}
\bar{D}_j \left[\frac{1}{\bar{\sigma}}(\bar{L}W)^{ij}\right] +
\bar{D}_j\bar{M}^{ij} - \frac{2}{3}\psi^6\bar{D}^i K = 8\pi\psi^{10} j^i .
\ee
Both Eqs.~(\ref{hamCTT}) and (\ref{momCTT}) are elliptic equations. The
weighting factor is often simply set to $\bar{\sigma}=1$. In this case we
obtain the standard CTT decomposition often called the 
York-Lichnerowicz decomposition~\cite{Lichnerowicz44,York71,York72}.
However, other choices are possible. For example, for
$\bar{\sigma}=\psi^{-6}$ we obtain the so called physical CTT 
decomposition~\cite{Cook00a}.

Let us now discuss how Eqs.~(\ref{hamCTT}) and (\ref{momCTT})
are used in practice. We start with some reasonable guess or approximation
for the physical situation we want to describe. For example, if we want to
find initial data for two black holes we could simply take the superposition
of the 3-metrics and extrinsic curvatures for two single black hole
solutions~\footnote{The superposition of two asymptotically flat 3-metrics can
be obtained by adding them and then subtracting the flat metric. For the
extrinsic curvature the superposition is a simple sum.}.
Since the superposition principle does not hold in a non-linear theory like
General Relativity, this superposition will at best be an approximate solution
that we denote here by $\gamma_{ij}^{approx}$ and $K^{ij}_{approx}$.
Hence $\gamma_{ij}^{approx}$ and $K^{ij}_{approx}$ will not satisfy the
Hamiltonian and momentum constraints. However, using the above
CTT decomposition we can now construct a
$\gamma_{ij}$ and $K^{ij}$ that will satisfy the constraints if we set
\begin{eqnarray}
\bar{\gamma}_{ij} &=& \gamma_{ij}^{approx} , \nonumber \\
K		&=& K_{approx} =
		\gamma_{ij}^{approx}K^{ij}_{approx} , \nonumber\\
\label{approx_as_free_data}
\bar{M}^{ij}	&=& K^{ij}_{approx} - 
		\frac{1}{3}\gamma^{ij}_{approx} K_{approx} ,
\end{eqnarray}
and then solve Eqs.~(\ref{hamCTT}) and (\ref{momCTT}) for
$\psi$ and $W^i$. Once we have these solutions we can use
Eq.~(\ref{conf-metric}) and Eqs.~(\ref{tracefreeA}), (\ref{A-rescale})
and (\ref{Abar-split}) to obtain a $\gamma_{ij}$ and
$K^{ij}$ that are now guaranteed to satisfy the Hamiltonian and momentum
constraints.

%%%%%%%%%%%%%%%%%%%%%%%%%%%%%%%%%%%%%%%%%%%%%%%%%%%%%%%%%%%%%%%%%%%
\subsection{The conformal thin sandwich approach}
%%%%%%%%%%%%%%%%%%%%%%%%%%%%%%%%%%%%%%%%%%%%%%%%%%%%%%%%%%%%%%%%%%%
\label{CTS_decomp}

A form of the conformal thin sandwich formalism (called the Wilson-Mathews
approach~\cite{Wilson95,Wilson:1996ty}) was first introduced under the
assumption that the conformal metric is flat, i.e.
$\bar{\gamma}_{ij}=\delta_{ij}$. Here we relax this assumption and present
the more general conformal thin sandwich (CTS) decomposition proposed by
York~\cite{York99}. Like the CTT decomposition, it starts again with the
conformal metric decomposition in Eq.~(\ref{conf-metric}). However, in order
to have a direct handle on time derivatives one also introduces
\be
\label{bar_u-def}
\bar{u}_{ij} := \partial_t \bar{\gamma}_{ij}
\ee
and demands that
\be
\label{conf_det_const}
\partial_t \bar\gamma = 0 ,
\ee
i.e. that the time evolution of conformal factor be chosen such
that the determinant of the conformal metric $\bar\gamma$ is
instantaneously constant. The latter condition yields
\be
\bar{u}_{ij} = \psi^{-4} (\partial_t\gamma_{ij} -
\frac{1}{3}\gamma_{ij}\partial_t\ln\gamma) ,
\ee
where $\gamma = \det(\gamma_{ij})$.
Together with $\partial_t\ln\gamma = \gamma^{ij}\partial_t\gamma_{ij}$,
Eqs.~(\ref{K-of-g}) and (\ref{tracefreeA}) we obtain
\be
\label{bar_u1}
\bar{u}_{ij} = \psi^{-4} (-2\alpha A_{ij} + (L\beta)_{ij}) .
\ee
Using Eq.~(\ref{LW-confLW}) for $W^i=\beta^i$, 
Eq.~(\ref{bar_u1}) can be rewritten as 
\be
\label{A_CTS}
\bar{A}^{ij} = \frac{1}{2\bar{\alpha}}
\left[ (\bar{L}\beta)^{ij} - \bar{u}^{ij} \right] ,
\ee
where we have defined
\be
\label{alpha_bar}
\bar{\alpha} = \psi^{-6} \alpha .
\ee

Inserting Eqs.~(\ref{R-barR}), (\ref{tracefreeA})
and (\ref{A_CTS}) into the Hamiltonian constraint (\ref{ham0})
yields again Eq.~(\ref{hamCTT}), but this time with $\bar{A}^{ij}$
defined according Eq.~(\ref{A_CTS}).
When we insert Eqs.~(\ref{tracefreeA})
and (\ref{A_CTS}) into the momentum constraint (\ref{mom0})
we find
\be
\label{momCTS}
\bar{D}_j \left[\frac{1}{2\bar{\alpha}}
\left\{(\bar{L}\beta)^{ij} -\bar{u}^{ij}\right\}\right] - 
\frac{2}{3}\psi^6\bar{D}^i K = 8\pi\psi^{10} j^i .
\ee
From Eqs.~(\ref{A_CTS}) and (\ref{momCTS}) we can see that the 
CTS equations can be obtained from the CTT equations by setting
\begin{eqnarray}
\bar{\sigma}	&=& 2\bar{\alpha} , \nonumber \\
W^i 		&=& \beta^i , \nonumber \\
\bar{M}^{ij} &=& -\frac{\bar{u}^{ij}}{2\bar{\alpha}} .
\end{eqnarray}

Up to this point the CTS approach may not seem to
differ very much from the conformal transverse traceless approach. Once we
specify $\bar{\gamma}_{ij}$, $\bar{u}^{ij}$, $K$ and $\bar{\alpha}$, we can
solve Eqs.~(\ref{hamCTT}) and (\ref{momCTS}) for $\psi$ and $\beta^i$ and
then use Eq.~(\ref{conf-metric}) and Eqs.~(\ref{tracefreeA}) and
(\ref{A_CTS}) to obtain a $\gamma_{ij}$ and $K^{ij}$ that are now guaranteed
to satisfy the Hamiltonian and momentum constraints. Notice, however, that
we do not just obtain initial data for the 3-metric and extrinsic
curvature in this way. We also obtain a preferred
lapse $\alpha = \psi^{6}\bar{\alpha}$ and shift $\beta^i$.

So far $\bar{\alpha}$ was an arbitrarily chosen function. It is possible to
relate it to the time derivative of $K$.
From Eqs.~(\ref{K_ij-evo}) and (\ref{ham0}) we find that
\be
\label{K-evo}
\partial_t K = \beta^i D_i K - D_i D^i \alpha +
\alpha \left(A^{ij} A_{ij} + \frac{1}{3}K^2\right) +4\pi\alpha (S+\rho) .
\ee
Thus the lapse $\alpha$ is related to $\partial_t K$. Using
Eqs.~(\ref{alpha_bar}) and (\ref{hamCTT}), Eq.~(\ref{K-evo}) can be
rewritten as an elliptic equation for $\bar{\alpha}$. We find
\begin{eqnarray}
\label{alphabar_CTS}
\bar{D}_i\bar{D}^i \bar{\alpha} &+&
\bar{\alpha}\left[ 
\frac{3}{4}\bar{R} - \frac{7}{4\psi^8}\bar{A}^{ij}\bar{A}_{ij} +
\frac{\psi^4}{6}K^2 + 42 (\bar{D}_i\ln\psi)(\bar{D}^i\ln\psi) 
\right] \nonumber \\
&& +14\bar{D}_i\bar{\alpha}\bar{D}^i\ln\psi +
\psi^{-2} (\partial_t K - \beta^i \bar{D}_i K ) 
= 4\pi\bar{\alpha}\psi^4 (S+4\rho) .
\end{eqnarray}
With the latter it is now possible to specify $\partial_t K$ instead
of $\bar{\alpha}$. The fact that we can directly specify the time 
derivatives $\bar{u}^{ij}$ and $\partial_t K$
is usually exploited when one wants to construct initial data in
quasi-equilibrium situations, where a coordinate system exists in which the
time derivatives of 3-metric and extrinsic curvature either vanish or are
small in some way, as we will discuss below.

Often Eq.~(\ref{alphabar_CTS}) for $\bar{\alpha}$ is rewritten as an
equation for $\psi\alpha$ and thus $\alpha$.
Using Eqs.~(\ref{alpha_bar}), (\ref{hamCTS}) and (\ref{alphabar_CTS})
we obtain
\begin{eqnarray}
\label{alpha_CTS}
\bar{D}_i\bar{D}^i (\psi\alpha) &-&
\psi\alpha\left[
\frac{1}{8}\bar{R} + \frac{5}{12}\psi^4 K^2 + 
\frac{7}{8}\psi^{-8}\bar{A}^{ij}\bar{A}_{ij} +
2\pi\psi^{4}(\rho+2S) \right]
\nonumber\\
&& = -\psi^{5} (\partial_t K - \beta^i \bar{D}_i K ) .
\end{eqnarray}

%%%%%%%%%%%%%%%%%%%%%%%%%%%%%%%%%%%%%%%%%%%%%%%%%%%%%%%%%%%%%%%%%%%
\subsection{Boundary conditions at spatial infinity}
%%%%%%%%%%%%%%%%%%%%%%%%%%%%%%%%%%%%%%%%%%%%%%%%%%%%%%%%%%%%%%%%%%%
\label{BCs_at_inf}

The conformal decompositions discussed above result in elliptic equations
that require boundary conditions. In the case of the CTT
decomposition we have to solve Eqs.~(\ref{hamCTT}) and (\ref{momCTT}) for
$\psi$ and $W^i$.
This is usually done in asymptotically inertial coordinates, i.e.
coordinates such that the 4-metric approaches
the Minkowski metric $\mbox{diag}(-1,1,1,1)$ at spatial infinity
which we denote by $r \to \infty$. In this case we obtain
\be
\label{inertialMink}
\lim_{r \to \infty} \alpha = 1, \ \ \ 
\lim_{r \to \infty} \beta^i = 0, \ \ \
\lim_{r \to \infty} \gamma_{ij} = \delta_{ij}, \ \ \
\lim_{r \to \infty} K^{ij} = 0 .
\ee
Usually we also choose
\be
\lim_{r \to \infty}\bar{\gamma}_{ij}=\delta_{ij} ,
\ee
so that Eq.~(\ref{conf-metric}) implies
\be
\label{psi_BC_inf}
\lim_{r \to \infty} \psi = 1 .
\ee
To ensure $\lim_{r \to \infty} K^{ij} = 0$ we usually choose
\be
\lim_{r \to \infty}\bar{M}^{ij} = 0 = \lim_{r \to \infty}K ,
\ee
which together with Eqs.~(\ref{tracefreeA}), (\ref{A-rescale})
and (\ref{Abar-split}) yields
\be
\label{W_BC_inf}
\lim_{r \to \infty} W^i = 0 .
\ee
Once Eqs.~(\ref{hamCTT}) and (\ref{momCTT}) are supplemented by the
boundary conditions (\ref{psi_BC_inf}) and (\ref{W_BC_inf}) at infinity we
will obtain unique solutions for $\psi$ and $W^i$.

In the case of the CTS
decomposition we have to solve Eqs.~(\ref{hamCTS}), (\ref{momCTS}) and 
(\ref{alphabar_CTS}) for $\psi$, $\beta^i$ and $\bar{\alpha}$.
If we again work in asymptotically inertial coordinates which result in
Eq.~(\ref{inertialMink}) at spatial infinity, we find again
Eq.~(\ref{psi_BC_inf}), as well as 
\be
\label{alphabar_BC_inf}
\lim_{r \to \infty} \bar{\alpha} = 1
\ee
and
\be
\label{beta_BC_inf}
\lim_{r \to \infty} \beta^i = 0 .
\ee
However, sometimes it is convenient to work in a corotating frame, which is
obtained by changing the spatial coordinates such that they rotate with a
constant angular velocity $\Omega$ with respect to the asymptotically inertial
coordinates at infinity that are used in Eq.~(\ref{inertialMink}).
When we change coordinates to this
corotating frame the 4-metric components change such that now
\be
\label{corotMink}
\lim_{r \to \infty} \alpha = 1, \ \ \
\lim_{r \to \infty} \beta^i = \lim_{r \to \infty} \Omega\Phi^i, \ \ \
\lim_{r \to \infty} \gamma_{ij} = \delta_{ij}, \ \ \
\lim_{r \to \infty} K^{ij} = 0 .
\ee
Here $\Phi^i$ is an asymptotic rotational Killing vector
at spatial infinity.
Hence the boundary conditions for $\psi$ and $\bar{\alpha}$
are unchanged, while the boundary condition for the shift becomes
\be
\label{beta_BC_inf_rot}
\lim_{r \to \infty} \beta^i = \Omega\Phi^i .
\ee
Together with these boundary conditions the elliptic equations 
(\ref{hamCTS}), (\ref{momCTS}) and (\ref{alphabar_CTS}) have
unique solutions, provided no other boundaries are present.

Notice that the shift condition~(\ref{beta_BC_inf_rot}) can be problematic
for some numerical codes since
$\beta_i \beta^i \to \Omega^2 r^2$ as $r \to \infty$,
so that the shift becomes infinite at spatial infinity.
This problem is usually avoided by replacing the shift $\beta^i$ by
\be
\beta^i = B^i + \Omega\Phi^i
\ee
in equations such as (\ref{A_CTS}) and (\ref{momCTS}).
These equations contain the shift only in the form $(\bar{L}\beta)^{ij}$,
and since
\be
(\bar{L}\Phi)^{ij} = 0
\ee
when $\bar{\gamma}_{ij}$ is flat, Eqs.~(\ref{A_CTS}) and (\ref{momCTS})
still have the same form, only with $\beta^i$ replaced by $B^i$.
The boundary condition for $B^i$ then is simply
\be
\label{B_BC_inf_rot}
\lim_{r \to \infty} B^i = 0 .
\ee

%%%%%%%%%%%%%%%%%%%%%%%%%%%%%%%%%%%%%%%%%%%%%%%%%%%%%%%%%%%%%%%%%%%
\subsection{ADM quantities at spatial infinity and the Komar integral}
%%%%%%%%%%%%%%%%%%%%%%%%%%%%%%%%%%%%%%%%%%%%%%%%%%%%%%%%%%%%%%%%%%%

The ADM mass $M^{ADM}_{\infty}$ and angular momentum $J^{ADM}_{\infty}$
at spatial infinity ($r \rightarrow \infty$) are given by
\begin{equation}
\label{M_ADM}
M^{ADM}_{\infty} 
= \sqrt{ -P^{ADM}_{\mu \ \infty} P^{ADM}_{\nu \ \infty} \eta^{\mu\nu} } ,
\end{equation}
\begin{equation}
\label{E_ADM}
P^{ADM}_{0 \ \infty} = \frac{1}{16\pi}\int_{S_{\infty}}
                    (\gamma_{ij,i}-\gamma_{ii,j} )d{S}^j ,
\end{equation}
\begin{equation}
\label{P_ADM}
P^{ADM}_{i \ \infty} = \frac{1}{8\pi}\int_{S_{\infty}}    
                 ({K}_{ij} - K\eta_{ij} )d{S}^j ,                    
\end{equation}
and
\begin{equation}
\label{J_ADM}
J^{ADM}_{\infty}= \frac{1}{8\pi}\int_{S_{\infty}} 
               ({K}_{ij}\Phi^j - K \Phi_i)d{S}^i .
\end{equation}
Here $\Phi^i$ is again the asymptotic rotational Killing vector
at spatial infinity.
Note that these definitions are valid only in coordinates where the
4-metric approaches the Minkowski metric at spatial infinity.
In this case lower and upper spatial indices do not need to be distinguished,
so that summation is implied over any repeated spatial indices.
The integrals here are over a closed sphere (denoted by $S_{\infty}$)
at spatial infinity. In spherical coordinates $d{S}^j$ is given by
$4\pi r^2
(\sin\theta\cos\phi, \sin\theta\sin\phi, \cos\theta)
d\theta d\phi$.
$M^{ADM}_{\infty}$ and $J^{ADM}_{\infty}$ tell us how much mass and angular
momentum is contained in the entire
spacetime~\footnote{Analogously to the ADM momentum $P^{ADM}_{i \ \infty}$,
it is also possible to compute the components of ADM angular momentum from
$
J^{ADM}_{i \ \infty} = \frac{1}{8\pi}\int_{S_{\infty}}
\epsilon_{ijk} x^j (K_{kl} - K\eta_{kl} )d{S}^l
$.
}.
In the case where $\gamma_{ij}$ in Eq.~(\ref{conf-metric}) is conformally
flat (and if $P^{ADM}_{i \ \infty} = 0$) the ADM mass simplifies
to
\be
M^{ADM}_{\infty} = -\frac{1}{2\pi}\int_{S_{\infty}} 
\partial_j\psi d{S}^j .
\ee

If the spacetime possess a Killing vector $\xi^{\mu}$, i.e.
a vector satisfying
\be
\pounds_{\xi} g_{\mu\nu} = \nabla_{(\mu} \xi_{\nu)} = 0 ,
\ee
then the Komar integral~\cite{Komar59,Ashtekar79a,Wald84}
\begin{equation}
\label{Komar}
I_K(\xi,S) = -\frac{1}{8\pi} 
\oint_S {\epsilon}_{\alpha\beta\mu\nu} \nabla^{\mu} \xi^{\nu}
dx^{\alpha} \wedge dx^{\beta}
\end{equation}
integrated over any closed 3-surface $S$ containing the sources
yields a value that is independent of $S$.

For a time-like Killing vector that is normalized to 
$\xi^{\mu}\xi_{\mu}=-1$ at spatial infinity 
$I_K(\xi,S) = M^{ADM}_{\infty}$, 
while for an axial Killing vector that approaches the asymptotic
rotational Killing vector $\Phi^{\mu}$ at spatial
infinity $I_K(\xi,S) = -2J^{ADM}_{\infty}$~\cite{Ashtekar79a}.
We now consider a helical Killing vector $\xi^{\mu}$ that
is given by
\be
\xi^{\mu} = \alpha T^{\mu} + \Omega\Phi^{\mu}
\ee
at spatial infinity, where $T^{\mu}$ and $\Phi^{\mu}$ are the asymptotic
time-translation and rotational Killing vectors at spatial infinity and
$\Omega$ is a constant. If $\xi^a$ is normalized such that
\begin{equation}
\label{xi_norm}
\lim_{r \rightarrow \infty} \xi^a n_a = -1 ,
\end{equation}
then the Komar integral over a sphere at $r \rightarrow \infty$
becomes~\cite{Ashtekar79a,Tichy:2003zg}
\begin{equation}
\label{Komar_ADM}
I_K(\xi,S) = M^{ADM}_{\infty} - 2\Omega J^{ADM}_{\infty} .
\end{equation}

%%%%%%%%%%%%%%%%%%%%%%%%%%%%%%%%%%%%%%%%%%%%%%%%%%%%%%%%%%%%%%%%%%%
\subsection{Quasi-equilibrium assumptions for the metric variables
of binary systems}
\label{Quasiequil_geometry}
%%%%%%%%%%%%%%%%%%%%%%%%%%%%%%%%%%%%%%%%%%%%%%%%%%%%%%%%%%%%%%%%%%%

We now make some additional simplifying assumptions to better deal
with quasi-equilibrium situations. For concreteness, let us discuss
a binary system made up of two orbiting objects that could be stars or 
black holes.
Such a system will radiate gravitational waves that carry away energy. Thus
over time the two objects will spiral towards each other.
According to post-Newtonian predictions~\cite{Peters:1963ux,Peters:1964} the
orbits circularize on a timescale which is much shorter than the inspiral
timescale due to the emission of gravitational waves. The same
circularization effect has also been shown in the extreme mass ratio limit
in Kerr spacetime~\cite{Ryan:1995xi,Kennefick:1995za,Kennefick:1998ab}.
Hence when we set up initial data we often assume that the binary is in an
approximately circular orbit.
% and that the spins of each object remain
% approximately constant over one orbit.
The circular orbital motion then can be transformed away
by changing to a corotating coordinate system, i.e. a coordinate system
that rotates with the orbital angular velocity of the binary
with respect to asymptotically inertial coordinates.
In this corotating system the two
objects do not move, which means $\partial_t g_{\mu\nu}$ vanishes
at least approximately. This implies that the time evolution 
vector $t^{\mu}$ is an approximate symmetry of the spacetime, 
which in turn implies the existence of an approximate helical
Killing vector $\xi^{\mu}$ (see e.g.~\cite{Tichy:2003zg,Tichy:2003qi}) with 
\be
\label{approxKV}
\pounds_{\xi} g_{\mu\nu} \approx 0 .
\ee
Here $\pounds_{\xi}$ is the Lie derivative (see \ref{Why-LieDerivs})
with respect to the vector $\xi^{\mu}$.
Corotating coordinates are understood as the coordinates
chosen such that $t^{\mu}$ lies along $\xi^{\mu}$, such that
\begin{equation}
\label{hKV}
\xi^{\mu} = t^{\mu} = {\alpha} n^{\mu} + \beta^{\mu} .
\end{equation}
Since we are considering orbiting binaries, this Killing vector would have to
be a helical Killing vector, which means that at spatial infinity
($r \rightarrow \infty$)
\begin{equation}
\label{asymptKV}
n^{\mu} = T^{\mu}  \ \  \mbox{and} \ \ \beta^{\mu}=\Omega\Phi^{\mu} ,
\end{equation}
where $T^{\mu}$ and $\Phi^{\mu}$ are the asymptotic
time-translation and rotational Killing vectors at spatial infinity and
$\Omega$ is the angular velocity with which the binary rotates.
In asymptotically inertial coordinates we have $T^{\mu}=(1,0,0,0)$ and
$\Phi^{\mu}=(0, -x^2+x_{CM}^2, x^1-x_{CM}^1, 0)$, where we have chosen
the rotational Killing vector to correspond to a
rotation about the a line through the center of mass $x_{CM}^i$
parallel to $x^3$-axis.

An approximate helical Killing vector with 
$\pounds_{\xi} g_{\mu\nu} \approx 0$ implies that the Lie derivatives
with respect to $\xi^{\mu}$ of all 3+1 quantities similarly vanish
approximately. For the initial data construction using the CTS
decomposition we will use
\begin{equation}
\label{geom_equil}
\pounds_{\xi} \bar{\gamma}_{ij} \approx \pounds_{\xi} K \approx 0 .
\end{equation}
In a corotating coordinate system where the time evolution vector $t^{\mu}$
lies along $\xi^{\mu}$, the time derivatives of these metric variables
are then equal to zero. 
From $\partial_t \bar{\gamma}_{ij} = 0$ it follows that $\bar{u}^{ij}$ in
Eq.~(\ref{A_CTS}) vanishes.
The assumption $\partial_t K = 0$ can be used in Eq.~(\ref{alphabar_CTS}) to
find $\bar{\alpha}$.

As we have seen above, the approximate helical Killing vector has the
form $\xi^{\mu} = \alpha T^{\mu} + \Omega\Phi^{\mu}$ at spatial 
infinity, where $\Omega$ is the orbital angular velocity of the binary.
It is clear that circular orbits are only possible for one particular value
of $\Omega$, for a given distance between the two components of the binary.
We will now discuss a method that can be used to find
this $\Omega$. If we insert $\xi^a$ as given in Eq.~(\ref{hKV}) 
into Eq.~(\ref{Komar}) we obtain
\begin{eqnarray}
\label{Komar2}
I_K(\xi,S) &=& I_K({\alpha}n,S)+I_K(\beta,S)  \nonumber \\ 
&=& \frac{1}{4\pi}\int_{S}\bar{\nabla}_i{\alpha} d\bar{S}^i
   -\frac{1}{4\pi}\int_{S} \bar{K}_{ij} \beta^i d\bar{S}^j .
\end{eqnarray}
Using Eq.~(\ref{asymptKV}) we find that the second term 
integrated at $r \rightarrow \infty$ is
\begin{equation}
\label{Komarbeta_infty}
I_K(\beta,S_{\infty})= -2\Omega \frac{1}{8\pi}\int_{S_{\infty}}
                          \bar{K}_{ij}\Phi^j d\bar{S}^i
          = -2\Omega J^{ADM}_{\infty} .
\end{equation}
Therefore combining 
Eqs.~(\ref{Komar_ADM}), (\ref{Komar2}) and (\ref{Komarbeta_infty}) 
the condition
\begin{equation}
\label{MK_MADM}
I_K({\alpha}n,S_{\infty}) = M^{ADM}_{\infty} 
\end{equation}
must hold if the helical Killing vector of Eq.~(\ref{hKV})
exists. This condition can then be used to fix the value 
of $\Omega$. 

The term $I_K({\alpha}n,S_{\infty})$ is often called the Komar mass. For
asymptotically flat stationary spacetimes the equality of Komar and ADM mass
has already been shown in~\cite{Beig1978}. It is closely related to the
virial theorem in general relativity~\cite{Gourgoulhon1994} and was first
used in~\cite{Gourgoulhon02,Grandclement02} to determine the orbital angular
velocity $\Omega$ for binary black hole initial data.

%%%%%%%%%%%%%%%%%%%%%%%%%%%%%%%%%%%%%%%%%%%%%%%%%%%%%%%%%%%%%%%%%%%
\subsection{Apparent horizons of black holes}
%%%%%%%%%%%%%%%%%%%%%%%%%%%%%%%%%%%%%%%%%%%%%%%%%%%%%%%%%%%%%%%%%%%

As is well known, black holes are enclosed in so called event horizons. An
event horizon is defined as the boundary of a spacetime region from where
nothing can escape to spatial infinity. Thus in order to find the event
horizon we need to know the entire future spacetime. Hence this definition
is not very useful when we want to construct initial data on a single
spatial slice. For this reason the concept of apparent horizons has been
introduced. One considers outgoing photons and considers the spatial
2-surface $\cal{S}$ that is given by the location of these outgoing photons at
time $t$. In flat spacetime this surface will expand, so that its area
increases (if the photons are outgoing). An apparent horizon is defined
as the outermost closed
surface where this expansion is zero. This concept captures the idea of
no escape from a black hole, but this time the definition involves only a
single spatial slice.
Let $s^{\mu}$ be the outward pointing spatial normal vector 
(with $s^{\mu}n_{\mu}=0$) of the surface $\cal{S}$, normalized to
\be
s^{\mu}s_{\mu} = 1 .
\ee
The induced metric on $\cal{S}$ is then given by 
\be   
h_{\mu\nu} = \gamma_{\mu\nu} - s_{\mu}s_{\nu} .
\ee
If we denote its determinant by $h$, the surface area of $\cal{S}$ is given
by integrating $\sqrt{h}$ over $\cal{S}$. 
We now define the outgoing lightlike vector
\be
\label{ougoing_light}
l^{\mu} = n^{\mu} + s^{\mu} .
\ee
The expansion can then be defined as
\be
\Theta = \frac{\pounds_l \sqrt{h}}{\sqrt{h}}
= -\frac{1}{2} h^{\mu\nu} \pounds_l h_{\mu\nu} .
\ee
Using $h^{\mu\nu} \pounds_s h_{\mu\nu} = -2 D_i s^i$ and
$h^{\mu\nu} \pounds_n h_{\mu\nu} = 2 K_{ij} s^i s^j - 2 K $, we find
\be
\label{AHexpansion}
\Theta = (\gamma^{ij} - s^i s^j)(D_i s_j - K_{ij}) .
\ee
The apparent horizon is now defined as a surface on which
$\Theta = 0$. In order to find this surface it is convenient to define
it as the location where a function $F(x^i)=0$. The normal vector then
becomes
\be
s^i = D^i F / u , \ \ \mbox{with } u=\sqrt{(D_i F) (D^i F)} .
\ee
Inserting into Eq.~(\ref{AHexpansion})
we obtain
\be
\label{AHeqn}
\Theta = 
\left[\gamma^{ij} - \frac{(D^i F)(D^j F)}{u^2}\right]
\left(\frac{D_i D_j F}{u} - K_{ij}\right) = 0 
\ee
as the equation that determines the location of the apparent 
horizon~\cite{Alcubierre_book}. A review on algorithms for
finding apparent horizons can be found in~\cite{ThornburgLRR}.
Given a spacetime with an event horizon and given a particular slicing,
we can try to find the apparent horizon on each slice by solving
Eq.~(\ref{AHeqn}). When they exist, the apparent horizons at different times
$t$ form a world tube through spacetime, that will either be inside or
coincide with the event horizon~\cite{Hawking73a}.

%%%%%%%%%%%%%%%%%%%%%%%%%%%%%%%%%%%%%%%%%%%%%%%%%%%%%%%%%%%%%%%%%%%
\subsection{Quasi-equilibrium boundary conditions at black hole horizons}
%%%%%%%%%%%%%%%%%%%%%%%%%%%%%%%%%%%%%%%%%%%%%%%%%%%%%%%%%%%%%%%%%%%

As we will see below (when we discuss puncture initial data)
it is possible to construct black hole initial data without any
inner boundaries at or inside the black hole horizons.
However, in many cases such as e.g. Misner or Bowen-York,
or superimposed Kerr-Schild initial data
one needs a boundary condition at each black hole as well.
Here we will discuss quasi-equilibrium boundary conditions 
introduced by Cook et al.~\cite{Cook:2001wi,Cook:2004kt,Caudill:2006hw}
that can be imposed at the apparent horizon of each black hole
within the CTS approach. As clarified in~\cite{Jaramillo:2004uc},
the assumptions required for these quasi-equilibrium boundary conditions are
essentially the same as those required of an isolated
horizon~\cite{Ashtekar00a,Dreyer02a,Ashtekar03a}.

One starts by picking some surface $\cal{S}$ (e.g. a coordinate sphere) that
one wishes to make into an apparent horizon in equilibrium.
To ensure that $\cal{S}$ is indeed an apparent horizon
one imposes 
\be
\label{zeroExp}
\Theta |_{\cal{S}} = 0 .
\ee
Since this horizon is supposed to be in equilibrium, one further
demands that the shear of outgoing null rays on $\cal{S}$ should
be~\cite{Ashtekar03a,Jaramillo:2004uc}
\be
\label{zeroShear}
\sigma_{\mu\nu} |_{\cal{S}} = 0 .
\ee
Finally, coordinates adapted to equilibrium
should be chosen such that the apparent horizon
does not move as we evolve from the initial slice at $t$ to the next
one at $t+dt$. 
A photon moving along the outgoing light ray given by $l^{\mu}$ of
Eq.~(\ref{ougoing_light}) moves along the curve 
$\frac{dx^{\mu}}{d\lambda} = l^{\mu}$, where $\lambda$ is the
parameter along the curve.
This curve intersects the slice at $t+dt$ for 
$dx^{0} = l^{0} d\lambda = dt$, and thus at $n^0 d\lambda = dt$, which
yields $d\lambda = \alpha dt$. So in the time interval $dt$ a photon
moves in the direction $\alpha l^{\mu}$ while a point with constant spatial
coordinates moves in the direction $t^{\mu}$. We could now choose our
coordinates such that $t^{\mu}-\alpha l^{\mu} =0$, which would ensure that
all photons that make up the surface $\cal{S}$ stay at the same spatial
coordinate. However, for the surface $\cal{S}$ to remain unchanged it is
sufficient when each photon only moves along the surface. So a less
restrictive condition is to demand that
\be
s_{\mu} (t^{\mu}-\alpha l^{\mu}) |_{\cal{S}} = 0 ,
\ee
which implies that this difference has no component perpendicular
to $\cal{S}$. It is equivalent to 
\be
\label{betaperp_BC_BH}
s_i \beta^i |_{\cal{S}}  = \alpha |_{\cal{S}}
\ee
and is interpreted as a boundary condition on the component of the
shift perpendicular to $\cal{S}$.

Following~\cite{Caudill:2006hw} we now introduce a conformally rescaled
normal vector and surface metric by defining
\begin{eqnarray}
h_{ij} &=& \psi^4 \bar{h}_{ij} , \\
s^i    &=& \psi^{-2} \bar{s}^i .
\end{eqnarray}
Combining Eqs.~(\ref{AHexpansion}), (\ref{zeroExp}) and the CTS
definitions for $K_{ij}$ in Eqs.~(\ref{tracefreeA}), (\ref{A-rescale})
and (\ref{A_CTS}) we arrive at~\cite{Cook:2001wi,Cook:2004kt,Pfeiffer:2007yz}
\be
\label{psi_BC_BH}
\bar{s}^i \bar{D}_i \ln\psi |_{\cal{S}} =
\left\{
-\frac{1}{4}\bar{h}^{ij} \bar{D}_i \bar{s}_j + 
\frac{K}{6}\psi^2
-\frac{\psi^{-4}}{8\bar{\alpha}}
[(\bar{L}\beta)^{ij} - \bar{u}^{ij}] \bar{s}_i\bar{s}_j
\right\} \Bigg|_{\cal{S}} .
\ee
This is a Robin-type boundary condition on the conformal factor, derived
under the assumption~(\ref{zeroExp}) of vanishing expansion, that defines
the apparent horizon.

So far we have boundary conditions only for the conformal factor and
the component of the shift perpendicular to $\cal{S}$. If we define
the parallel component by
\be
\beta_{\parallel}^i = h^i_j\beta^j ,
\ee
it can be shown~\cite{Caudill:2006hw} that the vanishing shear 
assumption~(\ref{zeroShear}) leads to
\be
(\bar{d}^{(i}\beta_{\parallel}^{j)} - 
\frac{1}{2}\bar{h}^{ij}\bar{d}_k\beta_{\parallel}^{k})|_{\cal{S}}
= \frac{1}{2}
(\bar{h}^i_k \bar{h}^j_l -\frac{1}{2}\bar{h}^{ij}\bar{h}_{kl}) 
\bar{u}^{kl}|_{\cal{S}} ,
\ee
where $\bar{d}^{i}$ is the derivative operator compatible with the 
2-metric $\bar{h}_{ij}$.
So if $\bar{u}^{ij}=0$, $\beta_{\parallel}^i$ is a conformal Killing vector
of the metric $\bar{h}_{ij}$ on $\cal{S}$.
In practice one uses
\be
\label{beta_BC_BH}
\beta^i |_{\cal{S}} = (\alpha s^i + \Omega_r \chi^i) |_{\cal{S}}
\ee
as boundary condition for the shift. Here $\chi^i$ is a rotational
conformal Killing vector on $\bar{h}_{ij}$ with affine length of $2\pi$
and with $\chi^i s_i=0$. The constant $\Omega_r$ is can be chosen
to change the spin of the black hole.

Finally, we also need a boundary condition for $\bar{\alpha}$. Several
conditions will work~\cite{Caudill:2006hw}. The simplest choice
is to use a Dirichlet boundary condition where one prescribes a value for
$\alpha\psi$ on $\cal{S}$.

%%%%%%%%%%%%%%%%%%%%%%%%%%%%%%%%%%%%%%%%%%%%%%%%%%%%%%%%%%%%%%%%%%%
\section{Black hole initial data for single black holes}
%%%%%%%%%%%%%%%%%%%%%%%%%%%%%%%%%%%%%%%%%%%%%%%%%%%%%%%%%%%%%%%%%%%
\label{BH_data}

We will now discuss how one can construct initial data for single
black holes.
We will present several analytic black hole solutions to the Einstein
equations. These solutions represent single stationary black holes. Since
they are obtained by solving the Einstein equations they automatically
satisfy the constraints.

%%%%%%%%%%%%%%%%%%%%%%%%%%%%%%%%%%%%%%%%%%%%%%%%%%%%%%%%%%%%%%%%%%%
\subsection{Non-spinning black holes}
%%%%%%%%%%%%%%%%%%%%%%%%%%%%%%%%%%%%%%%%%%%%%%%%%%%%%%%%%%%%%%%%%%%

A non-spinning spherically symmetric black hole
of mass $m$ can be described by the Schwarzschild metric
\be
\label{Schwarzschild0}
ds^2  = -\left(1-\frac{2m}{r_s}\right)dt^2 + \frac{dr_s^2}{1-2m/r_s}
        +r_s^2 (d\theta^2 + \sin^2\theta d\phi^2) ,
\ee
where $0\leq r_s \leq \infty$.
The event horizon is located at $r_s=2m$. The region $r_s\leq 2m$ is inside
the black and at $r_s=0$ there is a physical singularity where 
curvature invariants are infinite.
However, this metric is rarely used in numerical relativity because it has a
coordinate singularity at $r_s=2m$.
We can change the radial coordinate by setting
\be
\label{SchwToIso}
r_s = r \psi^2 ,
\ee
where we have defined
\be
\label{SchwConfac}
\psi = 1 + \frac{m}{2r} .
\ee
From this we obtain the metric in isotropic coordinates
\be
\label{IsoSchw}
ds^2  = -\left(\frac{1-\frac{m}{2r}}{1+\frac{m}{2r}}\right)^2dt^2 
 + \psi^4 \left[dr^2 +r^2 (d\theta^2 + \sin^2\theta d\phi^2)\right] ,
\ee
where $0\leq r \leq \infty$. 
As we can see the 3-metric is now conformally flat. 
The event horizon is at $r=m/2$. Notice however, that
the new $r$ coordinate does not cover the inside of the black hole
since $r_s\geq 2m$ for any $r$. For $r \to \infty$ the metric becomes flat.
Indeed, there is another asymptotically flat region at $r=0$. This can be
seen by performing the coordinate transformation
\be
\label{Schw_isometry}
r = \left(\frac{m}{2}\right)^2 \frac{1}{r'} ,
\ee
which maps the region with $r<m/2$ into $r'>m/2$.
As one can easily verify, the 3-metric in Eq.~(\ref{IsoSchw}) 
remains conformally flat under this transformation. Its new conformal factor
is $\psi' = 1 + \frac{m}{2r'}$, i.e. it is invariant under this
transformation. The transformation in Eq.~(\ref{Schw_isometry}) is thus an
isometry. This shows that at $r=0$ (i.e. $r' \to \infty$) we obtain
another asymptotically flat region.

One can also switch to Cartesian spatial coordinates, which yields
\be
ds^2  = -\left(\frac{1-\frac{m}{2r}}{1+\frac{m}{2r}}\right)^2dt^2
 + \psi^4 \left( dx^2 + dy^2 + dz^2 \right) , 
\ee
where now $r=\sqrt{x^2 + y^2 + z^2}$.
If we apply the 3+1 split we find 
\be
\alpha = \frac{1-\frac{m}{2r}}{1+\frac{m}{2r}}, \ \ \
\beta^i=0, \ \ \
\gamma_{ij} = \psi^4 \delta_{ij}, \ \ \
K_{ij} = 0 ,
\ee
which is often used as initial data for a single non-spinning black hole.

%%%%%%%%%%%%%%%%%%%%%%%%%%%%%%%%%%%%%%%%%%%%%%%%%%%%%%%%%%%%%%%%%%%
\subsection{Black holes with spin}
%%%%%%%%%%%%%%%%%%%%%%%%%%%%%%%%%%%%%%%%%%%%%%%%%%%%%%%%%%%%%%%%%%%
   
In order to describe a spinning black hole one can use the Kerr metric
that is also an exact solution to Einstein's equations. The 4-metric in so
called Kerr-Schild coordinates (see e.g. \cite{Misner73}) is given by
\be
\label{eq:KSmetric}
ds^2 = (\eta_{\mu\nu}+ 2H k_{\mu} k_{\nu}) dx^{\mu} dx^{\nu} .
\ee
where we the $x^{\mu}$ stand for $(t,x,y,z)$, $\eta_{\mu\nu}$ is the
Minkowski metric, and
\begin{eqnarray}
\label{KS-H}
H &=& \frac{mr}{r^2 + a^2 (z/r)^2} \\
\label{KS-k}
k_{\mu} dx^{\mu} &=& - dt - 
\frac{r(x dx + y dy) -a(x dy - y dx)}{r^2 + a^2} - \frac{z dz}{r} .
\end{eqnarray}
Here $m$ is the mass and $a$ is the spin parameter which takes values in
the range $0 \leq a \leq m$.
The function $r$ here depends on $(x,y,z)$ and is given implicitly by
\be
\label{KS-CartCoords}
x^2 + y^2 = (r^2 + a^2) \left[1-\frac{z^2}{r^2}\right] .
\ee
The event horizon is located at $r=m+\sqrt{m^2-a^2}$. The physical curvature
singularity occurs on a ring given by $x^2 + y^2 = a^2 $ and $z=0$,
which is inside the event horizon. 
If $a$ is non-zero, no spatial slicing exists in which the 3-metric can be
written in a conformally flat way~\cite{Garat:2000pn}.
The Kerr-Schild metric in 3+1 form is given 
by~\cite{Matzner98a}
\be
\label{KS-3-metric}
\alpha = \frac{1}{\sqrt{1+2H k_0 k_0}}, \ \ \
\beta_i=2H k_0 k_i, \ \ \
\gamma_{ij} = \delta_{ij} + 2H k_i k_j,
\ee
while the extrinsic curvature
\be
\label{KS-Kij}
K_{ij} = \frac{D_i\beta_j + D_j\beta_i}{2\alpha} ,
\ee
can be obtained by using the derivative operator $D_i$ compatible
with the 3-metric.
The Kerr-Schild coordinates described here are horizon penetrating in the
sense that there is no coordinate singularity in $\gamma_{ij}$ and $K_{ij}$
at the horizon, and that they cover both the inside and outside of the black
hole.

%%%%%%%%%%%%%%%%%%%%%%%%%%%%%%%%%%%%%%%%%%%%%%%%%%%%%%%%%%%%%%%%%%%
\section{Initial data for black hole binaries}
%%%%%%%%%%%%%%%%%%%%%%%%%%%%%%%%%%%%%%%%%%%%%%%%%%%%%%%%%%%%%%%%%%%
\label{BBH_data}

The binary black hole problem is more complicated and no analytic solutions
to the Einstein equations are known. 
Here we will discuss several approaches to set up initial for a black hole
binary. We do not attempt to discuss every possible method,
rather we will focus on some of the most widely used approaches.

%%%%%%%%%%%%%%%%%%%%%%%%%%%%%%%%%%%%%%%%%%%%%%%%%%%%%%%%%%%%%%%%%%%
\subsection{Black holes at rest}
%%%%%%%%%%%%%%%%%%%%%%%%%%%%%%%%%%%%%%%%%%%%%%%%%%%%%%%%%%%%%%%%%%%

It is possible to find analytic initial data for $N$
black holes at rest. We start by observing that
\be
K_{ij} = 0
\ee
satisfies the momentum constraint (\ref{mom0}) for the case of vacuum where
$\rho=0=j^i$. We further choose the 3-metric to be conformally flat,
i.e. we set
\be
\label{conformally_flat}
\bar{\gamma}_{ij} = \delta_{ij}
\ee
for the conformal metric of Eq.~(\ref{conf-metric}). 
Thus the Hamiltonian constraint (\ref{hamCTT}) simplifies to
\be
\label{hamBL}
\bar{D}^2 \psi = 0 ,
\ee
where $\bar{D}_i=\partial_i$ is the flat space derivative operator. 
A solution that satisfies the boundary condition (\ref{psi_BC_inf})
is given by
\be
\label{psiBL}
\psi_{BL} = 1 + \sum_{A=1}^N \frac{m_A}{2r_A}
\ee
Here $A$ labels the black holes, 
so that $m_A$ is the mass of black hole $A$, and 
$r_{A} = \sqrt{(x-c_{A}^x)^2 + (y-c_{A}^y)^2 + (z-c_{A}^z)^2}$ is the
conformal distance from the black hole center located at
$(x,y,z)=(c_{A}^x, c_{A}^y, c_{A}^z)$. These initial data are known as
Brill-Lindquist initial data~\cite{Brill63,Lindquist63} and
result in $N$ black holes at rest that will fall toward each other if
evolved forward in time.
They are
a straightforward generalization of the Schwarzschild solution in
isotropic coordinates presented above.
Notice that the conformal factor in Eq.~(\ref{psiBL}) is a solution of
Eq.~(\ref{hamBL}) only on $\mathbb{R}^3 \setminus \{c_1^i,c_2^i,...\}$,
i.e. there are singular points that have to be removed from the manifold.
The total ADM mass of these data is given by
$M^{ADM}_{\infty} = \sum_{A=1}^N m_A$.

The main difference between Brill-Lindquist initial data and the
Schwarzschild metric in isotropic coordinates is that we now have $N+1$
different asymptotically flat regions. One is located at $r=\infty$, the
others are at $r_A=0$ for $A=1, ..., N$. An observer sitting near
$r=\infty$ will observe $N$ black holes, while an observer sitting in one of
the other asymptotically flat regions will see only one black horizon. Due
to this asymmetry the asymptotically flat regions are not isometric to each
other as was the case for a single Schwarzschild black hole.
However, an approach due to
Misner~\cite{Misner63} allows us to construct another solution to
Eq.~(\ref{hamBL}) where the 3-metric 
has two isometric asymptotically flat hypersurfaces connected by $N$
black holes. 
The case of two black holes that satisfies this isometry condition is known
as Misner initial data. This solution is written in terms of an infinite
series expansion. As in the case of Brill-Lindquist data there are singular
points, but this time there are an infinite number of singular points for
each black hole.

%%%%%%%%%%%%%%%%%%%%%%%%%%%%%%%%%%%%%%%%%%%%%%%%%%%%%%%%%%%%%%%%%%%
\subsection{Puncture initial data}
%%%%%%%%%%%%%%%%%%%%%%%%%%%%%%%%%%%%%%%%%%%%%%%%%%%%%%%%%%%%%%%%%%%

It is possible to generalize Brill-Lindquist data to an
arbitrary number of moving black holes that
have non-vanishing $K_{ij}$. We present here so called puncture initial
data which is one of the main initial data types used in numerical
relativity. We start again with the conformally flat
metric in Eq.~(\ref{conformally_flat}), but this time we only assume
\be
\label{maxSlicing}
K = 0 .
\ee
Furthermore, we use the CTT decomposition 
with $\bar{\sigma}=1$ and choose
\be
\bar{M}^{ij} = 0
\ee
in Eq.~(\ref{Abar-split}). Then Eq.~(\ref{momCTT}) simplifies to
\be
\bar{D}^2 W^i + \frac{1}{3} \bar{D}^i \bar{D}_j W^j = 0 .
\ee
A solution of this equation is~\cite{Bowen80}
\be
\label{Bowen-York0}
W^i =  -\frac{1}{4r}\left(7P^i + n^i n_{j}P^j\right)
+\frac{1}{r^2}\epsilon_{ijk} n^j S^k ,
\ee
where
$r = \sqrt{(x-c^x)^2 + (y-c^y)^2 + (z-c^z)^2}$,
$n^i = (x^i - c^i)/r$,
and $c^i$, $P^i$ and $S^i$ are constant vectors.
Using Eq.~(\ref{Abar-split}) this yields
the conformal Bowen-York extrinsic curvature~\cite{Bowen80}
\be
\label{Bowen-York1}
\bar{A}^{ij}_{BY} = 
 \frac{3}{2r^2} [P^i n^j + P^j n^i -(\delta^{ij}-n^i n^j)P^k n_k]
+\frac{3}{r^3}   (n^i \epsilon^{jkl} + n^j \epsilon^{ikl}) S_k n_l .
\ee
Since the momentum constraint in Eqs.~(\ref{mom0}) and (\ref{momCTT}) is
a linear equation, linear superpositions of several such solutions will
again be a solution. We can thus generalize Brill-Lindquist data 
by writing
\begin{eqnarray}
\label{Aij_Punc}
\bar{A}^{ij} &=& \sum_A \Bigg\{
\frac{3}{2r_A^2}
[P_A^i n_A^j + P_A^j n_A^i -(\delta^{ij}-n_A^i n_A^j)P_A^k n_{Ak}] \nonumber\\
&&+\frac{3}{r_A^3}
 (n_A^i \epsilon^{jkl} + n_A^j \epsilon^{ikl}) S_{Ak} n_{Al} 
\Bigg\} ,
\end{eqnarray}
where again $A$ labels the black holes,
\be
r_{A} = \sqrt{(x-c_{A}^x)^2 + (y-c_{A}^y)^2 + (z-c_{A}^z)^2} ,  \ \ \
n_A^i = (x^i - c_{A}^i)/r_{A}
\ee
and $c_{A}^i$, $P_A^i$ and $S_A^i$ are constant vectors.
One can show that the ADM momentum and angular momentum for this
extrinsic curvature are given by
\be
P^{ADM}_{i\ \infty} = \sum_A P_A^i, \ \ \
J^{ADM}_{i\ \infty}
 = \sum_A \left(\epsilon^{ijk} c_{Aj} P_{Ak} +  S_A^i \right)
\ee
at $r=\infty$. Thus the vectors $P_A^i$ and $S_A^i$ can be interpreted
as momentum and spin of black hole $A$.

The next step is to solve the Hamiltonian constraint (\ref{hamCTT}). As we
have seen in Eq.~(\ref{psiBL}), Brill-Lindquist data are singular
at each black hole. Inserting the ansatz
\be
\label{psiPuncAnsatz}
\psi =\psi_{BL} + u
\ee
into Eq.~(\ref{hamCTT}) we obtain
\be
\label{hamPunc}
\bar{D}^2 u + \eta \left( 1 +\frac{u}{\psi_{BL}}\right)^{-7} = 0 ,
\ee
where we have introduced
\be
\eta = \frac{1}{8\psi_{BL}^7} \bar{A}_{ij} \bar{A}^{ij} .
\ee
As we can see from Eq.~(\ref{psiBL}), $\psi_{BL} \sim 1/r_A$ 
near $r_A=0$. From Eq.~(\ref{Aij_Punc}) we find
$\bar{A}_{ij} \bar{A}^{ij} \sim 1/r_A^6$ near $r_A=0$. 
Thus $\eta \sim r_A$ near $r_A=0$. Therefore we expect
$u$ to be regular at $r_A=0$. Indeed, 
Brandt and Br\"ugmann~\cite{Brandt97b} have shown that
Eq.~(\ref{hamPunc}) can be solved on $\mathbb{R}^3$ without any points
removed. Then $u$ will be at least $C^2$ at $r_A=0$. This method is known
as the puncture approach. It constitutes a significant simplification
since it obviates the need for any boundary conditions at $r_A=0$, that
would need to be specified if we were to solve on 
$\mathbb{R}^3 \setminus \{c_1^i,c_2^i,...\}$ instead.
This approach is widely used by the numerical relativity community.
It allows us to set up $N$ black holes with both momentum and spin
and like Brill-Lindquist data it results in a solution with $N+1$
asymptotically flat regions.

Note that the ADM mass depends on $u$ and thus is not equal to the
sum of the mass parameters $m_A$, which therefore are referred to as
bare masses that do not have a direct physical meaning.

%%%%%%%%%%%%%%%%%%%%%%%%%%%%%%%%%%%%%%%%%%%%%%%%%%%%%%%%%%%%%%%%%%%
\subsection{Bowen-York initial data}
%%%%%%%%%%%%%%%%%%%%%%%%%%%%%%%%%%%%%%%%%%%%%%%%%%%%%%%%%%%%%%%%%%%

Puncture initial data do not have an isometry. But Bowen and 
York~\cite{Bowen80} have shown that the solution
to the momentum constraint in Eq.~(\ref{Bowen-York1}) can
be generalized to
\begin{eqnarray} 
\label{Bowen-York2}
\bar{A}^{ij}_{\pm} &=& 
\frac{3}{2r^2} [P^i n^j + P^j n^i -(\delta^{ij}-n^i n^j)P^k n_k]
\nonumber \\
&&\pm \frac{3b^2}{2r^4} 
[P^i n^j + P^j n^i +(\delta^{ij} - 5n^i n^j)P^k n_k] \nonumber \\
&&+\frac{3}{r^3}  (n^i \epsilon^{jkl} + n^j \epsilon^{ikl}) S_k n_l .
\end{eqnarray}
The $\bar{A}^{ij}_{\pm}$ represent two inversion symmetric solutions,
i.e. solutions that are isometric under the transformation $r=b^2/r'$.
Here $r=b$ is the radius that constitutes the fixed-point set of the
isometry. Since the momentum constraint is linear we can again
consider superpositions of this $\bar{A}^{ij}_{\pm}$ to generate multi
hole solutions. The process of constructing momentum constraint solutions
with isometry from $\bar{A}^{ij}_{\pm}$ is rather complex and again results
in an infinite-series sum. However, it is possible to evaluate
this sum numerically~\cite{Cook91}. With the solution for $\bar{A}^{ij}$
at hand, the Hamiltonian constraint (\ref{hamCTT}) becomes
\be
\label{hamBY}
\bar{D}^2 \psi + 
\frac{1}{8}\psi^{-7} \bar{A}_{ij} \bar{A}^{ij} = 0 ,
\ee
which is an elliptic equation that has to be solved numerically. The
boundary condition at infinity is again Eq.~(\ref{psi_BC_inf}). This time,
however, we also need a boundary condition at each black hole. It is derived
by demanding isometry at each black hole and reads
\be
N_A^i \bar{D}_i \psi \Big|_{r_A = b_A} 
= - \frac{\psi}{2r_A} \Bigg|_{r_A = b_A} .
\ee
It needs to be imposed at the surfaces given by $r_A = b_A$ which constitutes
the fixed-point set of the isometry at each hole.
Here $N_A^i$ is the normal to the surface $r_A = b_A$.
The presence of these inner boundaries complicates the numerical method,
when we compare with the puncture approach.
For this reason these initial data are rarely used in practice.
Furthermore, in many applications one is interested only in the spacetime
outside the black holes, so that it is not important whether the initial
data indeed possess an isometry.

%%%%%%%%%%%%%%%%%%%%%%%%%%%%%%%%%%%%%%%%%%%%%%%%%%%%%%%%%%%%%%%%%%%
\subsection{Superimposed Kerr-Schild initial data}
%%%%%%%%%%%%%%%%%%%%%%%%%%%%%%%%%%%%%%%%%%%%%%%%%%%%%%%%%%%%%%%%%%%
\label{SKS_approach}

As we have seen above, both puncture initial data as well as Bowen-York
initial data are generalizations of the Schwarzschild
solution in isotropic coordinates. Both use a conformally flat 3-metric and
make use of analytic solutions to the momentum constraints, the so called
Bowen-York extrinsic curvature given in Eqs.~(\ref{Bowen-York1}) and
(\ref{Bowen-York2}). Let us consider a single black hole with spin only,
i.e. the extrinsic curvature is given by Eq.~(\ref{Bowen-York1})
with $c^i=0=P^i$ and a non-zero $S^i$. Once we solve the
Hamiltonian constraint for the conformal factor $\psi$, we obtain valid
initial data. These data, however, cannot be a spatial slice of the
stationary Kerr spacetime,
since no spatial slicing of Kerr exists with a conformally flat
3-metric~\cite{Garat:2000pn,ValienteKroon:2003ux,ValienteKroon:2004gj}.
This means that these initial data do not
describe a stationary situation. Rather, we obtain a black hole that is
surrounded by gravitational waves,
of which some will move out and carry away some of
the angular momentum. In fact, if we increase the magnitude of $S^i$, 
the ADM mass also increases so that the dimensionless spin
ratio $|S^i|/M_{ADM}^2$ does not increase above about
$0.928$~\cite{Dain:2002ee,Lovelace:2008tw}. 
Hence we cannot come close to the
extreme Kerr limit of $|S^i|/M_{ADM}^2 =1$.
Analogous effects occur if we consider a single moving black hole with
$S^i=0$ and a non-zero $P^i$. Again the data contain gravitational waves
that themselves will contain some of the momentum.

For this reason another approach has been 
developed~\cite{Matzner98a,Marronetti:2000yk}
where one starts from the single spinning Kerr solution given in
Eq.~(\ref{eq:KSmetric}). In order to obtain a moving black hole,
the 4-metric of Eq.~(\ref{eq:KSmetric}) can be boosted. 
Due to its special structure the 4-metric in Eq.~(\ref{eq:KSmetric})
is invariant under boosts. After the boost it is given by
\be
\label{eq:KSmetric_boosted}
g_{\mu'\nu'} = \eta_{\mu'\nu'}+ 2H' k_{\mu'} k_{\nu'} .
\ee
where $H'$ and $k_{\mu'}$ can be obtained by transforming the expressions
in Eqs.~(\ref{KS-H}) and (\ref{KS-k}) like a scalar and a vector respectively.
Since $H k_{\mu} k_{\nu}$ in the 4-metric drops off away
from the black hole, a superposition of two boosted black holes of the form
\be
g_{\mu\nu} = \eta_{\mu\nu} 
+ 2H^1 k^1_{\mu} k^1_{\nu} + 2H^2 k^2_{\mu} k^2_{\nu} ,
\ee
will be an approximate solution to Einstein's equations, that does not
satisfy the constraints exactly.
This approximation can then be used, as explained after 
Eq.~(\ref{approx_as_free_data}), to generate constraint satisfying
initial data.
This idea has first been used within the context of the
CTT decomposition~\cite{Marronetti:2000yk}.
More recently it has been also used within the CTS
approach to generate very highly spinning black hole
binaries~\cite{Lovelace:2008tw}. 
Since all time derivatives are assumed to be zero in corotating coordinates,
one has to specify only the conformal 3-metric and the trace of the
extrinsic curvature. They are then superposed according to
\begin{eqnarray}
\label{SKS_gamma}
\bar{\gamma}_{ij} &=& \delta_{ij} + 
\sum_A e^{-r_A^2/w_A^2}({\gamma}^A_{ij} - \delta_{ij}) \\
\label{SKS_K}
K &=& \sum_A e^{-r_A^2/w_A^2} K^A ,
\end{eqnarray}
where ${\gamma}^A_{ij}$ and $K^A$ are the the conformal 3-metric and the
trace of the extrinsic curvature of the boosted spinning Kerr-Schild 
black hole $A$. 
Here $e^{-r_A^2/w_A^2}$ is a Gaussian weighting factor that depends on a
length scale $w_A$ and the conformal distance $r_A$ from black hole $A$.
This factor is similar to the attenuation factors
in~\cite{Marronetti:2000yk,Bonning:2003im}. It ensures that near each black
hole we have a metric that is very close to a single black hole solution.
The length scale $w_A$ is chosen to be larger than the size of 
black hole $A$, but smaller than the distance to the nearest neighboring
black hole.

As boundary conditions \cite{Lovelace:2008tw} uses
Eqs.~(\ref{psi_BC_inf}), (\ref{alphabar_BC_inf}) and
(\ref{beta_BC_inf_rot}) at infinity. The apparent horizon
$\cal{S}_A$ of black hole $A$ is chosen to be the coordinate 
location of the event horizon of the boosted spinning black hole $A$
in the superposition. For the conformal factor and the shift
Eqs.~(\ref{psi_BC_BH}) and (\ref{beta_BC_BH}) are used as boundary
conditions on $\cal{S}_A$. The lapse boundary condition there is
given by
\be
\alpha\psi |_{\cal{S}_A}
= 1 + \sum_A e^{-r_A^2/w_A^2} (\alpha^A - 1) |_{\cal{S}_A} .
\ee

With this approach it is possible to construct initial data with very highly
spinning black holes. In~\cite{Lovelace:2008tw} a dimensionless spin of
0.9997 is achieved. Further improvements for such data constructed within
the CTS approach are presented in~\cite{Ossokine:2015yla}.

% THIS PAPER SEEMS UNPUBLISHED:
%In~\cite{Ruchlin:2014zva} a similar approach is described where
%the CTT Eqs.~(\ref{hamCTT}) and (\ref{momCTT}) are solved for a
%superposition of two Kerr black holes in coordinates where the 3-metric
%is close to conformally flat. In this approach it is possible to use a
%puncture approach so that no inner boundaries are needed.
%In~\cite{Ruchlin:2014zva} a dimensionless spin of 0.9869 is reached.

%%%%%%%%%%%%%%%%%%%%%%%%%%%%%%%%%%%%%%%%%%%%%%%%%%%%%%%%%%%%%%%%%%%
\subsection{Conformally flat CTS binary black hole data}
%%%%%%%%%%%%%%%%%%%%%%%%%%%%%%%%%%%%%%%%%%%%%%%%%%%%%%%%%%%%%%%%%%%

The CTS approach can also be used with conformally flat 
initial
data~\cite{Cook:2004kt,Caudill:2006hw,Pfeiffer:2007yz,Buchman:2012dw}.
In this case we assume all time derivatives are zero and simply set 
\begin{eqnarray}
\bar{\gamma}_{ij} &=& \delta_{ij} \\
K &=& 0 .
\end{eqnarray}
As boundary conditions at infinity one uses
Eqs.~(\ref{psi_BC_inf}), (\ref{alphabar_BC_inf}) and
(\ref{beta_BC_inf_rot}). The fact that we are dealing with black holes
is incorporated into the initial data by specifying two conformal 
coordinate spheres $\cal{S}_A$ where one imposes the apparent horizon
boundary conditions (\ref{psi_BC_BH}) and (\ref{beta_BC_BH}) for the
conformal factor and the shift. Usually one uses
\be
\frac{d\alpha\psi}{dr} \Big|_{\cal{S}_A} = 0
\ee
for the lapse boundary condition at the apparent horizons.
Such conformally flat data are easier to construct than superimposed
Kerr-Schild type initial data. But they suffer from the same problems as
puncture initial data, in that one cannot generate highly spinning black
holes.

%%%%%%%%%%%%%%%%%%%%%%%%%%%%%%%%%%%%%%%%%%%%%%%%%%%%%%%%%%%%%%%%%%%
\subsection{Excision versus puncture approaches}
%%%%%%%%%%%%%%%%%%%%%%%%%%%%%%%%%%%%%%%%%%%%%%%%%%%%%%%%%%%%%%%%%%%   

When we construct puncture initial data, we have no inner boundary
conditions and thus obtain data everywhere. However, if we use
the apparent horizon
boundary conditions (\ref{psi_BC_BH}) and (\ref{beta_BC_BH})
to construct black holes, in either the superimposed Kerr-Schild 
or the conformally flat CTS approach, we obviously obtain initial data
only outside the apparent horizon. The lack of data inside the black holes
leads to complications when we evolve such data numerically.
First of all, one needs to use a more complicated topology for the numerical
domain, where two spheres have to be excised. 
This approach is called black hole excision. Once this is done,
one might think that it should be possible to evolve only the outside of the
black holes, since nothing physical can come out of a black hole horizon. 
This means we should impose no boundary condition on any physical field or
mode at the horizon, since no physical mode can enter 
the numerical domain outside the horizon. This approach
works well for evolution systems such as the generalized harmonic evolution
system~\cite{Pretorius:2004jg,Pretorius:2005gq} that have no 
superluminal modes, when using their standard gauges.
We note however, that the often used BSSNOK system~\cite{Baumgarte:1998te}
as well as the Z4c system~\cite{Bernuzzi:2009ex,Ruiz:2010qj} possess
superluminal gauge modes that can cross the horizons, when the usual moving
puncture formalism~\cite{Campanelli:2005dd,Baker:2005vv} is used.
Thus one needs boundary conditions for some modes, which further complicates
the numerics. For this reason, one sometimes fills the black hole
interiors~\cite{Etienne:2007hr,Brown:2007pg,Brown:2008sb,Reifenberger:2012yg}
with approximate data when using the BSSNOK system.
Both black hole excision and black hole filling techniques can work. 
Nevertheless puncture initial data are easier to evolve 
with the BSSNOK system using the moving puncture 
formalism, since neither excision nor filling are then required.
For evolutions with the generalized harmonic evolution
system, black hole excision is required even for puncture initial data,
since the system in its standard form
cannot handle the divergences in the conformal factor and
the Bowen-York extrinsic curvature.
Notice, however, that with a recent modification of the generalized harmonic
system, where the 3-metric is conformally rescaled as in the BSSNOK system,
it has been possible to evolve Brill-Lindquist initial data
without excision~\cite{Mosta:2015sga}.

%BH stuffing from http://lanl.arxiv.org/pdf/1504.00286v1:
%
%[36] Z. B. Etienne, J. A. Faber, Y. T. Liu, S. L. Shapiro,
%and T. W. Baumgarte, Phys. Rev. D76, 101503 (2007),
%arXiv:0707.2083 [gr-qc]. ~\cite{Etienne:2007hr}
%[37] J. D. Brown, O. Sarbach, E. Schnetter, M. Tiglio,
%P. Diener, et al., Phys. Rev. D76, 081503 (2007),
%arXiv:0707.3101 [gr-qc]. ~\cite{Brown:2007pg}
%[38] J. D. Brown, P. Diener, O. Sarbach, E. Schnetter,
%and M. Tiglio, Phys. Rev. D79, 044023 (2009),
%arXiv:0809.3533 [gr-qc]. ~\cite{Brown:2008sb}

%%%%%%%%%%%%%%%%%%%%%%%%%%%%%%%%%%%%%%%%%%%%%%%%%%%%%%%%%%%%%%%%%%%
\subsection{Quasi-circular orbits}
%%%%%%%%%%%%%%%%%%%%%%%%%%%%%%%%%%%%%%%%%%%%%%%%%%%%%%%%%%%%%%%%%%%   

Above we have discussed the examples of puncture and superimposed Kerr-Schild
initial data. In both cases we can freely choose the momentum of both black
holes.
As we have discussed already in Sec.~\ref{Quasiequil_geometry}, we expect
most binaries to be in circular orbits with a slowly shrinking radius.
The question thus arises how one should pick the black hole momenta
for such a quasi-circular configurations.

For punctures an answer has been given in~\cite{Tichy:2003zg,Tichy:2003qi}.
We construct puncture initial data (using the standard CTT decomposition) for
two black holes. We choose the center of mass to be at rest and let $P_1^i$
and $P_2^i$ be perpendicular to the line connecting the two black holes. The momenta
are then characterized by the parameter $P = |P_1^i| = |P_2^i|$. It is
assumed that an approximate helical Killing vector $\xi^{\mu}$ satisfying
Eq.~(\ref{approxKV}) exists for the correct choice of $P$. We choose
coordinates, i.e. a lapse and shift such that the time evolution vector
$t^{\mu}=\xi^{\mu}$. In these coordinates all time derivatives should
approximately be zero. We then use $\partial_t K=0$ to turn the evolution
equation (\ref{K-evo}) into an elliptic equation for the lapse. 
We find Eq.~(\ref{alpha_CTS}) with $K=0=\bar{R}$ and $\rho=0=S$ for 
punctures in vacuum. This
elliptic equation can be solved for $\alpha$, with the puncture ansatz
\be
\label{alphapsiPuncAnsatz}  	
{\alpha}\psi = 1 - \left( \frac{c_1 m_1}{2r_1} +
         			\frac{c_2 m_2}{2r_2} \right) + v ,
\ee
where $v$ is a function that is regular at $r_A=0$.
Then Eq.~(\ref{alpha_CTS}) becomes
\be
\label{puncture_v}
\nabla^2 v = \frac{7}{8} 
({\alpha}\psi) \psi^{-8} \bar{A}_{ij} \bar{A}^{ij} .
\ee
The value of the lapse needs to be specified at
$r\to\infty$ and in the other two asymptotically flat regions at $r_1=0$ and
$r_2=0$. For $r\to\infty$ we use $\alpha=1$. The other two values
$\alpha|_{r_1=0}=-c_1$ and $\alpha|_{r_2=0}=-c_2$ 
are adjusted together with $P$ until the equality of Komar integral and
ADM mass of Eq.~(\ref{MK_MADM}) is satisfied in all three asymptotically
flat regions. This yields a unique $P$.

For initial data constructed within the CTS decomposition it is somewhat
easier to find the black hole momenta corresponding to the quasi-circular
orbits, since we can directly set certain time derivatives to zero, as
required when the time evolution vector $t^{\mu}$ is chosen to lie along an
approximate helical Killing vector $\xi^{\mu}$. To then solve the CTS
equations for say the superimposed Kerr-Schild initial data, we need to
specify the value of $\Omega$ in the shift boundary condition
(\ref{beta_BC_inf_rot}). We can then simply adjust $\Omega$ (as well as the
corresponding boost velocities of the individual Kerr-Schild holes) until
Eq.~(\ref{MK_MADM}) is satisfied.

\begin{figure}
\includegraphics[scale=0.6,clip=true]{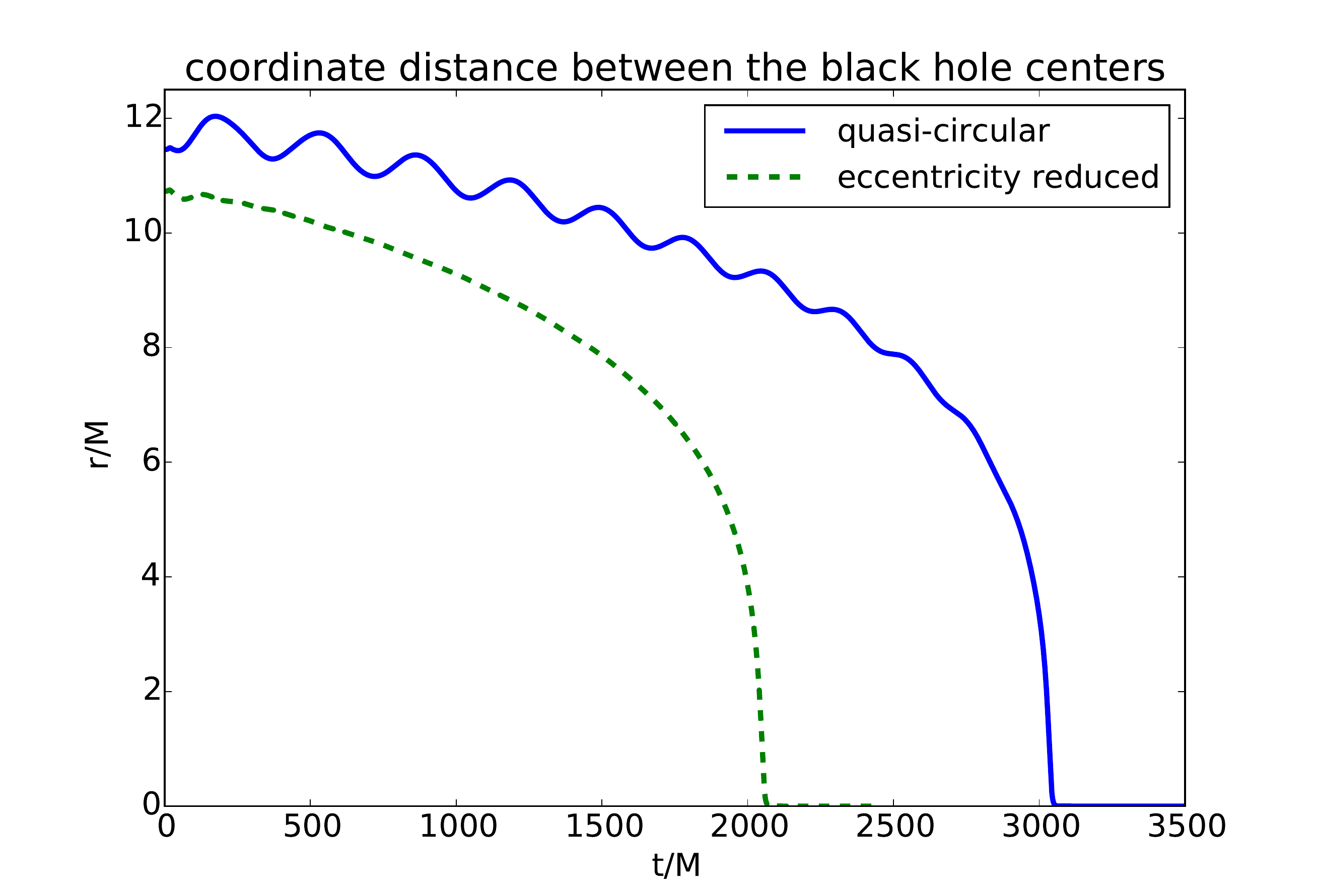}
\caption{\label{puncture_distance_evo}
% tgraph.py -v 0:12.5 om0.023q3S0.4_180_0_S0.6_0_0_sep.txt om0.025etmq3S0.4_0_0_S0.6_0_0_N96_sep_ecc.txt
% data is from:
% om0.023q3S0.4_180_0_S0.6_0_0_sep.txt
% om0.025etmq3S0.4_0_0_S0.6_0_0_N96_sep_ecc.txt
% the second one is ecc. reduced
This plot shows the coordinate distance $r$ for two different binary black
hole simulations versus time. In both cases the mass ratio is 3 and the 
spins are pointing along the orbital angular momentum with 
dimensionless spin magnitude of 0.4 for the smaller and 0.6 for the larger
black hole. Both simulations are 
starting from puncture initial data, but with slightly different initial
separations and black hole momenta.
The solid line corresponds to the quasi-circular setup where we have
only tangential momentum, chosen by the equality of Komar integral and
ADM mass. The broken line is for data where we have added a small radial
momentum and tuned the tangential momentum to obtain less eccentric 
orbits~\cite{Tichy:2010qa}.
}
\end{figure}
When we use Eq.~(\ref{MK_MADM}) to choose the black hole momenta, we obtain
quasi-circular initial data. These initial data can then be evolved to study
the orbits of the black hole binaries. For both punctures and the
superimposed Kerr-Schild initial data constructed within the CTS approach,
we find that the orbits are reasonably circular inspiral orbits. The
solid line in Fig.~\ref{puncture_distance_evo} shows the evolution of the
coordinate distance between the two black holes when we start from puncture
initial data and choose the initial momenta perpendicular to the line
connecting the two black holes, and use the equality of Komar integral and ADM mass
(discussed above) to determine the magnitude of the initial black hole
momentum parameter. As we can see the distance between the two black holes
is not monotonically decreasing during the inspiral due to some residual
eccentricity. This can be explained by the fact that in this case
the initial data are constructed in an approximation that assumes that the
orbits are exactly circular, i.e. both black holes have
only tangential momentum. Yet, in a real inspiral the black holes also have
a small inward momentum component. This means the initial momenta are not
the ones needed for a realistic inspiral situation.
The situation can be improved if we also allow for a radial momentum
component and adjust the tangential momentum magnitude. When this is done
according to the prescription described in~\cite{Tichy:2010qa} we obtain an
inspiral with much less eccentricity, as shown by the broken line in
Fig.~\ref{puncture_distance_evo}. By using such
methods~\cite{Marronetti:2007wz,Marronetti:2007ya,Brugmann:2008zz,
Husa:2007rh,Walther:2009ng,Tichy:2010qa} it is possible to achieve
eccentricities on the order of $10^{-4}$ for puncture initial data.

As already mentioned, for initial data constructed within the CTS approach
we also find eccentric orbits comparable to the solid line in
Fig.~\ref{puncture_distance_evo}, if we assume a helical Killing vector and
construct orbits with only tangential velocity. The reason is again that a
true inspiral does not have exactly a helical symmetry. However,
as described in~\cite{Pfeiffer:2007yz,Boyle:2007ft} one can replace
this helical symmetry by an inspiral symmetry, i.e. by assuming that the
approximate Killing vector has the form
\begin{eqnarray}
\label{inspiral_KV}
\xi_{insp}^0 &=& 1, \nonumber\\
\xi_{insp}^i &=& \Omega (-x^2+x_{CM}^2, x^1-x_{CM}^1, 0) +
\frac{v_r}{r_{12}} (x^i-x_{CM}^i) ,
\end{eqnarray}
where we have added a radial component to the helical vector and assume that
orbital angular velocity is along the $x^3$-direction.
Here $v_r$ is the radial velocity, $x_{CM}^i$ the center of mass position and
$r_{12}$ the distance between the two black holes. We can now adjust
$\Omega$ and $v_r$ to mimic true inspiral orbits.
In comoving coordinates we still have 
\be
\xi^{\mu} = t^{\mu} = \alpha n^{\mu} + \beta^{\mu} ,
\ee
so that at spatial infinity the shift now must have the form
\be
\label{beta_BC_inf_insp}
\lim_{r\to\infty}\beta^{i} = \xi_{insp}^i .
\ee
Within the CTS equations we can now use Eq.~(\ref{beta_BC_inf_insp}) as a
boundary condition on the shift at $r\to\infty$. Using an iterative
method~\cite{Pfeiffer:2007yz,Boyle:2007ft} it is possible to tune the values
of $\Omega$ and $v_r$ to achieve very low eccentricities on the order of
$5\times 10^{-5}$~\cite{Boyle:2007ft} for conformally flat CTS initial data.
Similar results can also obtained using superimposed Kerr-Schild CTS initial
data~\cite{Lovelace:2008tw}. In fact, this kind of eccentricity reduction
works better for CTS based initial data than for CTT puncture initial data.
The CTS decomposition gives us a direct handle on the shift (via
Eq.~(\ref{beta_BC_inf_insp})) and allows us to directly set certain time
derivatives to zero. It produces a preferred lapse and shift that when used
in the evolution, result in coordinates that are well adapted to
quasi-equilibrium. For CTT based initial data one does not have
such a preferred lapse and shift.
Hence oscillations in the coordinate distance $r$
are not due to real eccentricity alone, but also due to the
fact that the coordinates are still evolving as well. This
problem is quite visible after eccentricity reduction
in Fig.~\ref{puncture_distance_evo}.
The broken line shows a dip during the first $200M$ of the evolution.
This non-monotonic behavior makes it harder to define or measure
eccentricity, so that the reduction procedure works less well.

%%%%%%%%%%%%%%%%%%%%%%%%%%%%%%%%%%%%%%%%%%%%%%%%%%%%%%%%%%%%%%%%%%%
\section{Toward more realistic binary black hole initial data}
%%%%%%%%%%%%%%%%%%%%%%%%%%%%%%%%%%%%%%%%%%%%%%%%%%%%%%%%%%%%%%%%%%%   
\label{Realistic_BBH_data}

\begin{figure}
\begin{center}
\includegraphics[width=\textwidth]{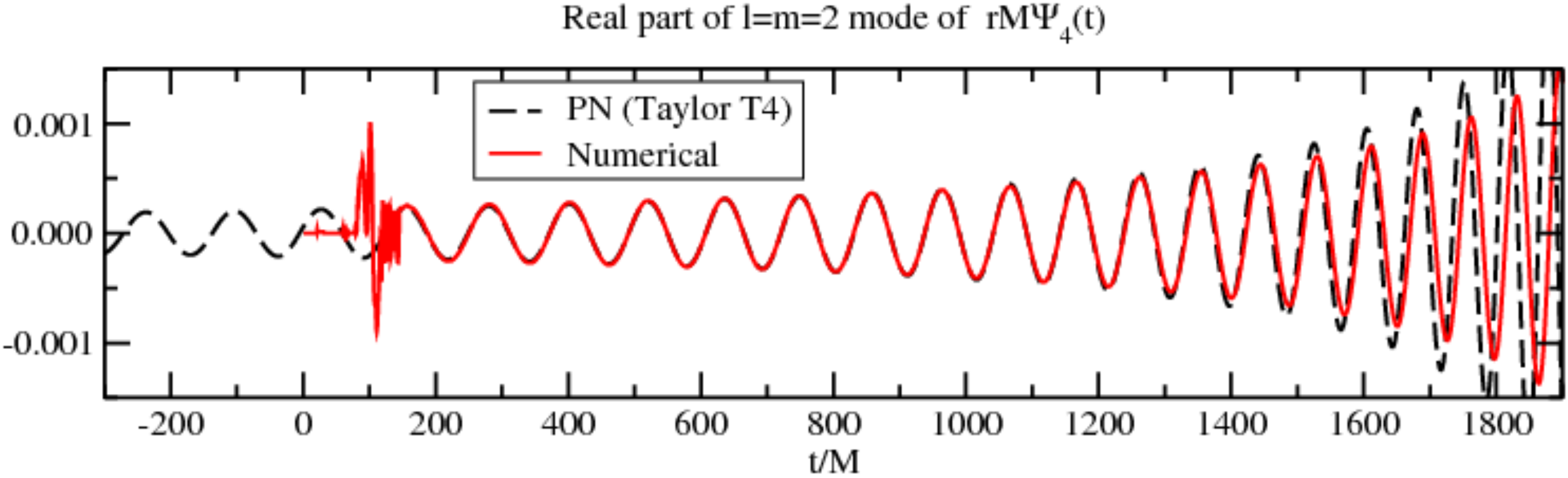}
\caption{ \label{NINJA2hybrid}
\small{The solid line shows the gravitational waveform for a binary black
hole simulation starting from puncture initial data with
masses $m_1=0.25M$, $m_2=0.75M$ and spins of $S_1=0.4M^2$, $S_2=0.6M^2$
aligned with the orbital angular momentum.
A post-Newtonian Taylor T4 waveform (broken line) has been
matched to the numerical
waveform by minimizing the difference in the 
time interval from $350M$ to $1100M$.
%The PN waveform can then be used instead of the numerical waveform at
%earlier times.
}
}
\end{center}
\vspace{-0.7cm}
\end{figure}

As discussed above, two orbiting black holes will emit gravitational
radiation. From post-Newtonian calculations we expect to see a so called
chirp signal during the inspiral phase,
i.e. a gravitational wave with slowly increasing amplitude and
frequency (see e.g. broken line in Fig.~\ref{NINJA2hybrid}).
The initial data should reflect this and contain such gravitational
waves.
However, when we construct conformally flat initial data using either
the CTT or CTS approach, the resulting conformal factor will be
monotonically decreasing as we move away from the black holes. Hence the
3-metric does not contain any wiggles. The same is true for the extrinsic
curvature. This means that the initial slice of our spacetime contains
no gravitational radiation at all, even though we have two orbiting black
holes. This is clearly unrealistic, even though the data satisfy the
constraints by construction. 
When such data are evolved in time the system starts to radiate
and over time fills the spatial slice with gravitational waves.
The results of a evolution starting from conformally flat puncture initial
data is shown in Fig.~\ref{NINJA2hybrid}. The solid line depicts the
dominant mode of the Weyl scalar $\Psi_4$ which encodes the emitted
gravitational waves. This mode has been extracted at a distance of about
$90M$ from the center of mass. As we can see, there is no gravitational
wave signal until disturbances from the black holes reach the extraction
radius at a time of about $90M$. At this time we see a strong transient
high frequency signal.
This signal is usually referred to as junk radiation, since it is an
artifact of the unrealistic initial data.
The broken line shows the same mode computed from a
corresponding post-Newtonian calculation, which contains no such junk
radiation.
Similar problems also occur for superimposed Kerr-Schild type data, because
each individual black hole in the superposition is a stationary black hole
solution of Einstein's equations that does not contain any gravitational
waves.

%%%%%%%%%%%%%%%%%%%%%%%%%%%%%%%%%%%%%%%%%%%%%%%%%%%%%%%%%%%%%%%%%%%
\subsection{Post-Newtonian based initial data}
%%%%%%%%%%%%%%%%%%%%%%%%%%%%%%%%%%%%%%%%%%%%%%%%%%%%%%%%%%%%%%%%%%%   

We know that post-Newtonian calculations can be highly accurate during the
inspiral phase before the black holes merge. This is also the regime in
which we would like to construct initial data. It thus seems natural that we
should try to incorporate post-Newtonian information into the initial data
construction. Notice, however, that post-Newtonian theory is an
approximation that assumes low velocities ($v/c \ll 1$) and weak gravity
($GM/r \ll 1$). While black holes may be moving slowly enough when they are
still well separated, their gravitational fields are always strong near each
black hole. For this reason post-Newtonian theory alone can give us reliable
initial data only away from each black hole. In~\cite{Tichy02,Kelly:2007uc}
post-Newtonian initial data in ADMTT gauge~\cite{Jaranowski98a} has
been investigated. In this gauge one can directly obtain the
3-metric and the extrinsic curvature as post-Newtonian expansions. It is
then possible to resum these expansion so that they take the form
\begin{equation}
\label{ADMTT-metric}
\gamma^{PN}_{ij} = \psi_{PN}^{4} \delta_{ij} + h_{ij}^{TT} 
\end{equation}
and 
\begin{eqnarray}
K_{PN}^{ij} = \psi_{PN}^{-10}\left( \bar{A}_{BY}^{ij} + k^{ij}_{TT} \right)
\end{eqnarray}
with 
\be
\label{psiPN}
\psi_{PN} = 1 + \sum_{A=1}^2 \frac{E_A}{2r_A} .
\ee
Here $E_A$ are the energies of the point particles used in the
post-Newtonian theory, $h_{ij}^{TT}$ and $k^{ij}_{TT}$ are 
higher order post-Newtonian terms that we do not write out here, and
$\bar{A}_{BY}^{ij}$ is the Bowen-York extrinsic curvature given in
Eq.~(\ref{Aij_Punc}) with $S_A^i=0$. 
As we can see the conformal factor is of Brill-Lindquist
form as in Eq.~(\ref{psiBL}). Thus these approximate initial
data look very similar to puncture initial data. The main difference is the
appearance of the extra terms $h_{ij}^{TT}$ and $k^{ij}_{TT}$, so that the
3-metric is no longer conformally flat.
It is thus clear that these data do indeed contain black holes, even though
the post-Newtonian expressions were derived using point particles.
Following the approach in~\cite{Tichy02,Kelly:2007uc} we can now construct
constraint satisfying initial data by using the CTT decomposition with
the free data
\begin{eqnarray}
\bar{\gamma}_{ij} &=& \psi_{PN}^{-4} \gamma^{PN}_{ij} \\
\bar{M}^{ij}      &=& \psi_{PN}^{10}
\left( K_{PN}^{ij} -\frac{1}{3}\gamma_{PN}^{ij} K_{PN} \right) \\
K                 &=& 0
\end{eqnarray}
and using the puncture ansatz 
\be
\label{PN_u}
\psi = \psi_{PN} + u
\ee
for the conformal factor. Inserting these expressions in Eqs.~(\ref{hamCTT})
and (\ref{momCTT}) we obtain elliptic equations for $u$ and $W^i$.
These equations can be solved if we impose the boundary conditions 
\be
\lim_{r \to \infty} u = 0
\ee
and Eq.~(\ref{W_BC_inf}). We then
obtain initial data that satisfy Hamiltonian and momentum
constraints. This was first done in~\cite{Tichy02} using a near zone
approximation for the term $h_{ij}^{TT}$ in Eq.~(\ref{ADMTT-metric}), but
the full $h_{ij}^{TT}$ was later calculated in~\cite{Kelly:2007uc}, and used
to solve the constraints in~\cite{Reifenberger2013}.
As we can see from Eq.~(\ref{PN_u}) the conformal factor gets modified when
we solve for the constraint Eqs.~(\ref{hamCTT})
and (\ref{momCTT}). This leads in general to an increase in the ADM mass.
Such changes are common when we use one of the conformal decompositions to
find constraint satisfying data from approximate data.
In \cite{Tichy02} and \cite{Reifenberger2013} it was found that 
for the case of the post-Newtonian data in ADMTT gauge, this change
in the mass can be prevented by also modifying the post-Newtonian
conformal factor to
\be
\label{qfac}
\psi_{PN} \to \psi_{PN}
 - q \frac{m_1 m_2}{2r_{12}}\left(\frac{1}{2r_1}+\frac{1}{2r_2}\right) .
\ee
This modification (parameterized by the number $q$) counters
the effect of the term $u$ in Eq.~(\ref{PN_u}) on the ADM mass.
The parameter $q$ should be chosen such that the resulting initial data
after solving the constraints in Eqs.~(\ref{hamCTT}) and (\ref{momCTT})
are as close as possible to the original post-Newtonian data. 
How exactly one best quantifies this closeness is still an open issue.
A partial answer has been given e.g. in~\cite{Tichy02} where $q$ was chosen
such that the binding energy after solving the constraints is as close as
possible to the post-Newtonian binding energy.

\begin{figure}
\vspace{-0.4cm}
\begin{center}
\rotatebox{-90}{
\includegraphics[width=0.7\textwidth]{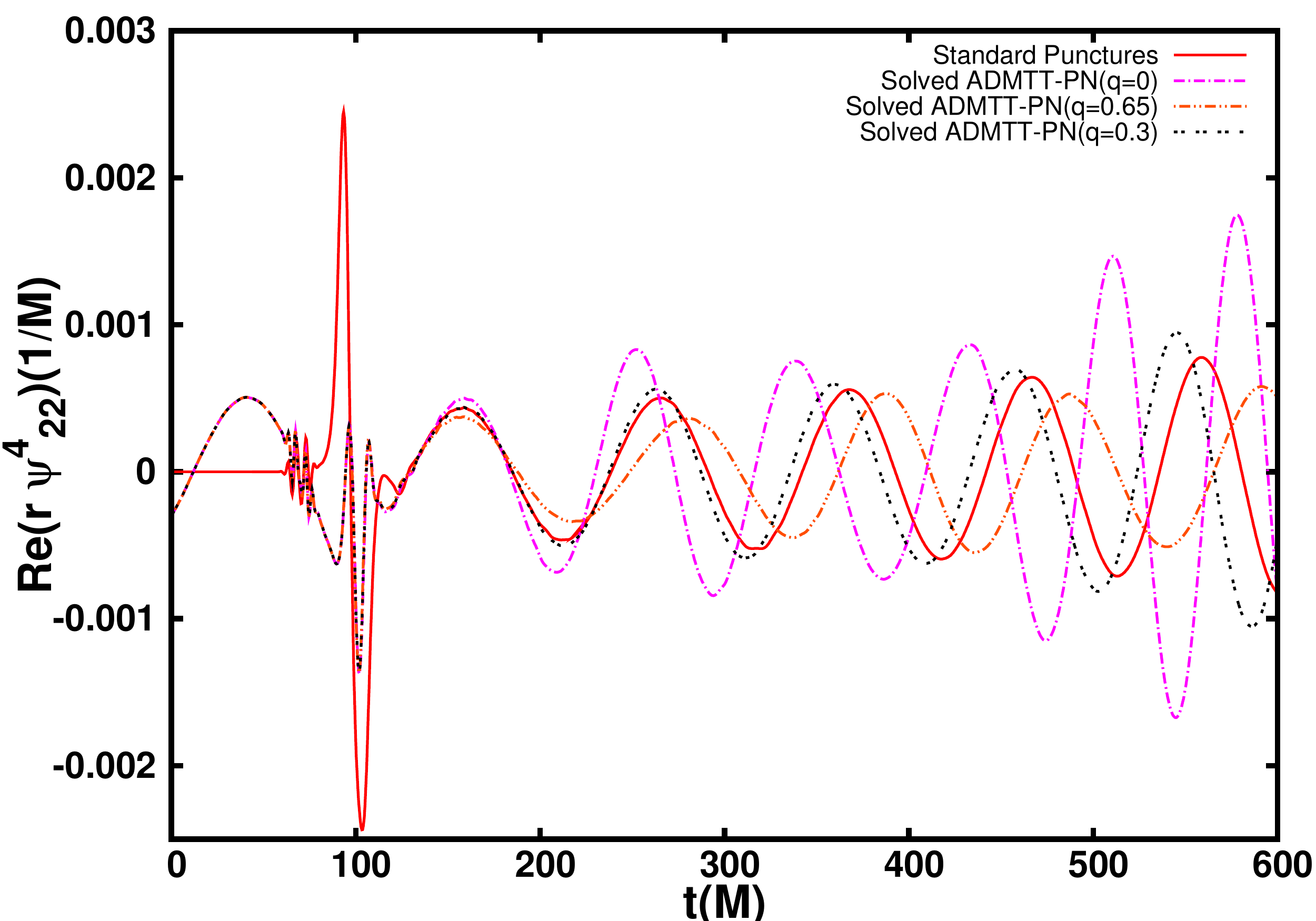}
}
\caption{ \label{junk_in22}
\small{
Results from the simulation of two non-spinning equal mass black holes
starting from an initial coordinate separation of $10M$.
The solid line shows the gravitational waveform for a binary black
hole simulation starting from standard puncture initial data.
Up to a time of about $t=90M$ we see no signal, because no waves are
built into puncture data. After this time we see a strong junk signal.
The other lines show the gravitational waveform when we start with
post-Newtonian based initial data 
(with different values for the parameter $q$),
where waves are built in from the start.
These signals still contain junk radiation but with a reduced amplitude.
}
}
\end{center}
\vspace{-0.5cm}
\end{figure}

In Fig.~(\ref{junk_in22}) we show the dominant $l=m=2$ mode of the
gravitational waveform for different initial data.
As we can see, the ADMTT based initial data leads to better waveforms with
reduced junk radiation.
Similar improvements in the waveform even exist if we do not solve the
constraint equations and just directly use the approximate post-Newtonian
data~\cite{Kelly:2009js,Reifenberger:2012yg}.

%%%%%%%%%%%%%%%%%%%%%%%%%%%%%%%%%%%%%%%%%%%%%%%%%%%%%%%%%%%%%%%%%%%
\subsection{Approximate initial data from matching}
%%%%%%%%%%%%%%%%%%%%%%%%%%%%%%%%%%%%%%%%%%%%%%%%%%%%%%%%%%%%%%%%%%%   

As we have seen, post-Newtonian theory can give very useful input when we
want to construct initial data with gravitational waves already
built into the initial slice. Yet, as discussed in the previous subsection,
post-Newtonian theory is not valid near a black hole. In fact one had to
judiciously resum the post-Newtonian expressions to even obtain black holes.
These resummations were guided by the aim to make the post-Newtonian data
similar to puncture initial data, because the latter contains well
understood black holes. Of course this resummation is ad hoc, and it is not
clear if we really obtain a good approximation near the black holes and
whether the black holes are in equilibrium.
In fact, since there is still some
amount of junk radiation, we do know that they are not really equilibrium.
For this reason an alternative method based on asymptotic matching has been
proposed~\cite{Alvi99,Yunes:2005nn,Yunes:2006iw,JohnsonMcDaniel:2009dq}. In
this method one matches a post-Newtonian 4-metric to two different perturbed
black hole metrics (one for each black hole) to obtain an approximate
solution everywhere. Then no ad hoc resummations are necessary, and we obtain
an approximation that is valid everywhere. The most sophisticated data of
this kind to date has been presented in~\cite{JohnsonMcDaniel:2009dq},
where a post-Newtonian 4-metric for circular orbits
containing outgoing gravitational waves has been matched to two tidally   
perturbed Schwarzschild metrics.
When we evolve initial data coming from this approximate 4-metric without
solving the constraint equations, we find that the data have waves built in
from the start and that there is less junk radiation than for puncture
initial data~\cite{Reifenberger:2012yg}. However, somewhat surprisingly the
junk radiation in the dominant $l=m=2$ mode has still about the same size as
the one coming from the resummed ADMTT gauge post-Newtonian data discussed
in the previous subsection, at least for a black hole separation of $10M$.
However, further studies are necessary as it is likely that the situation
will improve once we move the black holes further apart, since the 
then all the approximations used in the matching will improve. 

In~\cite{Chu:2013nkx} the two tidally deformed black hole 4-metrics
from~\cite{JohnsonMcDaniel:2009dq} (but not the post-Newtonian 4-metric)
have been used to construct a superposition as in Eqs.~(\ref{SKS_gamma}) and
(\ref{SKS_K}). This superposition thus contains black holes with the correct
tidal deformations, but without any realistic gravitational waves, since the
matched post-Newtonian 4-metric is not included. The constraint equations are
then solved for this superposition following the methods described in
Sec.\ref{SKS_approach} within the CTS approach. The only difference is that
as boundary condition at each black hole, \cite{Chu:2013nkx} uses
a Dirichlet condition for the shift, since tidally deformed black
holes do not have a vanishing shear, so that Eq.~(\ref{beta_BC_BH})
may not be appropriate. When these data are evolved,
junk radiation is reduced when compared to conformally flat
or superimposed Kerr-Schild data.

% Should I add something about George's quasi-circ?: No. It's a detail.

% % This seems not so important as well:
% 65. arXiv:1108.3550 [pdf, other]
% Puncture black hole initial data in the conformal thin-sandwich formalism
% Thomas W. Baumgarte
% This is a single trumpet puncture

%%%%%%%%%%%%%%%%%%%%%%%%%%%%%%%%%%%%%%%%%%%%%%%%%%%%%%%%%%%%%%%%%%%
\section{Matter equations for perfect fluids}
%%%%%%%%%%%%%%%%%%%%%%%%%%%%%%%%%%%%%%%%%%%%%%%%%%%%%%%%%%%%%%%%%%%
\label{Matter}

So far we have been discussing how one can create initial data for vacuum
solutions such as black holes. When we need initial data for situations
where matter is involved, we have to consider additional equations that
describe this matter. 
Here we will concentrate on matter that can be described as a perfect fluid.
This is sufficient for e.g. neutron stars.

%%%%%%%%%%%%%%%%%%%%%%%%%%%%%%%%%%%%%%%%%%%%%%%%%%%%%%%%%%%%%%%%%%%
\subsection{Perfect fluids}
%%%%%%%%%%%%%%%%%%%%%%%%%%%%%%%%%%%%%%%%%%%%%%%%%%%%%%%%%%%%%%%%%%%

For a perfect fluid the stress-energy tensor is given by
\begin{equation}
T^{\mu\nu} = [\rho_0(1+\epsilon) + P] u^{\mu} u^{\nu} + P g^{\mu\nu}.
\end{equation}
Here $\rho_0$ is the mass density (which is proportional the number
density of baryons), $P$ is the pressure, $\epsilon$ is the internal energy
density divided by $\rho_0$ and $u^{\mu}$ is the 4-velocity of the fluid.
The matter variables in Eq.(\ref{mattervars}) are then
\begin{eqnarray}
\label{fluid_matter}
\rho   &=& \alpha^2 [\rho_0(1+\epsilon) + P] u^0 u^0 - P \nonumber \\
j^i    &=& \alpha[\rho_0(1+\epsilon) + P] u^0 u^0
           (u^i/u^0 + \beta^i) \nonumber \\
S^{ij} &=& [\rho_0(1+\epsilon) + P]u^0 u^0 
           (u^i/u^0 + \beta^i) (u^j/u^0 + \beta^j) + P \gamma^{ij} ,
\end{eqnarray}
where $u^0 = -n_{\mu}u^{\mu}/\alpha$.

From $\nabla_{\nu} T^{\mu\nu} =0 $ we obtain the relativistic
Euler equation
\begin{equation}
\label{Euler0}
[\rho_0(1+\epsilon) + P] u^{\nu} \nabla_{\nu} u^{\mu} 
= -(g^{\mu\nu} + u^{\mu} u^{\nu}) \nabla_{\nu} P, 
\end{equation}
which together with the continuity equation
\begin{equation}
\label{continuity0}
\nabla_{\nu} (\rho_0 u^{\nu}) = 0
\end{equation}
governs the fluid. 
The five fluid Eqs.~(\ref{Euler0}) and (\ref{continuity0}) are not enough
to determine the the six quantities $\rho_0$, $\epsilon$, $P$ and $u^i$.
We thus also need an equation of state of the form
\be
P = P(\rho_0, \epsilon)
\ee
to close the system of fluid equations.
In many cases it is also useful to introduce the specific enthalpy
\begin{equation}
\label{spec_enthalpy_def}
h = 1 + \epsilon + P/\rho_0 ,
\end{equation}
because then the Euler equation can be simplified to
\begin{equation}
u ^\nu \nabla_\nu (h u_{\mu}) + \nabla_\mu h = 0 .
\label{eq:Euler_h}
\end{equation}
Using $u^{\mu} u_{\mu}=-1$, the latter can also be written as
\be
\label{Euler_p}
\vec{u}\cdot d\ut{p} = 0 .
\ee
Here $\vec{u}$ is the fluid velocity 4-vector and the one-form 
$\ut{p}$ has the components
\be
\label{p-def}
p_{\mu} = h u_{\mu} .
\ee
The dot in Eq.~(\ref{Euler_p}) indicates contraction of the vector $\vec{u}$
with the first index of the two-form $d\ut{p}$.

%%%%%%%%%%%%%%%%%%%%%%%%%%%%%%%%%%%%%%%%%%%%%%%%%%
\subsection{Expansion, shear and rotation of a fluid}
\label{exp_shear_rot}
%%%%%%%%%%%%%%%%%%%%%%%%%%%%%%%%%%%%%%%%%%%%%%%%%%

Using the projector
\begin{equation}
P_{\mu\nu} = g_{\mu\nu} + u_{\mu} u_{\nu} ,
\end{equation}
covariant derivatives of the fluid 4-velocity we can be split into 
\begin{equation}
\nabla_{\nu} u_{\mu} = \frac{1}{3}\theta P_{\mu\nu} + \varsigma_{\mu\nu} + 
\omega_{\mu\nu} - a_{\mu} u_{\nu} ,
\end{equation}
where the expansion, shear, rotation and acceleration of the fluid
are defined as
\begin{equation}
\theta = P^{\mu\nu} \nabla_{\mu} u_{\nu} ,
\end{equation}
\begin{equation}
\varsigma_{\mu\nu} = 
P_{\mu}^{\mu'} P_{\nu}^{\nu'} \nabla_{(\mu'} u_{\nu')}
- \frac{1}{3}\theta P_{\mu\nu} ,
\end{equation}
\begin{equation}
\label{fluid_rot1}
\omega_{\mu\nu} = P_{\mu}^{\mu'} P_{\nu}^{\nu'} \nabla_{[\nu'} u_{\mu']}
\end{equation}
and
\begin{equation}
a_{\mu} = u^{\nu} \nabla_{\nu} u_{\mu} .
\end{equation}
If the 4-velocity is of the form
\begin{equation}
\label{u_eq_futilde}
u^{\mu} = f \tilde{u}^{\mu} , 
\end{equation}
where $f$ is any scalar function, it immediately follows 
from $P_{\mu\nu}u^{\nu}=0$ that
\begin{equation}
\label{theta_f}
\theta = f P^{\mu\nu} \nabla_{\mu} \tilde{u}_{\nu} ,
\end{equation}
\begin{equation}
\label{varsigma_f}
\varsigma_{\mu\nu} =
f P_{\mu}^{\mu'} P_{\nu}^{\nu'} \nabla_{(\mu'} \tilde{u}_{\nu')}
- \frac{1}{3}\theta P_{\mu\nu} ,
\end{equation}
\begin{equation}
\label{fluid_rot2}
\omega_{\mu\nu} 
= f P_{\mu}^{\mu'} P_{\nu}^{\nu'} \nabla_{[\nu'} \tilde{u}_{\mu']} .
\end{equation}
From the latter is is clear that if the velocity is derived from a potential
$\phi$, i.e. if $h\ut{u} = \ut{p} = d\phi$, we immediately obtain
$\omega_{\mu\nu}=0$, which characterizes an irrotational fluid.

%%%%%%%%%%%%%%%%%%%%%%%%%%%%%%%%%%%%%%%%%%%%%%%%%%%%%%%%%%%%%%%%%%%
\section{Single neutron stars}
%%%%%%%%%%%%%%%%%%%%%%%%%%%%%%%%%%%%%%%%%%%%%%%%%%%%%%%%%%%%%%%%%%% 
\label{NS_data}

%%%%%%%%%%%%%%%%%%%%%%%%%%%%%%%%%%%%%%%%%%%%%%%%%%%%%%%%%%%%%%%%%%%
\subsection{Non-spinning neutron stars}
%%%%%%%%%%%%%%%%%%%%%%%%%%%%%%%%%%%%%%%%%%%%%%%%%%%%%%%%%%%%%%%%%%% 

Let us consider a static spherically symmetric star.
In this case the Einstein equations can be solved exactly.
The metric outside the star is given by the Schwarzschild metric of
Eq.~(\ref{Schwarzschild0}), but with $m$ replaced by
$m(r_{s*})$, where $r_{s*}$ is the star radius (in standard Schwarzschild
coordinates) and
\be
m(r_s) = \int_0^{r_s} dr' 4\pi r'^2 \rho_E
\ee
is the gravitational mass inside radius $r_s$ and
\be
\rho_E = \rho_0 (1 + \epsilon) .
\ee
Inside the star the metric is given by 
\be
ds^2 = -e^{2\phi} dt^2 + (1 - 2m(r_s)/r_s)^{-1} dr_s^2 +
        r_s^2 (d\theta^2 + \sin^2\theta d\phi^2) ,
\ee
where $m(r_s)$, $\Phi(r_s)$ and also $P(r_s)$ are found by integrating the
Tolman-Oppenheimer-Volkoff (TOV) equations~\cite{Tolman39,Oppenheimer39b}
\begin{eqnarray}
\label{TOV}
\frac{dm}{dr_s} &=& 4\pi r_s^2 \rho_E  \\
\frac{dP}{dr_s} &=& -\frac{(\rho_E + P) (m + 4\pi r_s^3 P)}{r_s (r_s-2m)} \\
\frac{d\Phi}{dr_s} &=& \frac{m + 4\pi r_s^3 P}{r_s (r_s-2m)}
\end{eqnarray}
together with some equation of state that allows us to obtain $\rho_E$ from
$P$. The TOV equations are ordinary differential equations that
have to be integrated out from $r_s=0$ to the radius where
$P=0$, which is the location of the star surface ($r_s=r_{s*}$). 
At $r_s=0$ we start with $m=0$ and some
value of $P=P_c$ which determines the core pressure and thus the total mass
of the star. We also have to start with some particular $\Phi$ at $r_S=0$.
We can set this value to 1 at first. The final $\Phi(r)$
can be obtained by adding a constant to it such that
$2\Phi(r_{s*})  = \ln(1 - 2m(r_{s*})/r_{s*})$. 
This shift in $\Phi$ ensures that the metric is continuous at
the star surface. It is possible, and often convenient, to transform the TOV
metric to isotropic coordinates by using the transformation given in
Eqs.~(\ref{SchwToIso}) and (\ref{SchwConfac}).
From the TOV metric it is then easy to obtain the 3-metric and extrinsic
curvature.

%%%%%%%%%%%%%%%%%%%%%%%%%%%%%%%%%%%%%%%%%%%%%%%%%%%%%%%%%%%%%%%%%%%
\subsection{Spinning neutron stars}
%%%%%%%%%%%%%%%%%%%%%%%%%%%%%%%%%%%%%%%%%%%%%%%%%%%%%%%%%%%%%%%%%%% 

The metric for a spinning neutron star is no longer spherically symmetric.
Instead one assumes a stationary and axisymmetric metric of the form
\be
%  Cook, Shapiro & Teukolsky (1994), but
% gamma=A, rho=B, alpha=C
ds^2 = -e^{A+B} dt^2 + e^{2C}(dr^2 + r^2 d\theta^2) + 
e^{A-B} r^2 \sin^2\theta (d\phi- \omega dt)^2 ,
\ee
where $A$, $B$, $C$ and $\omega$ are functions of $r$ and $\theta$ only. If
matter is again treated as a perfect fluid and an equation of state is
provided, one obtains a system of three field equations and one equation
expressible as a line integral~\cite{Ansorg01b}, both inside the star and in
the vacuum region on the outside. This system of equations is rather
complicated and we will not go through its derivation here. It can only be
solved numerically. Several groups have worked on this problem in the past.
A freely available public code (called RNS) that can solve the system of
equations has been developed by Stergioulas and
Friedman~\cite{Stergioulas95}. However, the most accurate results by far
have been achieved by Ansorg et al.~\cite{Ansorg01b,Ansorg:2003br}. It is
also worth mentioning that the metric outside a spinning star is not given
by the Kerr metric that is valid for a rotating black hole. A review about a
single spinning neutron stars can be found in~\cite{Stergioulas03}.

%%%%%%%%%%%%%%%%%%%%%%%%%%%%%%%%%%%%%%%%%%%%%%%%%%%%%%%%%%%%%%%%%%%
\section{Binary neutron star initial data}
%%%%%%%%%%%%%%%%%%%%%%%%%%%%%%%%%%%%%%%%%%%%%%%%%%%%%%%%%%%%%%%%%%% 
\label{BNS_data}

Binary neutron stars, like binary black holes, will be on approximately
circular orbits if they have been inspiraling already for a long time. We
can thus use many of the same methods as for black holes when we construct
binary neutron star initial data. Thus again we can assume that an
approximate helical or inspiral Killing vector $\xi^{\mu}$ as in
Eq.~(\ref{inspiral_KV}) should exist. So again we can choose coordinates
where Eq.~(\ref{hKV}) holds, and in which time derivatives of metric
variables can be approximated by zero. As mentioned previously, 
such a situation can be captured best by using the CTS decomposition,
which we will use almost exclusively for the binary neutron star initial data
described below. We will also restrict ourselves to the case of a
conformally flat metric as in Eq.~(\ref{conformally_flat}). Since gravity in
neutron stars is weaker and rotational velocities are lower than in black
holes, conformal flatness is generally a good approximation for neutron
stars. Notice however, that a very interesting non-conformally
flat method, named the waveless formulation, has been
developed~\cite{Bonazzola:2003dm,Shibata:2004qz,Yoshida:2006cn,Uryu:2009ye}.
In this formulation one obtains the same fluid equations as described below.
The equations for the metric variables are essentially the CTS equations,
plus one additional elliptic equation for the conformal metric. This extra
equation is derived from the evolution equation 
for the extrinsic curvature by using
a particular gauge to rewrite $R_{ij}$ in Eq.~(\ref{K_ij-evo}) as an
elliptic operator acting on the conformal metric. Thus the conformal metric
does not need to be assumed to be of a particular form, rather it is
computed from the additional elliptic equation. Instead one needs to specify
as free data the time derivative of a conformally rescaled tracefree
extrinsic curvature, in addition to the time derivative of the conformal
metric.

The main difference between binary neutron star initial data
and binary black hole initial data is that we now also have to find
quasi-equilibrium configurations for the matter, when we solve the Euler and
continuity equations in Eqs.~(\ref{Euler0}) and (\ref{continuity0}). Notice,
however, that the stars have only a finite extent so that we will use the
same boundary conditions for the metric variables as in
Sec.~\ref{BCs_at_inf} at infinity.

As in the black hole case, we will not discuss every possible method to
construct binary neutron star initial data here. Rather we will
concentrate on conceptual issues, as well as on describing the most widely
used methods.

%%%%%%%%%%%%%%%%%%%%%%%%%%%%%%%%%%%%%%%%%%%%%%%%%%%%%%%%%%%%%%%%%%%
\subsection{Corotating binary neutron stars}
%%%%%%%%%%%%%%%%%%%%%%%%%%%%%%%%%%%%%%%%%%%%%%%%%%%%%%%%%%%%%%%%%%% 

In order to construct initial data for two neutron stars in a corotating
configuration, we assume again a helical Killing vector $\xi^{\mu}$. The
fact that the two stars are corotating means that the orbital period and
both star spin periods all have the same value. Thus each fluid element
is at rest in corotating coordinates. This idea is captured by
assuming that the fluid in each star flows along the Killing vector
$\xi^{\mu}$, i.e. that the fluid 4-velocity is
\be
\label{corot0}
u^{\mu} = u^0 \xi^{\mu} .
\ee
In this case it is easy to show that the continuity equation
(\ref{continuity0}) is identically satisfied. 
Furthermore one can show that the Euler equation leads
to (see e.g. problem 16.17 in~\cite{Lightman75})
\begin{equation}
[\rho_0(1+\epsilon) + P] d \ln(u_{\mu}\xi^{\mu}) = -dP ,
\end{equation}
where $\xi^{\mu}$ is the assumed helical Killing vector.
With the help of the first law of thermodynamics
($d[\rho_0(1+\epsilon)] = [\rho_0(1+\epsilon) + P] d\rho_0 / \rho_0$)
this equation can be integrated to yield
\begin{equation}
\label{Bernoulli0}
u_{\mu}\xi^{\mu} = \frac{C\rho_0}{\rho_0(1+\epsilon) + P} ,
\end{equation}
where $C$ is a constant of integration that is different for each star.
Using Eqs.~(\ref{corot0}) and (\ref{spec_enthalpy_def}) we arrive at
\be
\label{Bernoulli0_corot}
h = -C u^0 ,
\ee
which tells us the value of the specific enthalpy for each point inside the
star. 

If we assume a polytropic equation of state 
\begin{equation}
\label{polytrop}
P = \kappa \rho_0^{1+1/n} ,
\end{equation}
we can express 
the mass density, the pressure and the internal energy in terms of $h$. We
obtain
\begin{eqnarray}
\label{rhoPeps_h_poly}
\rho_0   &=& \kappa^{-n} \left(\frac{h-1}{n+1}\right)^n \nonumber\\
P        &=& \kappa^{-n} \left(\frac{h-1}{n+1}\right)^{n+1} \nonumber\\
\epsilon &=& n \left(\frac{h-1}{n+1}\right) .
\end{eqnarray}
The constant $n$ here is known as the polytropic index, and $\kappa$
is a constant.

In order to construct binary neutron star initial data
we have to solve the five elliptic equations in 
Eqs.~(\ref{hamCTS}), (\ref{momCTS}) and (\ref{alpha_CTS})
with the matter terms given by Eqs.~(\ref{fluid_matter}),
(\ref{Bernoulli0_corot}) and (\ref{rhoPeps_h_poly}).
This is done through an iterative procedure~\cite{Tichy:2009yr} 
as described in subsection~\ref{BNS_iterations}.
%
% We can solve the whole set of equations by iterating over the following
% steps~\cite{Tichy:2009yr}:
% (i) We first come up with an initial guess for $h$ in 
% each star, in practice one often chooses
% Tolman-Oppenheimer-Volkoff solutions 
% for each. 
% (ii) Next we solve the 5 coupled elliptic equations
% in Eqs.~(\ref{hamCTS}), (\ref{momCTS}) and (\ref{alpha_CTS})
% for this given $h$.
% (iii) Then we use Eq.~(\ref{Bernoulli0_corot}) to update $h$ in each star.
% The constant $C$ in general has a different value for each star.
% We adjust the value for each star such that it has a prescribed rest
% mass. 
% After updating $h$ we go back to step (ii) and iterate until
% all equations are satisfied up to a given tolerance.

Examples of such corotating initial data can be found 
in~\cite{Baumgarte:1997xi,Baumgarte:1997eg,
Mathews:1997pf,Marronetti:1998xv,Tichy:2009yr}.
The problem with such data is that it is very unlikely for two neutron stars
to have spin periods that are both synchronized with the orbital period.
As pointed out by Bildsten and Cutler~\cite{Bildsten92},
the two neutron stars cannot be tidally locked,
because the viscosity of neutron star matter is too low.
Hence barring other effects, like magnetic dipole radiation, the spin of
each star remains approximately constant.
This means that initial data sequences of corotating configurations
for different separations cannot be used to approximate the inspiral of two
neutron stars.

%%%%%%%%%%%%%%%%%%%%%%%%%%%%%%%%%%%%%%%%%%%%%%%%%%%%%%%%%%%%%%%%%%%
\subsection{Irrotational binary neutron stars}
%%%%%%%%%%%%%%%%%%%%%%%%%%%%%%%%%%%%%%%%%%%%%%%%%%%%%%%%%%%%%%%%%%% 

Since corotating configurations are unrealistic, many groups use initial data
for irrotational stars. The advantage is that, sequences of irrotational
configurations can be used to approximate the inspiral of two neutron stars
without spin. Of course, for such configurations the fluid velocity will not
be along the helical Killing vector, which we will still assume to exist
approximately. Thus the fluid velocity is now written as
\begin{equation}
\label{V-def}   
u^{\mu} = u^0 \left( \xi^{\mu} + V^{\mu} \right) ,
\end{equation}
where $u^0 = -u^{\mu} n_{\mu}/\alpha$, and $V^{\mu}$ is a purely spatial
vector with $V^{\mu}n_{\mu}=0$
that describes the fluid velocity relative to the Killing vector
$\xi^{\mu}$.

In order to find the enthalpy for a quasi-equilibrium situation
we insert Eq.~(\ref{V-def}) into Euler Eq.~(\ref{Euler_p}) to obtain
\be
\label{xi_V_dp}
\vec{\xi}\cdot d\ut{p} + \vec{V}\cdot d\ut{p} = 0 .
\ee
Note that for any vector $\vec{v}$ and form $\ut{\omega}$ the 
Cartan identity
\be
\label{Cartan_id}
\pounds_{v}\ut{\omega} 
= \vec{v} \cdot d\ut{\omega} + d (\vec{v} \cdot \ut{\omega})
\ee
holds, where $\pounds_{v}\ut{\omega}$ is the Lie derivative of $\ut{\omega}$
along $\vec{v}$.
The dot in Eq.~(\ref{Cartan_id}) indicates contraction of a
vector with the first index of a form.
Using the Cartan identity to replace $\vec{\xi}\cdot d\ut{p}$,
Eq.~(\ref{xi_V_dp}) simplifies to
\be
\label{xi_V_dp_simp}
\pounds_{\xi}\ut{p} - d(\vec{\xi} \cdot \ut{p}) + \vec{V}\cdot d\ut{p} = 0 .
\ee
The first term in this equation term will be dropped since $\vec{\xi}$
is an approximate Killing vector, i.e.
\be
\label{Liexi_p_zero}
\pounds_{\xi}\ut{p} = \pounds_{t}\ut{p} \approx 0 ,
\ee
where we again have used corotating coordinates as in Eq.~(\ref{hKV}).

From Eq.~(\ref{Euler_p}) we can see that setting
\be
\label{p_from_velpot}
\ut{p} = d\phi
\ee
will automatically solve the Euler equation. The function $\phi$ is a
velocity potential from which we can then derive $p_{\mu}=h u_{\mu}$.
Notice that a fluid with a velocity coming from a potential,
according to $h u_{\mu} = \nabla_{\mu}\phi$, has zero
rotation (see Eqs.~(\ref{u_eq_futilde}) and (\ref{fluid_rot2})),
hence the name irrotational configuration for the initial data constructed
in this section.

If we insert Eq.~(\ref{p_from_velpot}) together with assumption
(\ref{Liexi_p_zero}) into Eq.~(\ref{xi_V_dp_simp}) we finally obtain
\be
d(\vec{\xi} \cdot \ut{p}) = 0 ,
\ee
which implies that 
\be
h u_{\mu} \xi^{\mu} = C ,
\ee
where $C$ is a constant that may be different in each star. Inserting
$\xi^{\mu} = u^{\mu}/u^0 - V^{\mu}$ from Eq.~(\ref{V-def}) and using
$u_{\mu}u^{\mu}=-1$ we arrive at
\be
\label{Euler_int_irr}
\frac{h}{u^0} + V^k D_k \phi = -C
\ee
As in the case of corotating configurations this last equation can be used
to find the specific enthalpy.
Note that $u^0$ can be obtained from $u_{\mu}u^{\mu}=-1$.

Using Eqs.~(\ref{V-def}) and (\ref{p_from_velpot}) one can show that
the continuity equation (\ref{continuity0}) 
becomes~\cite{Shibata98,Teukolsky98,Tichy:2011gw}
\be
\label{continuity_irrot}
D_i \left[ \frac{\rho_0 \alpha}{h} D^i \phi  
          -\rho_0 \alpha u^0 (\beta^i + \xi^i) \right] = 0 ,
\ee
where we have again dropped some Lie derivatives with respect to the Killing
vector $\vec{\xi}$. More details about this derivation will be provided in
the next section, where we consider the more general case of spinning stars.
Compared to the corotating configuration,
the irrotational case yields the additional elliptic 
Eq.~(\ref{continuity_irrot}), that we need to solve for the velocity
potential $\phi$ together we the other elliptic equations for the
metric variables in Eqs.~(\ref{hamCTS}), (\ref{momCTS}), (\ref{alpha_CTS}).
Such irrotational binary neutron stars have been produced by
many groups~\cite{Bonazzola:1998yq,Gourgoulhon:2000nn,Marronetti:1999ya,
Uryu:1999uu,Marronetti:2003gk,Taniguchi:2002ns,Taniguchi:2003hx,
Uryu:2005vv,Taniguchi:2010kj,Haas:2016cop}.

%%%%%%%%%%%%%%%%%%%%%%%%%%%%%%%%%%%%%%%%%%%%%%%%%%%%%%%%%%%%%%%%%%%
\subsection{Binary neutron stars with arbitrary spins}
%%%%%%%%%%%%%%%%%%%%%%%%%%%%%%%%%%%%%%%%%%%%%%%%%%%%%%%%%%%%%%%%%%% 
\label{BNS_with_spin}

We will now consider the general case of neutron stars with spin.
First we assume that the binary is in an approximately circular
orbit and that the spin of each star remains approximately constant.
As in the case of binary black holes (see
e.g.~\cite{Tichy:2003zg,Tichy:2003qi})
this implies the existence of an approximate helical Killing vector
$\xi^{\mu}$ with $\pounds_{\xi} g_{\mu\nu} \approx 0$.
In order to clarify the meaning of the approximate sign we now briefly
discuss two cases.

If both star spins are parallel to the orbital angular momentum
we have $\pounds_{\xi} g_{\mu\nu} = O(P_o/T_{ins})$, 
where the inspiral timescale $T_{ins}$ is much 
longer than the orbital timescale $P_o$. I.e. in a corotating coordinate
system all metric time derivatives are of order $O(P_o/T_{ins})$
and thus small.
For arbitrary spins the situation becomes more complicated. We can again
use corotating coordinates, but in this coordinate system the spin vectors
will be precessing on an orbital timescale $P_o$. This means there are
matter currents that change on a timescale $P_o$, and thus 
$\pounds_{\xi} u_{\mu}$ cannot be zero~\cite{Tichy:2011gw}
even approximately! On the other hand, the matter distribution
itself only changes on the much longer inspiral timescale $T_{ins}$.
In this case it is useful to consider gravity to be made up of
gravitoelectric and gravitomagnetic fields~\cite{Braginsky:1976rb,Misner73}.
The gravitoelectric parts of the metric are sourced by the matter
distribution and thus change only on the timescale $T_{ins}$, while
the gravitomagnetic parts of the metric are sourced by matter currents
and thus change on the shorter timescale $P_o$.
However, the gravitomagnetic parts are smaller 
than the gravitoelectric parts by $O(v/c)$~\cite{Braginsky:1976rb,Misner73}.
Thus we now have $\pounds_{\xi} g_{\mu\nu} = O(v/c) \approx 0$,
where we assume that the orbital velocity $v$ is smaller than the speed of
light.

We start again by writing the fluid velocity as in Eq.~(\ref{V-def}) and use
again the Cartan identity to bring the Euler equation into the form in
Eq.~(\ref{xi_V_dp_simp}). However, this time we cannot assume that
$\pounds_{\xi}\ut{p}$ is zero. Also, since both stars are spinning,
the fluid flow is no longer purely irrotational, so that now we write
\be
\label{p_dphi_w}
h\ut{u} = \ut{p} = d\phi + \ut{w} ,
\ee
where $d\phi$ and $\ut{w}$ denote the irrotational and rotational
pieces of the velocity. Here $\ut{w}$ is chosen to be purely
spatial, i.e. such that $\vec{n}\cdot\ut{w}=0$. It is clear that 
$d\phi$ describes the orbital or center of mass motion of the star,
while $\ut{w}$ is due to the star spin.

Using Eq.~(\ref{p_dphi_w}) together with the Cartan identity
(\ref{Cartan_id}), the last term in Eq.~(\ref{xi_V_dp_simp}) becomes
\be
\vec{V}\cdot d\ut{p} = \vec{V}\cdot d\ut{w}
= \pounds_{V}\ut{w} - d(\vec{V} \cdot \ut{w}) 
\ee
so that Eq.~(\ref{xi_V_dp_simp}) is equivalent to
\be
\label{Euler_d0}
d(\vec{\xi} \cdot \ut{p} + \vec{V} \cdot \ut{w}) =
\pounds_{\xi} (d\phi + \ut{w}) + \pounds_{V}\ut{w} =
\pounds_{\xi} d\phi  + \pounds_{\xi+V}\ut{w} .
\ee
Notice that according to Eqs.~(\ref{V-def}) and (\ref{p_dphi_w})
\be
\vec{\xi} + \vec{V} = \frac{1}{hu^0}(\vec{\nabla}\phi + \vec{w})
\ee
so that Eq.~(\ref{Euler_d0}) finally becomes
\be
\label{Euler_d1}
d(\vec{\xi} \cdot \ut{p} + \vec{V} \cdot \ut{w}) =
\pounds_{\xi} d\phi  + 
\pounds_{\frac{\nabla\phi}{hu^0}}\ut{w} +
\pounds_{\frac{w}{hu^0}}\ut{w}.
\ee

Following~\cite{Tichy:2011gw},
we will now argue that all three terms on the right hand side of 
Eq.~(\ref{Euler_d1}) are small.
Since the center of mass motion described by $d\phi$ should be unchanged
along the approximate symmetry vector $\vec{\xi}$, we will assume
that 
\be
\label{matter_assumpt1}
\pounds_{\xi} d\phi \approx 0 .
\ee
As we have explained, the spin and thus $\ut{w}$ changes along
$\vec{\xi}$. However, if we go along the center of mass 
direction given by $\vec{\nabla}\phi$, we expect
the spin to be unchanged. Thus we assume
\be 
\label{matter_assumpt2}
\pounds_{\frac{\nabla\phi}{hu^0}}\ut{w} \approx 0 .
\ee
Furthermore we assume that
\be 
\label{matter_assumpt3}
\pounds_{\frac{w}{hu^0}}\ut{w} \approx 0 ,
\ee
which tells us that the spinning motion of the star does not change if we go
along the direction of this spinning motion.

\begin{figure}
%\vspace{-0.4cm}
\begin{center}
\includegraphics[width=0.4\textwidth]{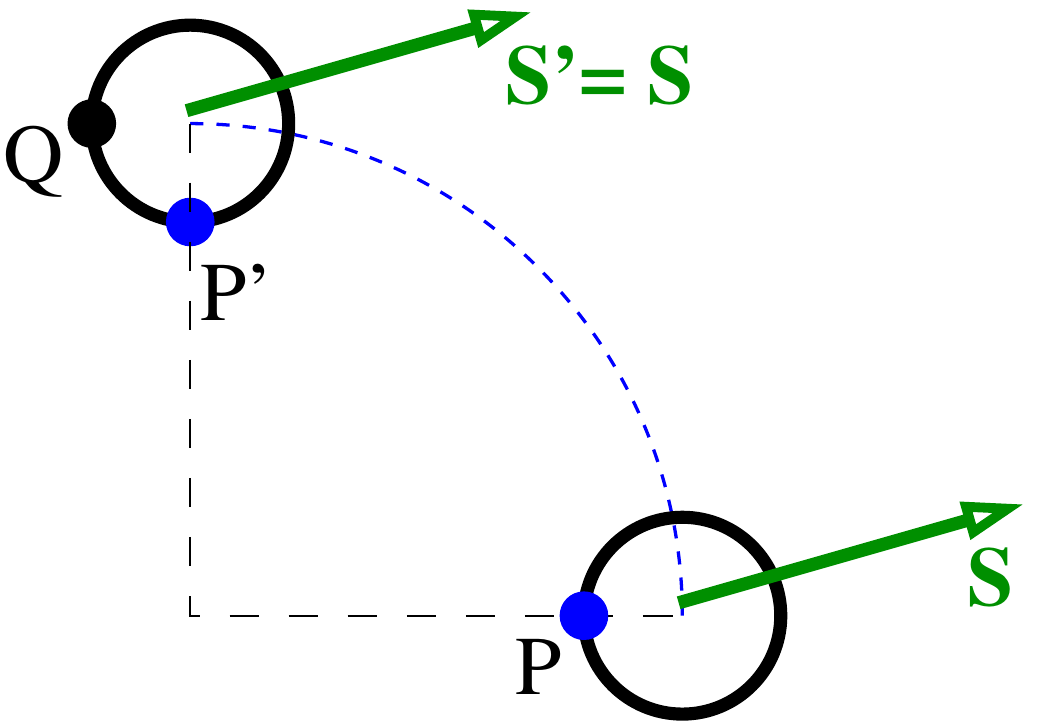}
\caption{ \label{constspin-figure}
\small{
Schematic picture of a star and its spin $S$:
The star starts on the bottom and then moves along a circular arc. Its
center follows the Killing vector $\vec{\xi}$.
The star spin $S$ is approximately unchanged over an orbital timescale.
The point $P$ becomes point $P'$ if we follow the orbit of the Killing
vector, while it becomes point $Q$ if we follow the center of mass
motion parallel to $\vec{\nabla}\phi$. The orbital fluid speed (excluding
the spin) at $P$ and $P'$ are the same. Yet, the fluid motion due to the spin
at $P'$ is not the same as at $P$. Rather it should be approximately the
same as at $P$ and $Q$.
}
}
\end{center}
\vspace{-0.5cm}
\end{figure}
The foregoing argument is also illustrated in Fig.~\ref{constspin-figure},
which schematically depicts the star over a quarter orbit. 
As the star moves, its spin stays approximately constant and
its center moves along the Killing vector $\vec{\xi}$.
If we start at point $P$ and follow $\vec{\xi}$ we end up at point $P'$. If
instead we follow the center of mass motion parallel to $\vec{\nabla}\phi$
we arrive at $Q$. As we can see the orbital fluid speed (excluding
the spin) at $P$ and $P'$ are the same, which is the origin of
assumption~\ref{matter_assumpt1}. However, the fluid motion due to the spin
at $P$ and $P'$ are not the same. Rather we expect the spin motion
to be approximately the same at point $P$ and $Q$, which supports
assumption~\ref{matter_assumpt2}.

Using the three assumptions in Eqs.~(\ref{matter_assumpt1}),
(\ref{matter_assumpt2}) and (\ref{matter_assumpt3}), we can integrate
Eq.~(\ref{Euler_d1}) and obtain
\be
\label{Euler_int}
C
= \vec{\xi} \cdot \ut{p} + \vec{V} \cdot \ut{w}
= \left(\frac{\vec{u}}{u^0}-\vec{V}\right)\cdot h\ut{u} +\vec{V} \cdot \ut{w}
= -\frac{h}{u^0} - V^i D_i\phi ,
\ee
where $C$ is the constant of integration.
This last equation is again used to find $h$. It reduces to
Eq.~(\ref{Bernoulli0_corot}) in the corotating case where $\vec{V}=0$, and
has exactly the same form as Eq.~(\ref{Euler_int_irr}) for the irrotational
case. In order to to find $u^0$ we use
\be
-1=\vec{u}\cdot\ut{u}=(\gamma^{\mu\nu}-n^{\mu}n^{\nu})u_{\mu}u_{\nu} =
\gamma^{\mu\nu}u_{\mu}u_{\nu} -(n_0 u^0)^2 .
\ee
We now multiply by $h^2$, use Eq.~(\ref{p_dphi_w}) and solve for
$u^0$. We find
\be
\label{uzero}
u^0 = \sqrt{h^2 - (D_i\phi +w_i)(D^i\phi +w^i)}/(h\alpha) .
\ee
If we insert Eq.~(\ref{uzero}) into the
integrated Euler equation~(\ref{Euler_int}) we can solve for $h$.
This yields
\begin{equation}
\label{h_from_Euler}
h = \sqrt{L^2 - (D_i \phi + w_i)(D^i \phi + w^i)},
\end{equation}
where we use the abbreviations
\begin{equation}
\label{h_from_Euler_L}
L^2 = \frac{b + \sqrt{b^2 - 4\alpha^4 [(D_i \phi + w_i) w^i]^2}}{2\alpha^2}
\end{equation}
and
\begin{equation}
\label{h_from_Euler_b}
b = [ (\xi^i+\beta^i)D_i \phi - C]^2 + 2\alpha^2 (D_i \phi + w_i) w^i .
\end{equation}

We now rewrite the continuity Eq.~(\ref{continuity0}) as
\begin{eqnarray}
0 &=& g^{\mu\nu}\nabla_{\mu}[\rho_0 u^0(\xi_{\nu}+V_{\nu})]
=
(\xi^{\mu}+V^{\mu})\nabla_{\mu}(\rho_0 u^0)
+ \rho_0 u^0 
[g^{\mu\nu}(\nabla_{\mu}\xi_{\nu}+\nabla_{\mu}V_{\nu})] \nonumber\\
&=&
\pounds_{\xi}(\rho_0 u^0) + V^i D_i (\rho_0 u^0)
+ \rho_0 u^0 
[g^{\mu\nu}\pounds_{\xi}g_{\mu\nu} + 
 (\gamma^{\mu\nu}-n^{\mu}n^{\nu})\nabla_{\mu}V_{\nu}] \nonumber\\
&=&
\pounds_{\xi}(\rho_0 u^0) + V^i D_i (\rho_0 u^0)
+ \rho_0 u^0 
[g^{\mu\nu}\pounds_{\xi}g_{\mu\nu} + D_i V^i 
+ V^{\nu}n^{\mu}\nabla_{\mu}n_{\nu} .
\end{eqnarray}
Using Eq.~(\ref{nabla_n}) and collecting terms we obtain
\be
\label{continuity1}
D_i(\alpha\rho_0 u^0 V^i) + 
\alpha\pounds_{\xi}(\rho_0 u^0) +
\alpha\rho_0 u^0 g^{\mu\nu}\pounds_{\xi}g_{\mu\nu}
= 0 .
\ee
Since $\vec{\xi}$ is an approximate Killing vector we can drop the last
term. We also assume that
\be
\label{matter_assumpt4}
\pounds_{\xi}(\rho_0 u^0) \approx 0
\ee
since the matter density should be constant along the symmetry vector
$\vec{\xi}$.
As we can see Eq.~(\ref{continuity1}) is identically satisfied in the
corotating case where $V^i=0$. For all other cases Eq.~(\ref{continuity1})
can be turned into an elliptic equation for $\phi$. To do this, first notice
that combining Eqs.~(\ref{V-def}) and (\ref{p_dphi_w}) yields
\be
\vec{V} = \frac{\vec{\nabla}\phi + \vec{w}}{hu^0} - \vec{\xi} .
\ee
If we project this equation onto the spatial slice using
$\gamma^{\mu}_{\nu}$ we find
\be
V^i = \frac{D^i\phi + w^i}{hu^0} - (\xi^i + \beta^i) .
\ee
When the latter is inserted into Eq.~(\ref{continuity1}) we obtain
\begin{equation}
\label{continuity4}
D_i \left[ \frac{\rho_0 \alpha}{h}(D^i \phi + w^i) 
          -\rho_0 \alpha u^0 (\beta^i + \xi^i) \right] = 0 . 
\end{equation}
Notice that this expression reduces to Eq.~(\ref{continuity_irrot})
in the irrotational case where $w^i=0$.

Equation~(\ref{continuity4}) is only valid inside the star and we have to
provide a boundary condition at the star surface where $\rho_0 = 0$.
This condition can be
obtained directly from Eq.~(\ref{continuity4}) in the limit of
$\rho_0 \to 0$ but $D_i \rho_0 \neq 0$. We find~\cite{Tichy:2011gw}
\begin{equation}
\label{starBC}
(D^i \phi)D_i \rho_0 + w^i D_i \rho_0 
= h u^0 (\beta^i + \xi^i) D_i \rho_0 .
\end{equation}

Notice that the rotational velocity piece $w^i$ can be freely chosen.
In~\cite{Tichy:2012rp} it is demonstrated that 
\be
\label{w-choice}
w^i = \epsilon^{ijk}\omega^j (x^k - x^k_{C*})
\ee
is a good choice, that results in almost rigidly rotating fluid 
configurations with low expansion and shear.
Here $x^k_{C*}$ denotes the location of the star center and $\omega^j$ is an
arbitrarily chosen vector that determines the star spin.
Summation over the repeated indices $j$ and $k$ is implied

To close the system of equations we need again an equation of state. The
simplest choice is the polytrope given in Eq.~(\ref{polytrop}), which again
results in Eqs.~(\ref{rhoPeps_h_poly}). However, other choices are
possible. For example, \cite{Dietrich:2015pxa} uses piecewise polytropic
equations of state to approximate several realistic equations of state at
zero temperature.

\begin{figure}
\vspace{-0.4cm}
\begin{center}
\includegraphics[width=\textwidth]{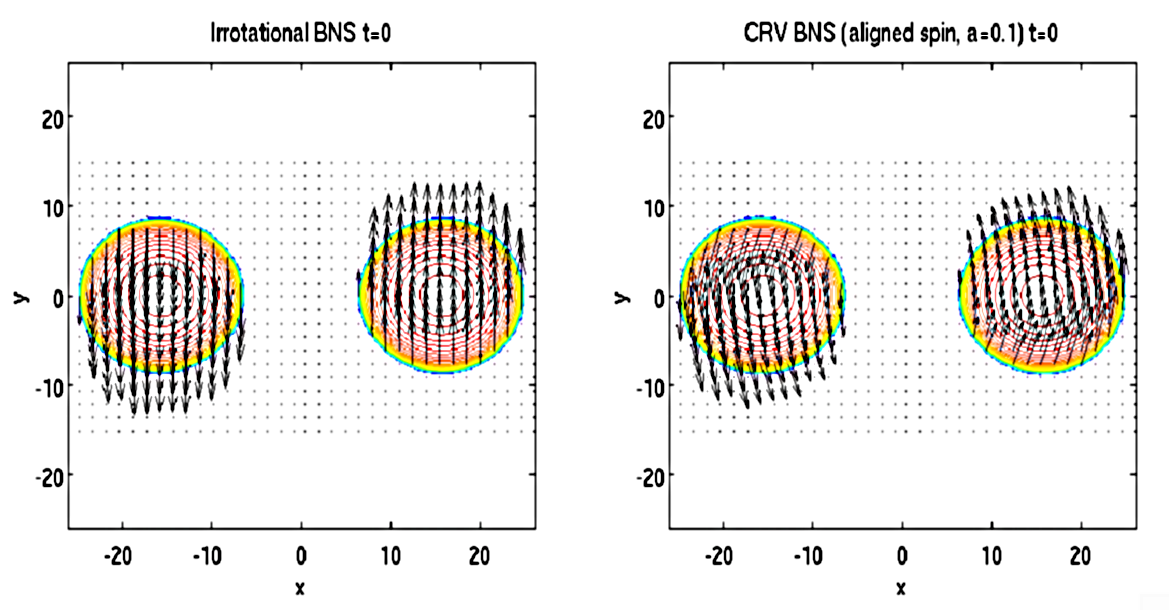}
\caption{ \label{CRVexample}
\small{
Two examples of binary neutron star initial data. Shown are contour lines of
the density and the velocity vectors of fluid elements in the orbital plane.
For an irrotational binary (left plot) the fluid velocity is perpendicular
to the line connecting the star centers. For counterclockwise spinning
stars (right plot) the fluid velocity is modified due to the additional
rotational piece $w^i$ in Eq.~(\ref{p_dphi_w}).
}
}
\end{center}
\vspace{-0.5cm}
\end{figure}
Figure~\ref{CRVexample} shows velocity vectors of fluid elements
in the orbital plane for two binaries. In the irrotational case (on the
left) the fluid velocity is in the orbital direction. For the case where
both stars are spinning (right plot) the fluid velocity has an additional
rotational contribution due to $w^i$ in Eq.~(\ref{p_dphi_w}).

The same set of equations discussed above has also been used
in~\cite{Tsokaros:2015fea,Tacik:2015tja}. Notice, however, that
\cite{Tsokaros:2015fea} erroneously assumes $\pounds_{\xi}\ut{p} = 0 =
\pounds_{V} \ut{w}$. But as shown in the appendix of~\cite{Tichy:2011gw},
$\pounds_{\xi}\ut{p}$ cannot be approximately zero for generic spins.
Fortunately the two errors incurred by the latter two assumptions cancel, so
that the right hand side of Eq.~(\ref{Euler_d0}) is again zero, and all
other equations follow as shown above.

Spinning binary neutron initial data constructed with the method described
above are evolved in~\cite{Tacik:2015tja}, and the star spins are monitored
using a newly defined quasi-local neutron star spin. It is found that the
star spins follow post-Newtonian predictions during the inspiral of the two
stars. This indicates that the above formalism together with the choice for
$w^i$ in Eq.~(\ref{w-choice}) works and results in realistic star spins.

%%%%%%%%%%%%%%%%%%%%%%%%%%%%%%%%%%%%%%%%%%%%%%%%%%%%%%%%%%%%%%%%%%%   
\subsection{Iteration procedures required to construct binary neutron star
initial data}
%%%%%%%%%%%%%%%%%%%%%%%%%%%%%%%%%%%%%%%%%%%%%%%%%%%%%%%%%%%%%%%%%%% 
\label{BNS_iterations}

As we have discussed above, the construction of binary neutron star 
initial data requires us to solve the elliptic equations of the
CTS approach in Eqs.~(\ref{hamCTS}), (\ref{momCTS}) and (\ref{alpha_CTS})
coupled to the matter equations which are given by Eqs.~(\ref{continuity4})
and (\ref{h_from_Euler}) for stars in general rotation states or only
Eq.~(\ref{Bernoulli0_corot}) for corotating configurations.
Both Eq.~(\ref{h_from_Euler}) and Eq.~(\ref{Bernoulli0_corot}) are algebraic
equations that specify the specific enthalpy $h$ given the other fields that
obey elliptic equations. The situation is complicated by the fact that the
star surfaces are at the location where $h=1$, so that one cannot
know where the star surfaces will be before solving all coupled
equations. Thus we do not know in which regions we indeed
have to include matter terms in Eqs.~(\ref{hamCTS}), (\ref{momCTS}) and
(\ref{alpha_CTS}). Notice also that the values of the matter terms are
undefined if we blindly use the equation of state
in regions where $h<1$ (see e.g. Eqs.~(\ref{rhoPeps_h_poly})).

For these reasons the Eqs.~(\ref{hamCTS}), (\ref{momCTS}), (\ref{alpha_CTS}),
(\ref{continuity4}) and (\ref{h_from_Euler}) are solved using an iterative
procedure~\cite{Tichy:2012rp}, that can be outlined as follows:
(i) We first come up with an initial guess for $h$ in
each star (in practice we often choose Tolman-Oppenheimer-Volkoff solutions).
(ii) Next, we solve the 6 coupled elliptic equations
in Eqs.~(\ref{hamCTS}), (\ref{momCTS}), (\ref{alpha_CTS})
and (\ref{continuity4}) for this given $h$.
(iii) Then we use Eq.~(\ref{h_from_Euler}) to update $h$ in each star.
After updating $h$ we go back to step (ii) and iterate until
all equations are satisfied up to a given tolerance.

There is however one major ingredient missing in the above outline.
Our entire method assumes a symmetry vector $\vec{\xi}$ of the
form~(\ref{inspiral_KV}). Thus we need particular values for $\Omega$ 
and $x_{CM}^1$\footnote{We only need to specify $x_{CM}^1$ because we 
can always rotate our coordinate system such that the line connecting the
two star centers is along the $x^1$-axis so that $x_{CM}^2=0=x_{CM}^3$.}.
Now, the star surfaces move between iteration steps. If we simply specify
constant values for $\Omega$ and $x_{CM}^1$, it turns out that the
stars drift away and that the iteration becomes unstable.
We thus have to find a way to keep the stars from drifting.
This is usually done by adjusting $\Omega$ and $x_{CM}^1$ such that the
star center does not move when we update $h$ using Eq.~(\ref{h_from_Euler}).
Notice that Eq.~(\ref{h_from_Euler}) depends on $\Omega$ and $x_{CM}^1$
through $\xi^i$ which is contained in $b$ from Eq.~(\ref{h_from_Euler_b}).
The star centers $x_{C*1/2}^1$ are defined by the maxima of $h$ along the
$x^1$-axis, given by $\partial_1 h|_{x_{C*1/2}^1} =0$.
In~\cite{Tichy:2012rp,Dietrich:2015pxa} it is shown that the latter
can be expressed as
\be
\label{forcebalance}
\partial_{1} \ln
\left[
\alpha^2 - \left(\beta^i+\xi^i+\frac{w^i}{hu^0}\right)
           \left(\beta_i+\xi_i+\frac{w_i}{hu^0}\right)
\right]\Bigg|_{x_{C*1/2}^1} = -2\partial_{1}\ln\Gamma\big|_{x_{C*1/2}^1}
\ee
where
\begin{equation}
\Gamma = 
\frac{
\alpha u^0
\left[
1-\left(\beta^i+\xi^i+\frac{w^i}{hu^0}\right)\frac{D_i\phi}{\alpha^2 hu^0}
        - \frac{w_i w^i}{(\alpha hu^0)^2}\right]
}
{
\sqrt{ 1 -  \left(\beta^i+\xi^i+\frac{w^i}{hu^0}\right)
            \left(\beta_i+\xi_i+\frac{w_i}{hu^0}\right)\frac{1}{\alpha^2} }
} .
\end{equation}
We then update $\Omega$ and $x_{CM}^1$ by solving
Eq.~(\ref{forcebalance}) for $\Omega$ and $x_{CM}^1$ so that 
the star centers $x_{C*1/2}^1$ remain in the same location, when we update
$h$ according to Eq.~(\ref{h_from_Euler}).
Equation~(\ref{forcebalance}) is often called the force balance 
equation~\cite{Gourgoulhon:2000nn}.
One noteworthy caveat is that
we evaluate the derivative of $\ln\Gamma$ in Eq.~(\ref{forcebalance})
for the $\Omega$ and $x_{CM}^1$ before the update. 

Finally, the complete iteration procedure involves the following main 
steps~\cite{Tichy:2012rp}:
(1) We first come up with an initial guess for $h$ in
each star, in practice we often choose 
Tolman-Oppenheimer-Volkoff solutions 
for each. 
(2) Next, we solve the 6 coupled elliptic equations
in Eqs.~(\ref{hamCTS}), (\ref{momCTS}), (\ref{alpha_CTS})
and (\ref{continuity4}) for this given $h$.
(3) We adjust $\Omega$ and $x_{CM}^1$ using the force balance
Eq.~(\ref{forcebalance}).
(4) Then we use Eq.~(\ref{h_from_Euler}) to update $h$ in each star.
The constant $C$ has a different value for each star.
We adjust the value for each star such that it keeps a prescribed rest
mass. 
After updating $h$ we go back to step (2) and iterate until
all equations are satisfied up to a given tolerance.

Notice however, that sometimes other conditions are used instead of the
force balance Eq.~(\ref{forcebalance}). As first found
in~\cite{Taniguchi:2010kj}, for massive stars in can be helpful
to determine $x_{CM}^1$ by demanding that the ADM linear momentum be
zero~\cite{Foucart:2010eq,Henriksson:2014tba,Dietrich:2015pxa,Haas:2016cop}.

%%%%%%%%%%%%%%%%%%%%%%%%%%%%%%%%%%%%%%%%%%%%%%%%%%%%%%%%%%%%%%%%%%%   
\subsection{General orbits and low eccentricity orbits}
%%%%%%%%%%%%%%%%%%%%%%%%%%%%%%%%%%%%%%%%%%%%%%%%%%%%%%%%%%%%%%%%%%% 

If we evolve the quasi-circular (i.e. with $v_r=0$ in Eq.~(\ref{inspiral_KV}))
binary neutron star initial data described in the previous sections, we find
that the orbits are roughly circular inspiral orbits. However, as in the
case of black holes (see e.g. Fig.~\ref{puncture_distance_evo}), the
distance between the stars exhibits small oscillations, which are a sign of
residual eccentricity caused by our assumption of exactly circular orbits.
As for black holes, we would like to adjust $\Omega$ and $v_r$ in
Eq.~(\ref{inspiral_KV}), to obtain less eccentric orbits when we evolve such
initial data. There is, however, one immediate problem. The value of $\Omega$
is determined by the force balance Eq.~(\ref{forcebalance}), and thus cannot
be readily changed. 

In~\cite{Moldenhauer:2014yaa} a method has been developed that lets us
construct orbits with arbitrary eccentricity. The extra parameter $e$
describing this eccentricity can then be adjusted along with $v_r$ to find
low eccentricity orbits when the data are evolved. We now describe how the
symmetry vector $\xi^{\mu}$ can be modified to account for
non-circular orbits. We make the following two assumptions: 
(i) $\xi^{\mu}$ is along the motion of the star center.
(ii) Without inspiral, each star center moves along a segment of an 
elliptic orbit at apoapsis.
Since we only need a small segment of an elliptic orbit near apoapsis, we
will approximate this segment by the circle inscribed into the
orbit there. Then the radii of the inscribed circles are
\be
\label{eq:r_c_of_d}
r_{c{1/2}} = (1-e) d_{1/2} ,
\ee
where $d_1$ and $d_2$ are the distances of the particles from the
center of mass at apoapsis and $e$ is the eccentricity parameter for
the elliptic orbit~\cite{Moldenhauer:2014yaa}. These two inscribed circles
are centered on the points
\be
\label{eq:x_circlecenters}
x^1_{c{1/2}} = x^1_{C*1/2} \mp r_{c{1/2}} 
= x^1_{CM} + e (x^1_{C*1/2} - x^1_{CM}) , 
\ee
where we have used $d_{1/2} = |x^1_{C*1/2} - x^1_{CM}|$ 
and assumed that apoapsis occurs on the
$x^1$-axis. (The minus and plus sign in the middle term of 
Eq.~(\ref{eq:x_circlecenters}) corresponds to the
subscripts $1$ and $2$, respectively.)
Assumptions (i) and (ii) then tell us that the symmetry
vector for elliptic orbits with inspiral must have the form
\begin{eqnarray}
\label{ellinspiralKV}
\xi_{1/2}^0 &=& 1, \nonumber\\
\xi_{1/2}^i &=& \Omega (-x^2, x^1-x_{c1/2}^1, 0) +
\frac{v_r}{r_{12}} (x^i-x_{CM}^i) .
\end{eqnarray}
The eccentricity parameter $e$ that appears in $x^1_{c1/2}$ and the radial
velocity $v_r$ can now be freely chosen to obtain any orbit we want.
Using this new symmetry vector $\vec{\xi}_{1/2}$, we can still solve
the initial data equations with the same methods as described above.
In order to obtain a true inspiral orbit with low eccentricity, we have to
adjust both $e$ and $v_r$. A procedure (similar to the one for black holes
in~\cite{Pfeiffer:2007yz,Boyle:2007ft}) to minimize the eccentricity has been
developed in~\cite{Dietrich:2015pxa}.
A similar approach has been also been applied 
in~\cite{Kyutoku:2014yba} and \cite{Haas:2016cop} for purely 
irrotational stars. Notice, however, that in~\cite{Kyutoku:2014yba} and
\cite{Haas:2016cop} only quasi-circular orbits are considered and
$\Omega$ and $v_r$ are directly adjusted. 
Similarly~\cite{Tacik:2015tja} considers quasi-circular orbits using the
methods for spinning stars discussed above and also directly adjusts
$\Omega$ and $v_r$.

As we have seen, it is possible to construct binary neutron star initial
data for spinning stars on arbitrary orbits. Nevertheless, due the many
technical complications discussed above, other simpler approaches are
sometimes still used. For example, one can use simple superpositions of
single stars without solving the constraints and the matter
equations~\cite{Gold:2011df,Tsatsin:2013jca}, or possibly with only the
constraint equations solved~\cite{East:2012zn}, or one can modify the
velocity field of irrotational stars to introduce
spin~\cite{Kastaun:2013mv}. Of course such simple methods introduce
additional errors that are not well understood.

%%%%%%%%%%%%%%%%%%%%%%%%%%%%%%%%%%%%%%%%%%%%%%%%%%%%%%%%%%%%%%%%%%%
\section{Initial data for black hole - neutron star binaries}
%%%%%%%%%%%%%%%%%%%%%%%%%%%%%%%%%%%%%%%%%%%%%%%%%%%%%%%%%%%%%%%%%%% 
\label{BH-NS_data}

Binaries made up of one black hole and one neutron
star can be constructed by combining the methods for black holes and neutron
stars discussed above.
As we have seen, there are a considerable number of choices on how one can
construct both black holes and neutron stars, thus there are many possible
combinations of these choices when it comes to mixed binaries.

%%%%%%%%%%%%%%%%%%%%%%%%%%%%%%%%%%%%%%%%%%%%%%%%%%%%%%%%%%%%%%%%%%%
\subsection{Simple approaches}
%%%%%%%%%%%%%%%%%%%%%%%%%%%%%%%%%%%%%%%%%%%%%%%%%%%%%%%%%%%%%%%%%%% 

One of the simplest approaches is used in~\cite{Baumgarte:2004}, where the
black hole is assumed to be much more massive than the neutron star. The
black hole is described by using the Kerr-Schild metric. The matter
equations and constraints are then only solved in a small patch around the
neutron star, which is assumed to be irrotational. The rest of the spacetime
is approximated to be given by the Schwarzschild metric.

As with neutron star binaries it is possible to use superpositions
of single neutron star and single black hole solutions. 
In~\cite{Stephens:2011as,East:2011xa} a boosted TOV star solution
is superimposed with a boosted black hole metric to obtain highly eccentric
orbits. This superposition solves the constraint equations only
approximately, and the fluid is not in true equilibrium.
Yet if the one starts from large separations the errors
due to a simple superposition are small.
This approach has also been extended~\cite{East:2012zn}
to using this superposition as free data
in the CTS decomposition and then solving the constraint 
Eqs.~(\ref{hamCTS}) and (\ref{momCTS}), but without solving the equilibrium
fluid equations. This method has recently also been generalized
to spinning neutron stars by using a solution for a single 
spinning star in the superposition~\cite{East:2015yea}.

%%%%%%%%%%%%%%%%%%%%%%%%%%%%%%%%%%%%%%%%%%%%%%%%%%%%%%%%%%%%%%%%%%%
\subsection{Non-spinning black hole - neutron stars binaries}
%%%%%%%%%%%%%%%%%%%%%%%%%%%%%%%%%%%%%%%%%%%%%%%%%%%%%%%%%%%%%%%%%%% 

Several other more comprehensive approaches exist for general mass ratios.
In~\cite{Grandclement:2006ht} the 3-metric is assumed to be conformally
flat. Both black hole and neutron star are assumed to be non-spinning. Then
the CTS Eqs.~(\ref{hamCTS}), (\ref{momCTS}), (\ref{alpha_CTS}), together
with the matter Eqs.~(\ref{Euler_int_irr}) and (\ref{continuity_irrot}) can
be solved to set up an irrotational neutron star. The black hole is
introduced by imposing the quasi-equilibrium apparent horizon boundary
conditions (\ref{psi_BC_BH}) and (\ref{beta_BC_BH})
on a coordinate sphere. The
$\Omega_r$ in Eq.~(\ref{beta_BC_BH}) is adjusted such that the black hole
spin is zero. Again an iteration procedure, as described above for neutron
star binaries, is used. During the iterations, the center of mass $x^i_{CM}$
is fixed by demanding that the ADM momentum $P_{i \ \infty}^{ADM}=0$, while
$\Omega$ is computed from the force balance Eq.~(\ref{forcebalance})
at the center of the star. Similar methods are also used
in~\cite{Taniguchi:2007xm}. The main difference is that $\Omega_r$ in
Eq.~(\ref{beta_BC_BH}) has the same magnitude as $\Omega$ so that the black
hole has only approximately zero spin. This approximation is later improved
in~\cite{Taniguchi:2007aq}. In~\cite{Taniguchi:2006yt} the same methods as
in~\cite{Taniguchi:2007xm} are used, except that the conformal metric is set
to be the Kerr-Schild metric.

All the approaches that use the quasi-equilibrium apparent horizon boundary
conditions result in initial data only outside the black hole. For this
reason a puncture based approach has been developed
in~\cite{Shibata:2006bs,Shibata:2006ks}. One assumes a conformally flat
metric as in Eq.~(\ref{conformally_flat}) and sets $K=0$. 
Then Eq.~(\ref{hamCTS}) can again be used with the puncture ansatz in
Eqs.~(\ref{psiPuncAnsatz}) and (\ref{psiBL}) with the black hole bare mass
$m_1$ and $m_2=0$, to find the conformal factor $\psi$. Similarly, 
we solve Eq.~(\ref{alpha_CTS}) for the lapse with the ansatz in
Eq.~(\ref{alphapsiPuncAnsatz}) with $m_2=0$. 
The extrinsic curvature is written as
\be
\bar{A}^{ij} = (\bar{L}W)^{ij} + \bar{A}^{ij}_{BY} ,
\ee
where $\bar{A}^{ij}_{BY}$ is the Bowen-York extrinsic curvature given in
Eq.~(\ref{Bowen-York1}) containing the black hole momentum (the spin is set
to zero). The latter equation corresponds to the CTT decomposition with
$\bar{\sigma}=1$ and $\bar{M}^{ij}=\bar{A}^{ij}_{BY}$.
We then use Eq.~(\ref{momCTT}) to find $W^i$, where we note that
$\bar{D}_j\bar{M}^{ij}=0=K$.
As noted in~\cite{Shibata:2006bs,Shibata:2006ks},
the fluid equations contain the shift, which we thus need to find.
The shift is computed by rewriting
the momentum constraint (\ref{momCTS}) of the CTS decomposition as
\be
\bar{D}_j(\bar{L}\beta)^{ij} - 2\bar{A}^{ij}_{BY}\bar{D}_j\bar{\alpha}
= 16\pi\bar{\alpha}\psi^{10} j^i .
\ee
All terms in this equation are regular at the black hole center ($r_1=0$),
as are all terms in the other elliptic equations.
We can thus solve all elliptic equations for the metric variables
($u$ from Eq.~(\ref{psiPuncAnsatz}), $v$ from Eq.~(\ref{alphapsiPuncAnsatz}),
$W^i$ and $\beta^i$) everywhere and do not need to excise any regions.
This approach has been used together with the fluid
Eq.~(\ref{Bernoulli0_corot}) for corotating
stars~\cite{Shibata:2006bs,Shibata:2006ks} and also with
Eqs.~(\ref{Euler_int_irr}) and (\ref{continuity_irrot}) for irrotational
stars~\cite{Kyutoku:2009sp}. As in other approaches, $\Omega$
is determined from the force balance Eq.~(\ref{forcebalance}). The
black hole momentum parameter $P^i$ in $\bar{A}^{ij}_{BY}$ is picked such
that $P_{i \ \infty}^{ADM}=0$. In order to find the center of mass
$x^i_{CM}$, \cite{Kyutoku:2009sp} chooses $J_{ \infty}^{ADM}$
such that it corresponds to the post-Newtonian angular momentum.

%%%%%%%%%%%%%%%%%%%%%%%%%%%%%%%%%%%%%%%%%%%%%%%%%%%%%%%%%%%%%%%%%%%
\subsection{Spinning black hole - neutron stars binaries}
%%%%%%%%%%%%%%%%%%%%%%%%%%%%%%%%%%%%%%%%%%%%%%%%%%%%%%%%%%%%%%%%%%% 

In order to deal with highly spinning black holes it is useful to again
consider a conformal metric that is of Kerr-Schild form in the vicinity of
the black hole. This has been done~\cite{Foucart:2008qt,Foucart:2010eq}
within the CTS approach. Following~\cite{Foucart:2008qt,Foucart:2010eq}, we
again assume that all time derivatives such $\bar{u}_{ij}$ and $\partial_t
K$ can be set to zero. We then have to specify the conformal 3-metric and
the trace of the extrinsic curvature. They are set to
\begin{eqnarray}
\label{KS_gamma1}
\bar{\gamma}_{ij} &=& \delta_{ij} +
e^{-r_1^4/w^4}({\gamma}^1_{ij} - \delta_{ij}) \\
\label{KS_K1}
K &=& e^{-r_1^4/w^4} K^1 ,
\end{eqnarray}
where ${\gamma}^1_{ij}$ and $K^1$ are the the conformal 3-metric and the
trace of the extrinsic curvature of a boosted spinning Kerr-Schild 
black hole. The boost velocity is given by
\be
v^i_{BH} = \Omega \epsilon^{i3k} c^k_{BH} ,
\ee
where $c^i_{BH}$ is coordinate location of the black hole 
center with respect to the center of mass. The weighting factor $w$ 
is chosen such that the conformal metric is flat at the neutron star.
The exponential damping of the form $e^{-r_1^4/w^4}$ used here
is necessary~\cite{Foucart:2008qt} to avoid the deviations from
quasi-equilibrium reported in~\cite{Taniguchi:2007xm}.
When then solve the CTS Eqs.~(\ref{hamCTS}), (\ref{momCTS}) and
(\ref{alpha_CTS}) together with e.g. the fluid Eqs.~(\ref{Euler_int_irr})
and (\ref{continuity_irrot}) for an irrotational neutron star.

We assume that the approximate symmetry vector is given by
\begin{eqnarray}
\label{GT_inspiral_KV}
\xi_{sym}^0 &=& 1, \nonumber\\
\xi_{sym}^i &=& \Omega (-x^2+x_{CM}^2, x^1-x_{CM}^1, 0) +
\frac{v_r}{r_{12}} (x^i-x_{CM}^i) + v^i_{GT} ,
\end{eqnarray}
where we have added the term $v^i_{GT}$ corresponding to a Galilean
transformation of the helical inspiral vector, in order take into
account the fact that the system may move since it radiates linear momentum.
Again we assume that orbital angular velocity is along the $x^3$-direction.
Here $v_r$ is the radial velocity, $x_{CM}^i$ the center of mass position and
$r_{12}$ the distance between the black hole and the neutron star.
In comoving coordinates we still have 
$\xi_{sym}^{\mu} = t^{\mu} = \alpha n^{\mu} + \beta^{\mu}$,
so that at spatial infinity the shift now must have the form
\be
\label{beta_BC_inf_sym}
\lim_{r\to\infty}\beta^{i} = \xi_{sym}^i .
\ee
As boundary conditions for $\psi$ and $\alpha$ we again use
Eqs.~(\ref{psi_BC_inf}) and (\ref{alphabar_BC_inf}) at infinity. 

The apparent horizon $\cal{S}$ of the black hole is chosen to be the
coordinate location of the event horizon of the boosted black hole in the
superposition. For the conformal factor and the shift Eqs.~(\ref{psi_BC_BH})
and (\ref{beta_BC_BH}) are used as boundary conditions on $\cal{S}$. The
lapse boundary condition there is given by
\be
\alpha |_{\cal{S}}
= e^{-r_1^4/w^4} \alpha^1 |_{\cal{S}} ,
\ee
where $\alpha^1$ is the lapse of the boosted black hole.

This set of equations again has to be solved iteratively. When low
eccentricity orbits are desired, $\Omega$ and $v_r$ are computed by an
eccentricity reduction procedure as in~\cite{Pfeiffer:2007yz,Boyle:2007ft}.
We can thus not use the force balance equation (\ref{forcebalance}) to
center the star. Rather we fix the positions of both star and black hole
centers (say at given locations on the $x^1$-axis), and then
translate the specific enthalpy $h$ computed in each iteration
so that its maximum occurs at the fixed star center.
Finally, $P_{i \ \infty}^{ADM}=0$ is enforced by adjusting $v^i_{GT}$.
This is called boost control in~\cite{Henriksson:2014tba} and is
particularly useful for very compact stars. The details of this procedure
can be found in~\cite{Foucart:2008qt,Foucart:2010eq,Henriksson:2014tba}.

With this approach it is possible to construct initial data for neutron
stars that are orbiting highly spinning black holes.
So far this has only been done for irrotational stars, but the approach
could easily be generalized to spinning stars using the methods explained 
in Sec.~\ref{BNS_with_spin}.

%%%%%%%%%%%%%%%%%%%%%%%%%%%%%%%%%%%%%%%%%%%%%%%%%%%%%%%%%%%%%%%%%%%
\section{Numerical codes used to construct initial data for binaries}
%%%%%%%%%%%%%%%%%%%%%%%%%%%%%%%%%%%%%%%%%%%%%%%%%%%%%%%%%%%%%%%%%%% 
\label{Codes}

Many different computer codes have been used over the years to compute
initial data. Here we only mention the ones that are currently 
most widely used.

% LORENE

One of the most widely used codes is the publicly available \textsc{Lorene}
library~\cite{Grandclement-etal-2000:multi-domain-spectral-method}.
\textsc{Lorene} provides a domain decomposition into several spherical-like
shells. In each shell, spectral methods in spherical coordinates are employed
to solve the relevant elliptic equations. Spectral expansions in terms of
spherical harmonics and Chebyshev polynomials are used in the angular
coordinates and the radial coordinate, respectively. In order to impose
boundary conditions at infinity the outermost shell is compactified by using
the inverse of the radius as a coordinate. The shells can be deformed so
that a neutron star surface can be at a domain boundary, which allows for
easy imposition of surface boundary conditions. For a binary, two sets of
domains are used, that are centered around each compact object. These
domains are embedded in a larger domain that extends to infinity. The
\textsc{Lorene} library has been used in several codes. Among them are the
codes that were used to construct binary black hole
data~\cite{Grandclement02}, binary neutron star
data~\cite{Bonazzola:1998yq,Gourgoulhon:2000nn,Taniguchi:2002ns,
Taniguchi:2003hx,Taniguchi:2010kj} and also mixed
binaries~\cite{Grandclement:2006ht,Taniguchi:2006yt,Taniguchi:2007aq}.

% Ansorg's punc solver

Most numerical simulations starting from binary black hole initial data of
puncture type, use the spectral puncture solver described
in~\cite{Ansorg:2004ds}. This method uses a coordinate transformation that
maps all of space into a finite cube. After mapping, the punctures are
located along two parallel edges of one of the cube faces, while spatial
infinity is mapped to the opposing face. This means that in the interior
of the domain $\psi$ will be $C^{\infty}$. Equation~(\ref{hamPunc})
is then solved by means of a pseudo-spectral method, where expansions
in Chebyshev and Fourier polynomials are used. This solver is publicly
available as the \textsc{TwoPunctures} thorn of the
\textsc{Einstein Toolkit}~\cite{Loeffler2012}.

% Spells

Many initial data sets for binary black
holes~\cite{Caudill:2006hw,Pfeiffer:2007yz,Lovelace:2008tw,Buchman:2012dw}
mixed binaries~\cite{Foucart:2010eq,Henriksson:2014tba} and neutron star
binaries~\cite{Tacik:2015tja,Haas:2016cop} have been computed using the
\textsc{spells} framework~\cite{Pfeiffer:2002wt}. In this framework space is
covered by computational domains that are made up of hexahedra, spherical
shells, and cylindrical shells. Within each domain appropriate spectral
expansions are used. For example, Chebyshev polynomials are used in all
three directions in hexahedra, spherical harmonics are used in the angular
directions of shells and Chebyshev polynomials the radial direction, Fourier
polynomials are used in the angular direction of cylindrical shells. The
domains can be distorted by coordinate transformations, so that usually
grids with domains are used that touch but do not overlap. For neutron
stars, the star surface is put at a domain boundary.

% SGRID

The first evolutions of binary neutron stars with realistic
spins~\cite{Bernuzzi:2013rza,Dietrich:2015pxa} were performed using initial
data computed with the \textsc{SGRID}
code~\cite{Tichy:2009yr,Tichy:2012rp,Dietrich:2015pxa}. This code uses
coordinate transformations~\cite{Ansorg:2006gd} that compactify all of the
space outside the neutron stars into two cubical domains. In each of these
domains the star surface is one of the cube faces and spatial infinity is
mapped into an edge of the opposing face. Two more domains are introduced
for each star that cover the inside of each star. The outer one of these is
again cubical with the star surface located on one of the faces. The
coordinate transformations contain adjustable functions so that arbitrary
star surfaces are possible. As long as the matter distribution is
$C^{\infty}$ inside each domain, the solutions of elliptic equations such as
Eqs.~(\ref{hamCTS}), (\ref{momCTS}), (\ref{alpha_CTS}), and
(\ref{continuity4}) are expected to be $C^{\infty}$ as well. In
\textsc{SGRID} these equations are thus solved by means of a pseudo-spectral
method, which converges exponentially if the solutions are $C^{\infty}$.
This code has also been used to construct the post-Newtonian based 
binary black initial data studied in~\cite{Reifenberger2013}.

% William East's code:

The codes mentioned up to here use spectral methods which tend to be
more efficient than finite difference methods due to their faster rate of
convergence when the computed solutions are $C^{\infty}$.
However, the simulations of binary neutron stars and mixed
binaries on highly eccentric orbits in~\cite{East:2015yea} have
used a finite differencing code~\cite{East:2012zn} to construct the initial
data. This code uses a full approximation storage implementation of the
multigrid algorithm with adaptive mesh refinement as described
in~\cite{Pretorius:2005ua}.

% COCAL - Compact Object CALculator from Uryu

Recently the so called \textsc{COCAL} code has been
developed~\cite{Uryu:2011ky,Tsokaros:2012kp,Uryu:2012uh,Tsokaros:2015fea}.
It is different from the codes mentioned before and is based on the
Komatsu-Eriguchi-Hachisu method~\cite{Komatsu89} for computing equilibrium
spinning neutron star models. In this method elliptic equations are solved
by using Green's functions for the Laplace operator. Any additional or
non-linear terms in the equations are added to the source terms, and the
equations are then solved iteratively. The \textsc{COCAL} code uses several
overlapping spherical shells. The Green's functions include terms for
different boundary conditions. The integrals over the Green's functions is
performed numerically by expanding into spherical harmonics. A comparison of
\textsc{COCAL} and \textsc{Lorene} in~\cite{Tsokaros:2015fea} shows that
\textsc{COCAL} gives the same results up to a difference of 0.05\%.

%%%%%%%%%%%%%%%%%%%%%%%%%%%%%%%%%%%%%%%%%%%%%%%%%%%%%%%%%%%%%%%%%%%
\section{Conclusions}
%%%%%%%%%%%%%%%%%%%%%%%%%%%%%%%%%%%%%%%%%%%%%%%%%%%%%%%%%%%%%%%%%%%
\label{Conclusions}

The construction of initial data in General Relativity is a well established
field of study these days. As we have seen, such initial data have to
satisfy both the Hamiltonian and momentum constraints in Eqs.~(\ref{ham0})
and (\ref{mom0}). In order to solve these constraints, conformal
decompositions have been developed, that result in elliptic equations which
can be solved numerically. The two most common approaches are the CTT and
CTS decompositions discussed in Secs.~\ref{CTT_decomp} and \ref{CTS_decomp}.
In both cases one has to specify certain free data such as the conformal
metric and boundary conditions. Once specified one can then construct
constraint satisfying initial data. As we have seen both decompositions can
be and have been used to construct various types of initial data. The answer
to the question of which one is preferable depends on what one tries to
achieve. If we have a way to supply approximate initial data (e.g. using
post-Newtonian input) that almost satisfy the constraints one can use either
the CTT or CTS decomposition to construct constraint satisfying initial
data. In this case it does not matter much which decomposition one chooses,
except maybe that the CTT equations are slightly simpler. On the other hand,
if one tries to construct binary initial data solely from quasi-equilibrium
conditions (e.g. for two objects on quasi-circular orbits), it is better to
use the CTS decomposition because it allows us to directly specify certain
time derivatives (which would be zero for quasi-equilibrium). This is why
this method is used for most types of binary neutron star initial data, and
also for many binary black hole initial data.

% should I discuss this?: 
%list of best ini dat??? <- no
%astrophysical realism
%adding waves

The main ideas for constructing binary black hole or binary neutron star
data are very similar. However, neutron stars contain matter and cannot
simply be modeled by certain boundary conditions on the black hole horizon.
This fact adds extra complications that make the actual construction of
initial data containing neutron stars more challenging. More equations have
to be solved and the numerical iteration procedures become more complex
because the star surfaces change during the iterations. Nevertheless, the
quality (in terms of astrophysical realism) of current binary initial data
containing one or two neutron stars is just as good as the quality of binary
black hole initial data. One possible exception is the addition of
gravitational waves to the initial data, which thus far has only been
carried out for binary black hole initial data. This addition, however,
could in principle be done for neutron stars with the same methods as for
black holes, e.g. by using post-Newtonian input.

%%%%%%%%%%%%%%%%%%%%%%%%%%%%%%%%%%%%%%%%%%%%%%%%%%
%%%%%%%%%%%%%%%%%%%%%%%%%%%%%%%%%%%%%%%%%%%%%%%%%%

%\begin{acknowledgments}
\ack

The equations for neutron stars with arbitrary spins were first derived
in~\cite{Tichy:2011gw} within the 3+1 split. The expressions one has to deal
with in this derivation are rather long. The author wishes to credit
Charalampos M. Markakis with the more elegant derivation in terms of
4-vectors and one-forms that is presented in Sec.~\ref{BNS_with_spin} here.

The author is grateful to George Reifenberger for allowing to include a plot
from his thesis~\cite{Reifenberger2013} as Fig.~\ref{junk_in22} of this
article.
The author also wishes to thank Sebastiano Bernuzzi for providing
Fig.~\ref{CRVexample} for this article.

This work was supported by NSF grant PHY-1305387

%We also acknowledge TACC at UT Austin for providing
%HPC resources under allocation TG-PHY080019.

%\end{acknowledgments}

%%%%%%%%%%%%%%%%%%%%%%%%%%%%%%%%%%%%%%%%%%%%%%%%%%
\appendix
%%%%%%%%%%%%%%%%%%%%%%%%%%%%%%%%%%%%%%%%%%%%%%%%%%

%%%%%%%%%%%%%%%%%%%%%%%%%%%%%%%%%%%%%%%%%%%%%%%%%%
\section{Lie derivatives}
\label{Lie-appendix}
%%%%%%%%%%%%%%%%%%%%%%%%%%%%%%%%%%%%%%%%%%%%%%%%%%

We collect here some useful equations about Lie derivatives.

%%%%%%%%%%%%%%%%%%%%%%%%%%%%%%%%%%%%%%%%%%%%%%%%%%
\subsection{Lie derivative of a tensor}
%%%%%%%%%%%%%%%%%%%%%%%%%%%%%%%%%%%%%%%%%%%%%%%%%%

Let $T^{\alpha...}_{\ \ \ \ \beta...}$ be a tensor with any number of
indices. The Lie derivative of this tensor with respect to the
vector $v^{\mu}$ is then given by
\be
\label{Lie_v_T}
\pounds_{v} T^{\alpha...}_{\ \ \ \ \beta...} =
v^{\mu} \rhd_{\mu} T^{\alpha...}_{\ \ \ \ \beta...} -
T^{\mu...}_{\ \ \ \ \beta...} \rhd_{\mu} v^{\alpha} - ... +
T^{\alpha...}_{\ \ \ \ \mu...} \rhd_{\beta} v^{\mu} + ... ,
\ee
where $\rhd_{\mu}$ is any derivative operator such 
as $\partial_{\mu}$ or the covariant derivative $\nabla_{\mu}$.

Applied to the metric (using $\rhd_{\mu}=\nabla_{\mu}$) we thus obtain
\be
\pounds_{v} g_{\mu\nu} = \nabla_{\mu}v_{\nu} + \nabla_{\nu}v_{\mu} ,
\ee
which results in Killings equation $\nabla_{(\mu}v_{\nu)}=0$ 
if $g_{\mu\nu}$ has a symmetry such that $\pounds_{v} g_{\mu\nu}=0$.

%%%%%%%%%%%%%%%%%%%%%%%%%%%%%%%%%%%%%%%%%%%%%%%%%%
\subsection{Lie and covariant derivatives of a tensor density}
%%%%%%%%%%%%%%%%%%%%%%%%%%%%%%%%%%%%%%%%%%%%%%%%%%

Recall that for any invertible matrix $A$ we have
\be
\label{lndet_trln}
\ln|\det A| = \tr(\ln A) ,
\ee
where $\det$ and $\tr$ denote determinant and trace.
If we change $A$ by $\delta A$, its determinant $\det A$ thus changes
according to
\be
\label{dlndet}
\delta \ln|\det A| = \tr(A^{-1}\delta A) .
\ee
Applied to the metric $g_{\mu\nu}$ the latter can be rewritten as
\be
\label{delta_g}
\delta \sqrt{|g|} = \frac{1}{2}\sqrt{|g|} g^{\mu\nu}\delta g_{\mu\nu} ,
\ee
where $g=\det g_{\mu\nu}$.
We can use Eq.~(\ref{delta_g}) to find partial derivatives of $\sqrt{|g|}$
with the result
\be
\label{partial_g}
\partial_{\rho} \sqrt{|g|}
= \frac{1}{2}\sqrt{|g|} g^{\mu\nu}\partial_{\rho} g_{\mu\nu} .
\ee
We can also use Eq.~(\ref{delta_g}) to define the covariant derivative
and the Lie derivative of $\sqrt{|g|}$. These definitions yield
\be
\label{nabla_g}
\nabla_{\rho} \sqrt{|g|}
:= \frac{1}{2}\sqrt{|g|} g^{\mu\nu}\nabla_{\rho} g_{\mu\nu} = 0 .
\ee
and
\be
\label{Lie_v_g}
\pounds_v \sqrt{|g|}
:= \frac{1}{2}\sqrt{|g|} g^{\mu\nu}\pounds_v g_{\mu\nu}
=  \sqrt{|g|} \nabla_{\rho} v^{\rho} .
\ee

A tensor density of weight $w$ is defined as
\be
\mathfrak{T}^{\alpha...}_{\ \ \ \ \beta...} 
:= |g|^{w/2} T^{\alpha...}_{\ \ \ \ \beta...} ,
\ee
where $T^{\alpha...}_{\ \ \ \ \beta...}$ is a tensor.
From Eq.~(\ref{nabla_g}) we immediately find that
\be
\nabla_{\rho}\mathfrak{T}^{\alpha...}_{\ \ \ \ \beta...}
= |g|^{w/2} \nabla_{\rho} T^{\alpha...}_{\ \ \ \ \beta...}
= |g|^{w/2} \nabla_{\rho} 
  (|g|^{-w/2}\mathfrak{T}^{\alpha...}_{\ \ \ \ \beta...}) .
\ee
From Eq.~(\ref{Lie_v_g}) we obtain
\be
\label{Lie_tensordensity0}
\pounds_v\mathfrak{T}^{\alpha...}_{\ \ \ \ \beta...}
= |g|^{w/2} \pounds_v T^{\alpha...}_{\ \ \ \ \beta...} +
  w |g|^{w/2} T^{\alpha...}_{\ \ \ \ \beta...} \nabla_{\rho} v^{\rho} . 
\ee
Using Eq.~(\ref{Lie_v_T}) we can rewrite the latter as
\begin{eqnarray}
\label{Lie_tensordensity}
\pounds_v\mathfrak{T}^{\alpha...}_{\ \ \ \ \beta...}
&=& 
v^{\mu} \rhd_{\mu} \mathfrak{T}^{\alpha...}_{\ \ \ \ \beta...} -
\mathfrak{T}^{\mu...}_{\ \ \ \ \beta...} \rhd_{\mu} v^{\alpha} - ... +
\mathfrak{T}^{\alpha...}_{\ \ \ \ \mu...} \rhd_{\beta} v^{\mu} + ...
\nonumber\\ 
&&+
w \mathfrak{T}^{\alpha...}_{\ \ \ \ \beta...} \rhd_{\mu}v^{\mu} ,
\end{eqnarray}
where $\rhd_{\mu}$ stands for either $\nabla_{\mu}$ or $\partial_{\mu}$.
From Eq.~(\ref{Lie_tensordensity0}) we see that 
$\pounds_v\mathfrak{T}^{\alpha...}_{\ \ \ \ \beta...}$ is again
a tensor density of weight $w$.

%%%%%%%%%%%%%%%%%%%%%%%%%%%%%%%%%%%%%%%%%%%%%%%%%%
\subsection{Why symmetries are best expressed in terms of Lie derivatives}
%%%%%%%%%%%%%%%%%%%%%%%%%%%%%%%%%%%%%%%%%%%%%%%%%%
\label{Why-LieDerivs}

If a tensor density has a symmetry along some vector $\vec{\xi}$,
a coordinate system exists such that this symmetry is along one of the
coordinates lines, say the coordinate lines along which only the $x^0$
coordinate varies. In these coordinates we have
\be
\frac{\partial}{\partial x^0}\mathfrak{T}^{\alpha...}_{\ \ \ \ \beta...} = 0
\ee
if $\mathfrak{T}^{\alpha...}_{\ \ \ \ \beta...}$ has this symmetry.
Now note that in these coordinates 
\be
\vec{\xi} = \frac{\partial}{\partial x^0}
\ee
which implies
\be
\xi^{\alpha} = (1,0,...) ,
\ee
so that
\be
\partial_{\mu} \xi^{\alpha} = 0 .
\ee
Using the latter together with $\rhd_{\mu}=\partial_{\mu}$ 
in Eq.~(\ref{Lie_tensordensity}), we obtain
\be
\pounds_{\xi}\mathfrak{T}^{\alpha...}_{\ \ \ \ \beta...}
=\xi^{\mu} \partial_{\mu} \mathfrak{T}^{\alpha...}_{\ \ \ \ \beta...} + 0
=\frac{\partial}{\partial x^0} \mathfrak{T}^{\alpha...}_{\ \ \ \ \beta...}
=0 .
\ee
Thus the the Lie derivative of any tensor density vanishes if it has a
symmetry in the direction of $\vec{\xi}$. This gives us a coordinate
independent criterion we can use to describe symmetries.

%%%%%%%%%%%%%%%%%%%%%%%%%%%%%%%%%%%%%%%%%%%%%%%%%%
\subsection{Useful Lie derivatives in the 3+1 formalism}
%%%%%%%%%%%%%%%%%%%%%%%%%%%%%%%%%%%%%%%%%%%%%%%%%%
\label{LieDerivs_in_3+1}

From Eqs.~(\ref{tvecdef}) and (\ref{Lie_v_T}) using 
$\rhd_{\mu}=\partial_{\mu}$ it follows that
\be
\label{Lie_t}
\pounds_t T^{\alpha...}_{\ \ \ \ \beta...} 
= \partial_t T^{\alpha...}_{\ \ \ \ \beta...} .
\ee

In order to derive Eq.~(\ref{K-of-g}) it is useful to consider
\be
\pounds_{\alpha n} \gamma_{\mu\nu}
= \alpha n^{\rho}\rhd_{\rho} \gamma_{\mu\nu} +
\gamma_{\rho\nu}\rhd_{\mu}(\alpha n^{\rho}) +
\gamma_{\mu\rho}\rhd_{\nu}(\alpha n^{\rho})
= \alpha\pounds_n \gamma_{\mu\nu} ,
\ee
where we have used $\gamma_{\rho\nu}n^{\rho}=0$.
Thus 
\be
K_{\mu\nu} = -\frac{1}{2\alpha}\pounds_{\alpha n}\gamma_{\mu\nu}
= -\frac{1}{2\alpha}
  (\pounds_{t}\gamma_{\mu\nu}-\pounds_{\beta}\gamma_{\mu\nu})
\ee
which results in Eq.~(\ref{K-of-g}), once we use
\be
\pounds_{\beta}\gamma_{ij} = D_i \beta_j + D_j \beta_i 
\ee
and Eq.~(\ref{Lie_t}).

%%%%%%%%%%%%%%%%%%%%%%%%%%%%%%%%%%%%%%%%%%%%%%%

%%%%%%%%%%%%%%%%%%%%%%%%%%%%%%%%%%%%%%%%%%%%%%%%%%
\section{Elliptic Equations}
\label{Elliptic-appendix}
%%%%%%%%%%%%%%%%%%%%%%%%%%%%%%%%%%%%%%%%%%%%%%%%%%

We mention here some properties of elliptic equations that are important
for the construction of initial data.

%%%%%%%%%%%%%%%%%%%%%%%%%%%%%%%%%%%%%%%%%%%%%%%%%%
\subsection{Classification of second order partial differential equations}
%%%%%%%%%%%%%%%%%%%%%%%%%%%%%%%%%%%%%%%%%%%%%%%%%%

Consider a second order partial differential equation for the function
$u=u(x^l)$ of the form
\be
a^{ij} \partial_i \partial_j u + F(x^l, u, \partial_l u) = 0 ,
\ee
where $F$ is a function of $x^l$, as well as $u$ and its first derivatives,
and the matrix $a^{ij}$ is a function of $x^l$ and possibly $u$.
When $a^{ij}$ is independent of $u$ and $F$ is linear in $u$ the partial
differential equation is called linear.
If $a^{ij}$ also depends on $u$ but $F$ is linear in $u$ the equation is
sometimes called quasi-linear. If $F$ is not linear in $u$ the equation is
called non-linear.

The term $a^{ij} \partial_i \partial_j u$ with the highest
derivatives is referred to as the principal part of the equation, and
determines the type of the equation.
If the eigenvalues of $a^{ij}$ are all positive or all negative
for all $x^l$, the equation is called {\it elliptic}. 
If one or more eigenvalues are zero, the equation is called
{\it parabolic}. If no zero eigenvalues exist, but some eigenvalues are
positive while others are negative the equation is called {\it hyperbolic}.

The definition for elliptic equations can be extended to systems of
equations of the form
\be
a^{ij}_{AB} \partial_i \partial_j u^B + F_A(x^l, u^C, \partial_l u^C) = 0 ,
\ee
where $u^C$ now stands for a vector of functions.
The system of equations is called {\it elliptic} if~\cite{mclean2000strongly}
\be       
a^{ij}_{AB} \xi_i \xi_j \geq c |\xi_i|^2 ,
\ee
for all $\xi_i$, all $x^l$ and some constant $c>0$.

%%%%%%%%%%%%%%%%%%%%%%%%%%%%%%%%%%%%%%%%%%%%%%%%%%
\subsection{Finding solutions of elliptic equations}
%%%%%%%%%%%%%%%%%%%%%%%%%%%%%%%%%%%%%%%%%%%%%%%%%%

Many useful theorems about the existence and uniqueness of solutions, as
well as how one can construct solutions, are known for linear elliptic
equations (see e.g.~\cite{mclean2000strongly,evans2010partial}).
Unfortunately, much less is known about strongly coupled non-linear systems
of elliptic equations, such as the the CTS system given by
Eqs.~(\ref{hamCTS}), (\ref{momCTS}) and (\ref{alphabar_CTS}). For example,
in~\cite{Pfeiffer:2005jf} it is demonstrated that there are cases where this
system does not admit one unique solution, but instead has several branches
of solutions. It is beyond the scope of this article to detail the
conditions under which the CTT or CTS systems are known to have unique
solutions, and under which conditions not much is known. Note, however, that
some of these conditions are listed in~\cite{Bartnik:2002cw}, and that
important results can be found in~\cite{Cantor1977,Cantor81,Maxwell:2003av}.

Here we follow a more practical approach and simply note that in most cases
of interest to us, we can numerically construct solutions to both the CTT
and CTS systems. In order to do so we use the same methods as for linear
elliptic equations, i.e. we treat the elliptic equations as boundary value
problems. This means that in addition to the elliptic equations, we also have
to supply boundary conditions at the boundaries of the domain in which we
want to solve the elliptic equations. Only then can we obtain unique
solutions. These conditions are usually obtained from physical
considerations and are imposed at spatial infinity, at black hole horizons,
or at star surfaces. For computational reasons one sometimes further divides
the domain of interest into several adjacent computational domains. In this
case one also has to specify boundary conditions at these domain boundaries.
Since these boundaries are artificial, one simply demands continuity of the
functions $u^C$ and their normal derivatives across such computational
domain boundaries.

%%%%%%%%%%%%%%%%%%%%%%%%%%%%%%%%%%%%%%%%%%%%%%%
\vskip 1cm

%%%%%%%%%%%%%%%%%%%%%%%%%%%%%%%%%%%%%%%%%%%%%%%
% REFERENCES
%%%%%%%%%%%%%%%%%%%%%%%%%%%%%%%%%%%%%%%%%%%%%%%

\bibliographystyle{unsrt}
\bibliography{references}

\end{document}